%% file: semiblind0210-17pages-without-color.tex
\documentclass[10pt,journal,twocolumn,nofonttune]{IEEEtran}
\usepackage[cmex10]{amsmath}
\usepackage{setspace}
\usepackage{times}
\usepackage{bbm}
\usepackage{threeparttable}
\usepackage{amssymb}
\usepackage{graphicx,dblfloatfix}
\usepackage{cite}
\usepackage{subeqnarray}
\usepackage{xcolor}
\usepackage{stmaryrd}
\usepackage{multirow}
\usepackage{algorithm}
\usepackage{algorithmic}
\usepackage{diagbox}
\usepackage{url}
\usepackage{pifont}
\usepackage{hyperref}
\usepackage{upgreek}
\usepackage{booktabs}
\usepackage{wasysym}

\newcommand{\nc}{\newcommand}
\nc{\bsm}{\boldsymbol}
\nc{\mbb}{\mathbb}
\nc{\mbs}{\mathbbmss} 
\nc{\mbf}{\mathbf}
\nc{\mcl}{\mathcal}

\newcommand{\xiaowuhao}{\fontsize{9pt}{\baselineskip}\selectfont}

\input{defns.tex}

\newtheorem{pro}{Proposition}
\newtheorem{asum}{Assumption}

\begin{document}

\title{Semi-Blind Cascaded Channel Estimation for Reconfigurable Intelligent Surface \\ Aided Massive MIMO} 

\author{Zhen-Qing He, Hang Liu, Xiaojun Yuan\IEEEmembership{, Senior Member, IEEE}, Ying-Jun Angela Zhang\IEEEmembership{, Fellow, IEEE}, and Ying-Chang Liang\IEEEmembership{, Fellow, IEEE}   \vspace{-0.5cm}
\thanks{Z.-Q. He, X. Yuan, Y.-C. Liang are with the Center for Intelligent Networking and Communications and also with the National Key Laboratory of Science and Technology on Communications, University of Electronic Science and Technology of China, Chengdu 611731, China (e-mail: \{zhenqinghe; xjyuan; ycliang\}@uestc.edu.cn).}
\thanks{H. Liu and Y.-J. A. Zhang are with the Department of Information Engineering, The Chinese University of Hong Kong, Shatin, New Territories, Hong Kong SAR (e-mail: \{lh117; yjzhang\}@ie.cuhk.edu.hk).}
}

\maketitle

\begin{abstract}
Reconfigurable intelligent surface (RIS) is envisioned to be a promising green technology to reduce the energy consumption and improve the coverage and spectral efficiency of massive multiple-input multiple-output (MIMO) wireless networks. In a RIS-aided MIMO system, the acquisition of channel state information (CSI) is important for achieving passive beamforming gains of the RIS, but is also challenging due to the cascaded property of the transmitter-RIS-receiver channel and the lack of signal processing capability of the passive RIS elements. The state-of-the-art approach for CSI acquisition in such a system is a pure training-based strategy that depends on a long sequence of pilot symbols. In this paper, we investigate semi-blind cascaded channel estimation for RIS-aided massive MIMO systems, in which the receiver simultaneously estimates the channel coefficients and the partially unknown transmit signal with a small number of pilot sequences. Specifically, we formulate the semi-blind cascaded channel estimation as a trilinear matrix factorization task. Under the Bayesian inference framework, we develop a computationally efficient iterative algorithm using the approximate message passing principle to resolve the trilinear inference problem. Meanwhile, we present an analytical framework to characterize the theoretical performance bound of the proposed approach in the large-system limit via the replica method developed in  statistical physics. Extensive simulation results demonstrate the effectiveness of the proposed semi-blind cascaded channel estimation algorithm.
\end{abstract}

\vspace{-6pt}
\section{Introduction}

\subsection{Motivation}



\IEEEPARstart{M}{assive} multiple-input multiple-output (MIMO) technology, a key component of the fifth-generation (5G) wireless communications systems, has received  considerable research interests in both academia and industry in recent years \cite{marzetta2016fundamentals,bjornson2016massive, huawei}. Although massive MIMO has huge potential to improve the spectral efficiency and boost the system throughput, the wireless propagation links between the transmitter and the receiver might suffer from deep fading and shadowing due to unfavorable propagation conditions such as in urban areas or indoor environments. More recently, reconfigurable intelligent surface (RIS) \cite{huang2018large, di2020smart}, also known as intelligent reflecting surface \cite{wu2018intelligent}, has emerged as a promising technology to improve the link reliability of MIMO wireless networks as it can artificially reconfigure the wireless propagation environment. In contrast to a conventional amplify-and-forward relay, a RIS consists of a large number of low-cost and passive elements without radio frequency chains and additional thermal noise. Owing to the advances of programmable metamaterials \cite{cui2014coding}, it has become feasible to reconfigure the phase shifts of RISs in real time. As such, the RIS-aided massive MIMO has been envisioned as a key enabler for the next generation wireless communications systems to enhance network coverage and achieve smart radio environments \cite{liaskos2018new, di2019smart, di2020smart}.

Specifically, the passive beamforming gain of RIS-aided MIMO systems can be achieved by adjusting the phase shifts of the RIS elements so that the desired signals can be added constructively at the receiver. Extensive recent researches have verified the capacity advantage of RIS-aided MIMO systems by passive beamforming design under various criteria \cite{wu2018intelligent, huang2018large, yan2019, zappone2020overhead}. However, to realize the full advantage of RISs, the availability of the channel state information (CSI) plays a critical role. As a nearly-passive device, RISs are generally not equipped with any radio frequency chains and are not capable of performing any baseband processing functionality. Consequently, the transmitter-RIS and RIS-recevier cascaded channels cannot be estimated via traditional MIMO channel estimation methods. Currently, the design of the CSI acquisition method in RIS-aided MIMO systems is still in its infancy and the state-of-art approaches (see e.g., \cite{mishra2019channel, he2019cascaded, wang2019channel, liu2019matrix, jensen2020optimal, wei2020channel, elbir2020deep, liu2020deep}) are pure training-based schemes that require prohibitively long training pilots. This motivates us to explore new channel estimation techniques that reliably estimate the RIS cascaded channels with a reduced number of pilot overhead.


\vspace{-8pt}
\subsection{Contributions}
In this paper, we investigate semi-blind cascaded channel estimation for the RIS-aided uplink massive MIMO system, where  the base station (BS), with the help of a RIS, is equipped with a
large number of antennas to communicate with multiple single-antenna users. In particular, the BS can simultaneously estimate the channel coefficients and the transmitted data using a small number of pilots. We aim to design a computationally efficient semi-blind channel estimation algorithm and characterize its asymptotic performance in the large-system limit. The main contributions of our work are summarized as follows.

\begin{enumerate}
  \item We formulate the semi-blind cascaded channel estimation problem in the RIS-aided MIMO system as a trilinear estimation problem, where the received signal consists of the product of the RIS-BS channel matrix, the user-RIS channel matrix, and the transmit data matrix. Additionally, unlike the existing training-based works \cite{mishra2019channel, he2019cascaded, wang2019channel, liu2019matrix, jensen2020optimal, wei2020channel, elbir2020deep, liu2020deep} that estimate the channel coefficients and the transmitted signal separately, we consider a joint estimation of the both channels and the payload data with a small number of pilots.

  \item We develop a computationally efficient algorithm to approximately solve the trilinear estimation problem using the Bayesian minimum mean square error (MMSE) criterion. Specifically, we adopt the approximate message passing (AMP) framework \cite{rangan2011generalized,parker2014bilinear} to calculate the marginal posterior distributions of the channel coefficients and the unknown transmit signal. The proposed algorithm has low per-iteration complexity, since it only needs to update the means and variances of the messages and involves basic matrix-vector  multiplication.

  \item We analyse the asymptotic minimum mean square errors (MSEs) of the posterior mean estimators in the large-system limit by using the replica method developed in statistical physics \cite{SpinGlass, tanaka2002statistical, MP_BIGAMP3}. We show that with perfect knowledge of prior
      distributions of the channels and data, the MSEs of the posterior mean estimators can be determined by the fixed point of a set of scalar equations. Extensive numerical experiments under Rayleigh fading channels and quadrature phase-shift keying (QPSK) signal show that the proposed algorithm can well approach the asymptotic MSE bound, demonstrating the efficiency of the proposed message passing algorithm.

\end{enumerate}

\subsection{Related Work}

The cascaded channel estimation for RIS-aided MIMO systems has been investigated in a variety of recent works using different schemes. For example, in \cite{mishra2019channel} and \cite{he2019cascaded}, the on-off reflection pattern was used for the RIS cascaded channel estimation. Unlike \cite{mishra2019channel} using the linear minimum mean square error criterion, \cite{he2019cascaded} developed a two-stage approach that includes a sparse matrix factorization stage and a matrix completion stage. By appropriately designing the RIS reflection coefficients, \cite{wang2019channel} proposed a three-phase scheme based on the correlation among user-RIS-BS reflection channels of different users, whereas \cite{jensen2020optimal} proposed a Cram\'{e}r-Rao bound minimization based approach. Based on the channel sparsity, \cite{liu2019matrix} formulated the RIS cascaded channel estimation problem as a matrix-calibration based sparse matrix factorization problem. In \cite{wei2020channel}, the authors developed a PAEAllel FACtor (PARAFAC) decomposition based approach to unfold the cascaded channel model. Moreover, machine-learning-based RIS channel estimation methods were also introduced. For instance, \cite{elbir2020deep} designed a twin convolutional neural network architecture to estimate the RIS cascaded channels. In \cite{liu2020deep}, a deep denoising neural network was used to assist the estimation of the compressive RIS channels. Besides these, \cite{zheng2019intelligent} exploited the unit-modulus constraint to design a new reflection pattern at the IRS to aid the channel estimation of RIS-aided orthogonal frequency division multiplexing system with a single antenna both at the transmitter and the receiver.

The above mentioned works all belong to the training-based approach that needs a long duration of pilot sequences. In the training-based approach, each transmission frame is divided into two phases: a training phase and a data transmission phase. In the training phase, pilots are transmitted to facilitate the channel estimation at the receiver side. In the data transmission phase, unknown data are transmitted and the receiver performs detection based on the estimated channel in the training phase. Compared with the two separate phases for channel estimation and data detection, semi-blind channel estimation approach is able to improve the channel estimation performance because the estimated data can be utilized as soft pilots to enhance the channel estimation accuracy. This approach has proven successful in conventional massive MIMO semi-blind channel estimation \cite{wen2015bayes, prasad2015joint, nayebi2017semi}. Note that the semi-blind channel estimation problem in conventional massive MIMO systems is a special case of our work when the user-RIS-BS link is absent.

As a final remark, we note that the on-off reflection pattern design of the RIS in this work is similar to that in our prior work \cite{he2019cascaded}. However, this work significantly improves upon our prior work \cite{he2019cascaded}  in the following aspects. First, the direct and indirect link channels are estimated separately in \cite{he2019cascaded}, whereas we consider the joint estimation of both the direct and indirect link channels. Second, the work in \cite{he2019cascaded}  considers only the training-based channel estimation with all transmit data being as pilots by two separate stages: sparse matrix factorization and matrix completion. Instead, the current work treats only small partial transmit data being as pilots and simultaneously performs channel estimation and signal detection under the Bayesian MMSE criterion. The associated problem hence becomes a trilinear inference problem that is far more complicated than the bilinear inference problem in \cite{he2019cascaded}. Third, the asymptotic MMSE performance is analyzed in our work. Simulation results show that compared with the method in \cite{he2019cascaded}, the proposed algorithm has a significant performance improvement even with a much smaller number of pilots and is able to perform close to the asymptotic MMSE bound in various parameter settings.

\subsection{Organization and Notation}

The remaining sections are organized as follows. Sections \ref{Section-II} and \ref{Section-III} introduce the system model and the problem formulation, respectively. Section \ref{Section-IV} presents the semi-blind channel estimation algorithm, analyses the computationally complexity, and discusses the technical details. Section \ref{Section-V} elaborates the asymptotic performance bound by the replica method. Section \ref{Section-VI} evaluates the performance of the proposed algorithm via extensive numerical experiments. Finally, Section \ref{Section-VII} concludes the paper.

{\it Notation:} We use $\mbs{C}^{m \times n}$ to represent the space of $m \times n$ dimensional complex number. The superscripts ``$\ast$'',  ``$\mathsf{T}$'', and ``$\mathsf{H}$'' denote the conjugate, the transpose, and the conjugate transpose, respectively. For a vector $\bsm{x} \in \mbs{C}^n$, we use $x_i$ to represent its $i$-th element. For a matrix $\bsm{X} \in \mbs{C}^{m \times n}$, we use $x_{ij} $, $\bsm{x}_i^\mathsf{T} \in \mbs{C}^{n}$, $\bsm{x}_j \in \mbs{C}^m$, and ${\rm tr}(\bsm{X})$ to denote its $(i,j)$-entry, its $i$-th row, its $j$-th column, and its trace when $m=n$, respectively. For $x \in \mbs{C}$, $|x|$ and $\angle\,x $ denote its amplitude and phase, respectively. We use $\|\cdot\|_2$ and $\|\cdot\|_\mathsf{F}$ to denote the $\ell_2$-norm of a vector and the Frobenius norm of a matrix, respectively. We use $\bsm{x} \sim \mcl{N}(\bsm{x},\bsm{\mu}; \bsm{\Sigma})$ and $\bsm{x} \sim \mcl{CN}(\bsm{x},\bsm{\mu}; \bsm{\Sigma})$ denote that $\bsm{x}$ follows the real normal and complex circularly-symmetric normal distributions with mean $\bsm{\mu}$ and covariance $\bsm{\Sigma}$, respectively. In addition, we use $\varpropto$,  $\delta(\cdot)$, $\jmath \triangleq \sqrt{-1}$, $\bsm{0}$, $\bsm{I}_m$, $\rm{Re}(\cdot)$, $\odot $, $\mbs{E}\{\cdot\}$, and $\mbs{var}\{\cdot\}$ to represent the equality up to a constant multiplicative factor, the Dirac delta function, the imaginary unit, the all-zero vector/matrix with a proper size, the $m \times m$ identity matrix, the real operator, the element-wise multiplication, the expectation operator, and the variance operator, respectively.

\section{System Model}

\label{Section-II}

\begin{figure}[t]
    \centering
    \includegraphics[scale=0.8]{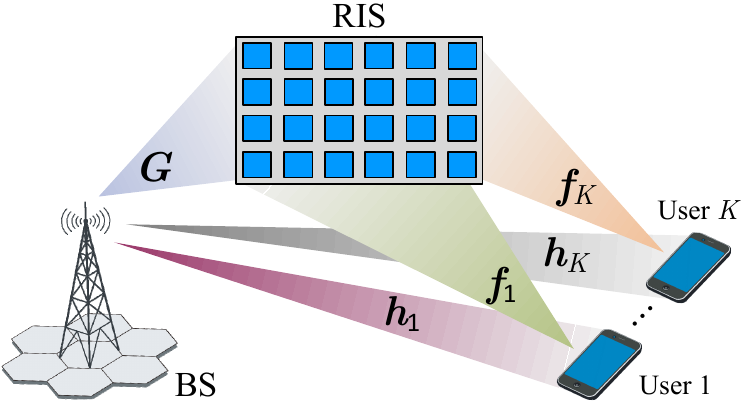}
    \vspace{-0.1cm}
    \caption{A RIS-aided massive MIMO system.}
    \label{fig1}
    \vspace{-0.01cm}
\end{figure}

We consider a RIS-aided uplink massive MIMO wireless system, as shown in Fig. \ref{fig1}, where the RIS consists of $N$ passive reflect elements and the BS with $M$ receive antennas serves $K$ single-antenna users. The RIS is deployed to enhance the communication quality between the users and the BS. We assume quasi-static block-fading channels with coherence time $T$, i.e., the channels remain approximately constant within each transmission block of length $T$. The baseband equivalent channels from the RIS to the BS, from the users to the RIS, and from the users to the BS are denoted by $\bsm{G} \in \mbs{C}^{M \times N} $, $\bsm{H} \triangleq [\bsm{h}_1,\ldots,\bsm{h}_K] \in \mbs{C}^{M \times K}$, and $\bsm{F} \triangleq [\bsm{f}_1,\ldots,\bsm{f}_K] \in \mbs{C}^{N \times K}$, respectively, where $\bsm{h}_k$ and $\bsm{f}_k$ denote the channel vectors from the $k$-th user to the RIS and to the BS, respectively. Under perfect carrier and timing recovery, we can model the discrete-time received signal at the BS as
\begin{align} \label{single-snap}
\bsm{y}^{(t)} \!= \!  \bsm{G} \big( \bsm{s}^{(t)} \odot (\bsm{F} \bsm{x}^{(t)} ) \big) \!+\! \bsm{H} \bsm{x}^{(t)} \!+\! \bsm{w}^{(t)},t =1,\ldots,T,
\end{align}
where $\bsm{x}^{(t)}$ and $\bsm{w}^{(t)}$ are the transmit symbols and the additive noise following from $\mcl{CN}(0,\sigma^2 \bsm{I}_M)$ at time instant $t$, respectively. In addition, $\bsm{s}^{(t)} \triangleq \big[|s_1^{(t)}| e^{\jmath \angle s_1^{(t)}},$ $\ldots, |s_N^{(t)}|  e^{\jmath \angle s_N^{(t)}} \big]^\mathsf{T}$ is the reflect vector of the RIS with $\angle s_n^{(t)} \in (0,2\pi]$ and $|s_n^{(t)}| \in \{0,1\}$ being the phase shift and the on/off-state of the $n$-th RIS reflect element at time instant $t$, respectively. By summarizing all the $T$ samples in \eqref{single-snap}, the received signal at the BS is compactly expressed as
\begin{align} \label{received-matrix}
\bsm{Y} = \bsm{G} \big( \bsm{S} \odot ( \bsm{F} \bsm{X} ) \big) + \bsm{H} \bsm{X} + \bsm{W},
\end{align}
where $\bsm{Y} \triangleq \big[ \bsm{y}^{(1)},\ldots,\bsm{y}^{(T)} \big] \in \mbs{C}^{M \times T}$, $\bsm{S} \triangleq [\bsm{s}^{(1)}, \ldots, $ $\bsm{s}^{(T)}\big] \in \mbs{C}^{N \times T}$, $\bsm{X} \triangleq \big[\bsm{x}^{(1)},\ldots, $ $\bsm{x}^{(T)}\big] \in \mbs{C}^{K \times T}$, and $\bsm{W} \triangleq \big[\bsm{w}^{(1)},\ldots, \bsm{w}^{(T)} \big]$ $ \in \mbs{C}^{M \times T}$. With the help of a programmable smart controller, the RIS is capable of dynamically rescattering the electromagnetic waves towards a desired user by adjusting $\{\bsm{s}^{(t)}\}$. The design of $\{\bsm{s}^{(t)}\}$ is usually referred to as passive beamforming \cite{wu2018intelligent, huang2018large, yan2019}.

Note that accurate information of $\bsm{G}$, $\bsm{F}$, and $\bsm{H}$ is essential to the passive beamforming design. To acquire the CSI, a two-stage approach in \cite{he2019cascaded} utilizes the on-off reflection patten of the RIS to get a sparsity structure of $\bsm{F} \bsm{X}$. In this paper, by following \cite{he2019cascaded}, we set the phases of the on-state RIS elements to be zero and assume that $\bsm{S}$ consists of independent and identically distributed (i.i.d.) random variables drawn from the Bernoulli distribution with Bernoulli parameter $\rho$.\footnote{We emphasize that the 0-1 matrix $\bsm{S}$ is deterministic to the receiver. That is, $\bsm{S}$ can be generated as a $0$-$1$  pseudo-random matrix, and is known to both the RIS and the receiver by sharing a common random seed.} Herein, $\rho$ is also called the sparsity level of $\bsm{S}$ to indicate the probability of turning on a RIS element (i.e., the sampling rate of the RIS). The $0$-$1$ pattern of $\bsm{S}$ has also been exploited for passive information transfer \cite{yan2019} and modulating information bits \cite{guo2019reflecting}. Furthermore, we assume that $\bsm{G}$, $\bsm{F}$, $\bsm{H}$, and $\bsm{X}$ are composed of i.i.d. variables with zero means and have the following separable probability density
functions (PDFs):
\begin{align}
p(\bsm{G}) & = \prod_{m=1}^M \prod_{n=1}^N  p(g_{mn}), \,
p(\bsm{F})  = \prod_{n=1}^N \prod_{k=1}^K   p(f_{nk}), \label{g-f-prior} \\
p(\bsm{H}) & = \prod_{m=1}^M \prod_{k=1}^K   p(h_{mk}),\, p(\bsm{X}) = \prod_{k =1}^K \prod_{t =1 }^{T}  p(x_{kt}). \label{h-x-prior}
\end{align}
This paper is concerned with the estimation of $\bsm{G}$, $\bsm{F}$, and $\bsm{H}$ with only a portion of the columns of $\bsm{X}$ being used as the pilots under the Bayesian inference framework, as will be addressed in the following sections.

\section{Problem Formulation}

\label{Section-III}

\subsection{Semi-Blind Scheme}

Suppose that for every $T$ symbol duration, the first $T_\mathsf{p}$ $(T_\mathsf{p} < T)$ symbol durations are utilized to transmit pilot sequences. The remaining $T_\mathsf{d}  = $\,$T - T_\mathsf{p}$ time is used for data transmission. This is equivalent to partitioning $\bsm{X}$ as
\begin{align}
\bsm{X} = [\bsm{X}_\mathsf{p},\bsm{X}_\mathsf{d}], \label{X-deco}
\end{align}
where $\bsm{X}_\mathsf{p} \in \mbs{C}^{K \times T_\mathsf{p}}$ and $\bsm{X}_\mathsf{d} \in \mbs{C}^{K \times T_\mathsf{d}}$ represent the pilot sequences and the unknown transmitted data, respectively. As a result, the PDF of $\bsm{X}$ can be expressed as
\begin{align}
p(\bsm{X}) = \underbrace{\left(\prod_{k=1}^{K} \prod_{t=1}^{T_\mathsf{p}} p(x_{\mathsf{p},kt}) \right)}_{\triangleq \,p(\bsm{X}_\mathsf{p})} \underbrace{\left(\prod_{k=1}^{K} \prod_{t=1}^{T_\mathsf{d}} p(x_{\mathsf{d},kt}) \right)}_{\triangleq \,p(\bsm{X}_\mathsf{d})}.
\end{align}
In particular, given a {\it known} pilot matrix $\overline{\bsm{X}}_\mathsf{p}$, we have \begin{align} \label{pilots-delta}
p(\bsm{X}_\mathsf{p}) = \delta(\bsm{X}_\mathsf{p} - \overline{\bsm{X}}_\mathsf{p}) = \prod_{k=1}^{K} \prod_{k=1}^{T_\mathsf{p}} \delta (x_{\mathsf{p},kt} - \overline{x}_{\mathsf{p},kt}).
\end{align}

This work is to simultaneously estimate $\bsm{G}$, $\bsm{F}$, $\bsm{H}$, and $\bsm{X}_\mathsf{d}$ with given $\bsm{S}$ and $\bsm{X}_\mathsf{p} = \overline{\bsm{X}}_\mathsf{p}$. The usage of the partially known pilots $\bsm{X}_\mathsf{p} = \overline{\bsm{X}}_\mathsf{p}$ for implementing the joint channel estimation of $\bsm{G}$, $\bsm{F}$, $\bsm{H}$ and data detection of $\bsm{X}_\mathsf{d}$ is called semi-blind channel estimation. Note that the semi-blind channel estimation scheme in conventional massive MIMO systems \cite{wen2015bayes, prasad2015joint,nayebi2017semi} is a special case of the problem considered here when the user-RIS-BS link does not exist (i.e., $\bsm{G} = \bsm{0}$ or $\bsm{F} = \bsm{0}$).

A close observation of \eqref{received-matrix} reveals that our semi-blind channel estimation problem belongs to the family of trilinear estimation problem, because \eqref{received-matrix} involves the multiplication of the three unknown matrices $\bsm{G}$, $\bsm{F}$, and $\bsm{X}_\mathsf{d}$. When the RIS is removed or turned off, i.e., $\bsm{G} = \bsm{F} = \bsm{0}$, the considered semi-blind channel estimation problem is reduced to a bilinear estimation problem encountered in conventional massive MIMO as only two unknown terms $\bsm{H}$ and $\bsm{X}_\mathsf{d}$ are involved.

\vspace{-0.1cm}
\subsection{Bayesian MMSE Estimation} \label{section3}


Using the mean square error metric,  the MMSE estimators of $\bsm{G}$, $\bsm{F}$, $\bsm{H}$, and $\bsm{X}_\mathsf{d}$ solve for
\begin{align}
& \min_{\widetilde{\bsm{G}}} \mbs{E} \left\{ \big\| \bsm{G} - \widetilde{\bsm{G}} \big\|_\mathsf{F}^2 \right\}, \min_{\widetilde{\bsm{F}}} \mbs{E} \left\{ \big\| \bsm{F} - \widetilde{\bsm{F}} \big\|_\mathsf{F}^2 \right\}, \label{min-G-F} \\
& \min_{\widetilde{\bsm{H}}} \mbs{E} \left\{ \big\| \bsm{H} - \widetilde{\bsm{H}} \big\|_\mathsf{F}^2 \right\},\min_{\widetilde{\bsm{X}}_\mathsf{d}} \mbs{E} \left\{ \big\| \bsm{X}_\mathsf{d} - \widetilde{\bsm{X}}_\mathsf{d}  \big\|_\mathsf{F}^2 \right\}, \label{min-H-X}
\end{align}
where the expectations are evaluated over the joint distribution $p(\bsm{Y},\bsm{G}, \bsm{F}, \bsm{H},\bsm{X}_\mathsf{d})$. According to Bayes' rule and the first-order optimal condition, the closed-form solutions of \eqref{min-G-F} and \eqref{min-H-X} are given by the following posterior mean estimators (also known as the MMSE estimators):
\begin{subequations} \label{postetior-all}
\begin{align}
\hat{g}_{mn} &  = \mbs{E} \{g_{mn} | \bsm{Y}\},
\hat{f}_{nk} = \mbs{E} \{ f_{nk} | \bsm{Y} \}, \\
\hat{h}_{mk} &  = \mbs{E} \{h_{mk} | \bsm{Y} \},
\hat{x}_{\mathsf{d},kt} = \mbs{E} \{ x_{\mathsf{d},kt} | \bsm{Y} \}, \label{mean-x}
\end{align}
\end{subequations}
where the expectations are taken with respect to the marginal posterior distributions $p (g_{mn}|\bsm{Y})$, $p (f_{nk}|\bsm{Y}) $, $p (h_{mk}|\bsm{Y}) $, and $p (x_{\mathsf{d}, kt}|\bsm{Y}) $, respectively. The averaged MSEs of $\widehat{\bsm{G}}  $, $\widehat{\bsm{F}}  $, $\widehat{\bsm{H}}  $, and $\widehat{\bsm{X}}_{\mathsf{d}}$ are defined as
\begin{subequations} \label{MSE-all}
\begin{align}
\text{MSE}_{\bsm{G}} &\triangleq \frac{1}{MN} \mbs{E} \left\{ \| \bsm{G} - \widehat{\bsm{G}} \|_\mathsf{F}^2 \right\} , \\
\text{MSE}_{\bsm{F}} &\triangleq \frac{1}{NK} \mbs{E} \left\{  \| \bsm{F} - \widehat{\bsm{F}} \|_\mathsf{F}^2\right\},  \\
\text{MSE}_{\bsm{H}} &\triangleq \frac{1}{MK} \mbs{E} \left\{  \| \bsm{H} - \widehat{\bsm{H}} \|_\mathsf{F}^2\right\}, \\
\text{MSE}_{\bsm{X}_\mathsf{d}} &\triangleq \frac{1}{KT_\mathsf{d}} \mbs{E} \left\{  \| \bsm{X}_\mathsf{d} - \widehat{\bsm{X}}_\mathsf{d} \|_\mathsf{F}^2\right\}, \label{MSE-X-o}
\end{align}
\end{subequations}
respectively, where the expectations are over the joint posterior distribution $p(\bsm{Y},\bsm{G}, \bsm{F}, \bsm{H},\bsm{X}_\mathsf{d})$.


%

Exact evaluation of the MMSE estimators in \eqref{postetior-all} are generally intractable because they involve high-dimensional integrations in the marginalization of the joint posterior distribution $p(\bsm{G}, \bsm{F}, \bsm{H},\bsm{X}_\mathsf{d} | \bsm{Y})$. In the subsequent section, we develop a computationally efficient iterative algorithm to approximately calculate the MMSE estimators \eqref{postetior-all} based on the AMP framework \cite{donoho2009message, rangan2011generalized,parker2014bilinear}. We will then analyse the asymptotic performance of the theoretical MSEs in \eqref{MSE-all} to evaluate the performance of the proposed algorithm.

\begin{figure}[t]
    \centering
    \includegraphics[scale=0.49]{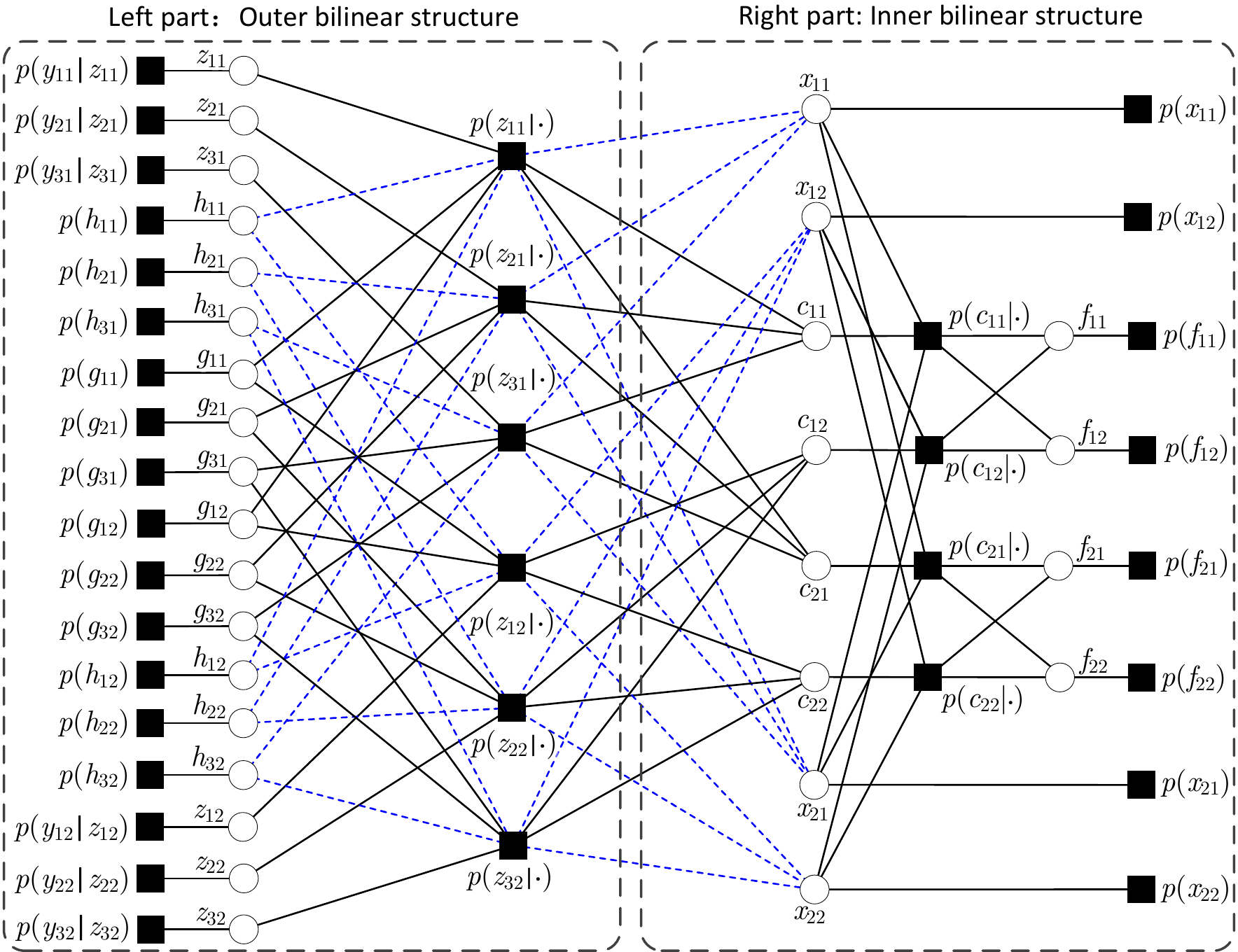}
    \vspace{-0.5cm}
    \caption{The factor graph representation of \eqref{fatorization1} for a toy example with $M=3$ and $N=T=K=2$.}
    \label{factor1}
\end{figure}

\section{Semi-Blind Channel Estimation via Trilinear Approximate Message Passing}

\label{Section-IV}


\subsection{Factor Graph Representation}


We begin with a factor graph representation of the posterior distribution of the unknown variables. By defining
\begin{align}
\bsm{C} \triangleq \bsm{S} \odot ( \bsm{F} \bsm{X} ),\,\bsm{Z} \triangleq \bsm{G}\bsm{C} + \bsm{H} \bsm{X}, \label{define-Z-C}
\end{align}
and using Bayes' theorem, the joint posterior distribution of $\bsm{G}$, $\bsm{F}$, $\bsm{H}$, $\bsm{X}$, $\bsm{Z}$, and $ \bsm{C}$ can be expressed by using the factorized distribution in \eqref{fatorization1} (shown at the bottom of the next page)
\begin{figure*}[!b]
\hrulefill
\begin{align}
p(\bsm{G}, \bsm{F}, \bsm{H},\bsm{X}, \bsm{Z}, \bsm{C} | \bsm{Y})
 &=\frac{1}{p(\bsm{Y})} p(\bsm{Y} | \bsm{Z})  p(\bsm{Z} | \bsm{G},\bsm{C},\bsm{H},\bsm{X}) p (\bsm{C} | \bsm{F},\bsm{X}) p(\bsm{G}) p(\bsm{F}) p(\bsm{H}) p(\bsm{X})  \notag \\
& = \frac{1}{p(\bsm{Y})} \! \left( \prod_{m=1}^M \prod_{t=1}^T p(y_{mt} | z_{mt} ) p\left(z_{mt} | \bsm{g}^\mathsf{T}_m,\bsm{c}_t ,\bsm{h}^\mathsf{T}_m,\bsm{x}_t \right)  \right) \! \left(\prod_{n=1}^N \prod_{t=1}^T p(c_{nt} | \bsm{f}^\mathsf{T}_n,\bsm{x}_t) \right) \!  \left(\prod_{m=1}^M \prod_{n=1}^N p(g_{mn}) \right) \notag \\
&~~~~~\times \left( \prod_{n=1}^N \prod_{k=1}^K  p(f_{nk}) \right) \left( \prod_{m=1}^M \prod_{k=1}^K p(h_{mk}) \right) \left(\prod_{k=1}^{K} \prod_{t=1}^{T_\mathsf{p}} p(x_{\mathsf{p},kt}) \right) \left(\prod_{k=1}^{K} \prod_{t=1}^{T_\mathsf{d}} p(x_{\mathsf{d},kt}) \right), \label{fatorization1}
\end{align}
\end{figure*}
where $p(\bsm{Y}) = \int_{\bsm{G}, \bsm{F}, \bsm{H},\bsm{X}, \bsm{Z}, \bsm{C}} p(\bsm{Y}, \bsm{G}, \bsm{F}, \bsm{H},\bsm{X}, \bsm{Z}, \bsm{C})$ is the normalization constant and
\begin{align}
p(y_{mt} | \cdot) & \triangleq p(y_{mt} | z_{mt}) = \mcl{CN} (y_{mt}; z_{mt}, \sigma^2) , \label{yz-likelihood} \\
p(z_{mt} | \cdot) & \triangleq p\left(z_{mt} | \bsm{g}^\mathsf{T}_m, \bsm{c}_t,\bsm{h}^\mathsf{T}_m, \bsm{x}_t \right) \notag \\
& = \delta\left(z_{mt}-\bsm{g}_m^\mathsf{T} \bsm{c}_t - \bsm{h}_m^\mathsf{T} \bsm{x}_t \right), \\
p(c_{nt}| \cdot) & \triangleq p(c_{nt} |\bsm{f}^\mathsf{T}_n,\bsm{x}_t) = \delta \left(c_{nt} - s_{nt} \bsm{f}_n^\mathsf{T} \bsm{x}_t \right).
\end{align}
The factorized distribution in \eqref{fatorization1} can be visualized by the factor graph shown in Fig. \ref{factor1}, which is divided into two subgraphs, i.e., the left part and the right part. Therein, the variables $\{y_{mt}\}$, $\{z_{mt}\}$, $\{c_{nt}\}$, $\{g_{mn}\}$, $\{f_{nk}\}$, $\{h_{mk}\}$, and $\{x_{kt}\}$ are represented by the ``variable nodes'' that appear as hollow circles. The distributions $\{p(y_{mt} | \cdot)\}$, $\{ p(z_{mt} | \cdot)\}$, $\{p(c_{nt} | \cdot)\}$,  $\{p(g_{mn})\} $, $\{p(f_{nk})\},$ $\{p(h_{mk})\}$, and $\{p(x_{kt})\}$ are represented by the ``factor nodes'' that appear as black filled squares. Each variable node is connected to its associated factor nodes. In Fig. \ref{factor1}, the dashed blue lines exist only when the direct link exists.

\subsection{Trilinear AMP Algorithm}

A direct alternative of computing \eqref{postetior-all} is to use the canonical  sum-product algorithm (SPA), which passes messages through the edges of the factor graph in Fig. \ref{factor1}. These messages describe the PDFs of the variables $\{z_{mt}\}$, $\{c_{nt}\}$,  $\{g_{mn}\}$, $\{f_{nk}\}$, $\{h_{mk}\}$, and $\{x_{kt}\}$. However, exact inference using SPA remains analytically intractable and computationally prohibitive due to the involved high-dimensional integration. To circumvent this, we resort to the AMP framework \cite{donoho2009message, rangan2011generalized, parker2014bilinear} to approximately calculate  the MMSE estimators in \eqref{postetior-all}. The approximations rest primarily on the central-limit-theorem (CLT) and Taylor-series arguments that become accurate under the large-system limit, i.e., $M$, $N$,  $K$, $T_\mathsf{p}$, $T_\mathsf{d} \to \infty$, with the ratios $M/K$, $N/K$, $T_\mathsf{p}/K$, $T_\mathsf{d}/K$ being fixed and finite. In the algorithm derivation, we assume the following scaling conventions: the elements of $\bsm{F}$ and $\bsm{H}$ both scale as the order of $\mcl{O}(1/\sqrt{K})$; the elements of $\bsm{G}$ scale as the order of $\mcl{O}(1/\sqrt{N})$; the elements of $\bsm{X}$ and $\bsm{W}$ both scale as the order of $\mcl{O}(1)$. As such, the elements of $\bsm{C}$, $\bsm{H} \bsm{X}$, $\bsm{G} \bsm{C}$, and $\bsm{Z}$ scale as the order of $\mcl{O}(1)$, i.e., the same order as those of $\bsm{W}$ and $\bsm{X}$.

Before going to the algorithmic description, we first provide an intuitive description of how the trilinear inference problem in \eqref{received-matrix} can be split into two bilinear inference problems. Based on \eqref{define-Z-C}, we rewrite \eqref{received-matrix} as
\begin{align}
\bsm{Y} & =  [\bsm{G},\,\bsm{H}]\,\big[\bsm{C}^\mathsf{T},\,\bsm{X}^\mathsf{T}\big]^\mathsf{T} + \bsm{W}, \label{left-bilinear} \\
\bsm{C} & = \bsm{S} \odot \bsm{F} \bsm{X}.  \label{right-bilinear}
\end{align}
Clearly, inferring $[\bsm{G},\bsm{H}]$ and $\big[\bsm{C}^\mathsf{T},\,\bsm{X}^\mathsf{T}\big]^\mathsf{T}$ with noisy observation $\bsm{Y}$ in \eqref{left-bilinear} and inferring $\bsm{F}$ and $\bsm{X}$ with missing observation $\bsm{C}$ in \eqref{right-bilinear} are both bilinear inference problems. More specifically, the bilinear inference problem \eqref{right-bilinear} is embedded into the bilinear inference problem \eqref{left-bilinear} and the two bilinear problems are connected via the common variables $\bsm{C}$ and $\bsm{X}$. Therefore, by following the general idea of the AMP framework \cite{donoho2009message, rangan2011generalized, parker2014bilinear}, we derive the approximate message flows of the trilinear inference problem from the left part (i.e., the outer bilinear structure) to the right part (i.e., the inner bilinear structure) of the factor graph in Fig. \ref{factor1}.

\subsubsection{AMP within the outer bilinear structure}
According to the sum-product rule, the message at iteration $i$ from the factor node $p(z_{mt}|\cdot)$ to the variable node $z_{mt}$ is given by
\begin{align}
& \Delta_{p(z_{mt}|\cdot) \rightarrow z_{mt}} (i, z_{mt}) \notag \\
& \propto \int_{\{g_{mn}\}_{n=1}^N,\{c_{nt}\}_{n=1}^N,\{h_{mk}\}_{k=1}^K,\{x_{kt}\}_{k=1}^K}  p\left(z_{mt} | \cdot \right) \notag \\
& ~~~~\,\times \prod_{n} \left( \Delta^{(i)}_{g_{mn} \rightarrow p(z_{mt}|\cdot)} (i, g_{mn})\,\Delta^{(i)}_{c_{nt} \rightarrow p(z_{mt}|\cdot)} (i, c_{nt}) \right) \notag \\
& ~~~~\,\times \prod_{k} \left( \Delta^{(i)}_{h_{mk} \rightarrow p(z_{mt}|\cdot)} (i, h_{mk})\, \Delta^{(i)}_{x_{kt} \rightarrow p(z_{mt}|\cdot)} (i, x_{kt}) \right).
\end{align}
Note that $z_{mt} = \sum_{n=1}^N g_{mn} c_{nt} + \sum_{k=1}^K h_{mk} x_{kt}$. Therefore, for large $N$ and $K$, the CLT motivates the approximation of $\Delta_{p(z_{mt}|\cdot) \rightarrow z_{mt}} (i, z_{mt}) $ as a Gaussian message \cite[Sec. II D]{parker2014bilinear}:
\begin{align}
\Delta_{p(z_{mt}|\cdot) \rightarrow z_{mt}} (i, z_{mt})  \approx  \mcl{CN}(z_{mt}; \hat{p}_{mt}(i), v^p_{mt}(i) ), \label{delta-pz-z}
\end{align}
where
\begin{align}
\hat{p}_{mt}(i) =& \sum_{n=1}^{N}\hat{g}_{t, mn}(i)\hat{c}_{m, nt}(i) + \sum_{k=1}^K \hat{h}_{t, mk}(i) \hat{x}_{m, kt}(i), \label{pmthat}
\end{align}
\begin{align}
v^p_{mt}(i) = & \sum_{n=1}^{N} \Big( |\hat{g}_{t, mn}(i)|^2 v^c_{m, nt}(i) + v^g_{t,mn}(i) |\hat{c}_{m, nt}(i)|^2 \notag \\
& + v^g_{t,mn}(i) v^c_{m, nt}(i) \Big) + \sum_{k=1}^{K} \Big( |\hat{h}_{t,mk}(i)|^2 v^x_{m,kt}(i) \notag \\
& + v^h_{t, mk}(i) |\hat{x}_{m, kt}(i)|^2 + v^h_{t, mk}(i) v^x_{m,kt}(i) \Big). \label{vpmthat}
\end{align}
Herein, $\{\hat{g}_{t, mn}(i),v^g_{t,mn}(i)\}$, $\{ \hat{c}_{m, nt}(i), v^c_{m, nt}(i)\}$, $\{\hat{h}_{t, mk}(i), $ $ v^h_{t, mk}(i)\}$, and $\{\hat{x}_{m, kt}(i), v^x_{m,kt}(i) \}$ stands for the means and the variances of $\Delta_{g_{mn} \rightarrow p(z_{mt}|\cdot)} (i, g_{mn})$, $\Delta_{c_{nt} \rightarrow p(z_{mt}|\cdot)} (i, c_{nt})$, $\Delta_{h_{mk} \rightarrow p(z_{mt}|\cdot)} (i, h_{mk})$, and $\Delta_{x_{kt} \rightarrow p(z_{mt}|\cdot)} (i, x_{kt}) $, respectively.

The message from $p(z_{mt}|\cdot)$ to $g_{mn}$ is given by
\begin{align}
& \Delta_{p(z_{mt}|\cdot) \rightarrow g_{mn}} (i, g_{mn}) \notag \\ &\propto \int_{\{g_{mn'}\}_{n'=1 (\neq n)}^N,z_{nt},\{c_{nt}\}_{n=1}^N,\{h_{mk}\}_{k=1}^K,\{x_{kt}\}_{k=1}^K} p(z_{mt}|\cdot) \notag \\
& \times \left( \Delta_{z_{mt} \rightarrow p(z_{mt}|\cdot)} (i, z_{mt} ) \right) \prod_{n' \neq n} \left( \Delta^{(i)}_{g_{mn} \rightarrow p(z_{mt}|\cdot)} (i, g_{mn}) \right) \notag \\
& \times \prod_{n} \left( \Delta_{c_{nt} \rightarrow p(z_{mt}|\cdot)} (i, c_{nt}) \right) \prod_{k} \left( \Delta^{(i)}_{i, h_{mk} \rightarrow p(z_{mt}|\cdot)} (i, h_{mk}) \right) \notag \\
& \times \prod_{k} \left( \Delta_{x_{kt} \rightarrow p(z_{mt}|\cdot)} (i, x_{kt}) \right).  \label{delta-pz-gmn}
\end{align}
Using the CLT argument again, we can approximate $z_{mt}$ conditioned on $g_{mn}$ as the Gaussian distribution
\begin{align} \label{con-gauss}
p(z_{mt} | g_{mn}) & \sim \mcl{CN} \Big( z_{mt}; g_{mn} \hat{x}_{m,kt} + \hat{p}_{n, mt}(i), \notag \\
& ~~~~~~~~~~~~~~\, |g_{mn}|^2v^{x}_{m,kt} + v^{p}_{n, mt}(i) \Big),
\end{align}
where
\begin{align}
&\hat{p}_{n, mt}(i) = \!\! \sum_{n'=1 (\neq n)}^{N}\hat{g}_{t, mn}(i)\hat{c}_{m, nt}(i) \notag \\
&~~~~~~~~~~~~~~~~~~ + \sum_{k=1}^K \hat{h}_{t, mk}(i) \hat{x}_{m, kt}(i), \label{ppnmt} \\
&v^p_{n, mt}(i)=\!\!  \sum_{n'=1 (\neq n)}^{N} \!\!\! \Big( |\hat{g}_{t, mn}(i)|^2 v^c_{m, nt}(i) \!+\! v^g_{t,mn}(i) |\hat{c}_{m, nt}(i)|^2 \notag \\
&~~~~~~~~~~~~~~~~~~ + v^g_{t,mn}(i) v^c_{m, nt}(i) \Big) + \sum_{k=1}^{K} \Big( |\hat{h}_{t,mk}(i)|^2 v^x_{m,kt}(i) \notag \\
&~~~~~~~~~~~~~~~~~~ + v^h_{t, mk}(i) |\hat{x}_{m, kt}(i)|^2 + v^h_{t, mk}(i) v^x_{m,kt}(i) \Big). \label{vpnmt}
\end{align}
With the conditional-Gaussian \eqref{con-gauss} and  $\Delta^{(i)}_{z_{mt} \rightarrow p(z_{mt}|\cdot)} (z_{mt}) = \mcl{CN} (y_{mt}; z_{mt}, \sigma^2)$, we can change the multiple integral in \eqref{delta-pz-gmn} into the single integral of $z_{mt}$:
\begin{align}
& \Delta_{p(z_{mt}|\cdot) \rightarrow g_{mn}} (i, g_{mn}) \notag \\
& \approx \int_{z_{mt}} \mcl{CN} \Big( z_{mt}; g_{mn} \hat{c}_{m,nt}(i) + \hat{p}_{n, mt}(i), \notag \\
&~~~~~~~~~~~~\,|g_{mn}|^2v^{c}_{m,nt}(i) + v^{p}_{n, mt}(i) \Big) \mcl{CN} \big( y_{mt}; z_{mt}, \sigma^2 \big). \label{delta-pzmt-gmn2}
\end{align}
By expanding the exponential of \eqref{delta-pzmt-gmn2} over $g_{mn}$ at its marginal posterior $\hat{g}_{mn} (i)$ through the second-order Taylor approximation \cite[Sec. II D]{parker2014bilinear}, we obtain the following Gaussian approximation
\begin{align}
& \Delta_{p(z_{mt}|\cdot) \rightarrow g_{mn}} (i, g_{mn}) \notag \\
& \approx \exp \Big\{ \big( \hat{u}_{mt} {(i)} \hat{c}_{m, nt}(i)  + v_{mt}^u(i) \left|\hat{c}_{nt}(i) \right|^2 \notag \\
&~~~~ \times \hat{g}_{mn}(i) \big) g_{mn} - \Big( v_{mt}^u(i) |\hat{c}_{nt}(i)|^2 - v^c_{nt}(i) \notag \\
&~~~~ \times \left(|\hat{u}_{mt} {(i)}|^2 - v_{mt}^u(i) \right) \Big) |g_{mn}|^2 \Big\},  \label{delta-pzmt-gmn3}
\end{align}
where
\begin{align}
& v_{mt}^u(i) = \frac{1}{v_{mt}^p(i)} \left( 1- \frac{v_{mt}^z (i)}{v_{mt}^p(i)} \right), \label{v-u-mt}\\
& \hat{u}_{mt}(i)=\frac{1}{v_{mt}^p(i)} \big( \hat{z}_{mt}(i)-\hat{p}_{mt}(i) \big). \label{m-u-mt}
\end{align}
Herein, $\{\hat{z}_{mt}(i),v_{mt}^z (i)\}$ and $\{\hat{c}_{nt}(i), v_{nt}^c (i)\}$ are the marginal posterior means and variances of $z_{mt}$ and $c_{nt}$ at iteration $i$, respectively; and their updates will be detailed in the subsequent Section \ref{marginal-all}.

Analogous to the approximation \eqref{delta-pzmt-gmn3}, the factor-to-variable messages $\Delta_{p(z_{mt}|\cdot) \rightarrow h_{mk}} (i,h_{mk})$ can be approximated as
\begin{align}
& \Delta_{p(z_{mt}|\cdot) \rightarrow h_{mk}} (i, h_{mk}) \notag \\
& \approx \exp \Big\{ \big( \hat{u}_{mt} {(i)} \hat{x}_{m, kt}(i)  + v_{mt}^u(i) \left|\hat{x}_{kt}(i) \right|^2 \notag \\
&~~~~ \times \hat{h}_{mk}(i) \big) h_{mk} - \Big( v_{mt}^u(i) |\hat{x}_{kt}(i)|^2 - v^x_{kt}(i) \notag \\
&~~~~ \times \left(|\hat{u}_{mt} {(i)}|^2 - v_{mt}^u(i) \right) \Big) |h_{mk}|^2 \Big\},  \label{delta-pzmt-hmk3}
\end{align}
respectively, where $\hat{x}_{kt}(i)$ and $v_{kt}^x (i)$ are the marginal posterior means and variances of $x_{kt}$ at iteration $i$, and their updates will be discussed in the subsequent Section \ref{marginal-all}.

Using the Gaussian message \eqref{delta-pzmt-gmn3} and following  \cite[Sec. II E and F]{parker2014bilinear}, the message from $g_{mn}$ to $p(z_{mt}|\cdot)$ at iteration $i+1$ can be approximated as
\begin{align}
& \Delta_{g_{mn} \rightarrow p(z_{mt}|\cdot)} (i+1, g_{mn}) \notag \\
& \propto p(g_{mn}) \prod_{t' \neq t} \Delta_{p(z_{mt'}|\cdot) \rightarrow g_{mn}} (i, g_{mn}) \notag \\
& \approx p(g_{mn}) \, \mcl{CN} \Big(g_{mn}; \hat{q}^g_{t, mn}(i), v_{t, mn}^{qg}(i)\Big), \label{appro-gmn-zmt}
\end{align}
where
\begin{align}
\hat{q}^g_{t, mn}(i) & \approx \hat{q}^g_{mn}(i) - v^{qg}_{mn}(i) \hat{c}_{nt}(i) \hat{u}_{mt}(i),  \\
v_{t, mn}^{qg}(i) & \approx v_{mn}^{qg}(i) \triangleq \left( \sum_{t=1}^{T}|\hat{c}_{nt}(i)|^2 v_{mt}^u (i) \right)^{-1},
\label{tmn-gq}
\end{align}
\begin{align}
\hat{q}^g_{mn}(i) & \triangleq \hat{g}_{mn}(i) \left(1- v_{mn}^{qg}(i) \sum_{t=1}^{T}v_{nt}^{c}(i) v_{mt}^u(i) \right) \notag \\
&~~~~ +v_{mn}^{qg}(i)\sum_{t=1}^{T} \hat{c}^\ast_{nt}(i) \hat{u}_{mt}(i). \label{g-g-mn}
\end{align}
Similarly, the variable-to-factor message $\Delta^{(i)}_{h_{mk} \rightarrow p(z_{mt}|\cdot)} (h_{mk})$ can be approximated as
\begin{align}
& \Delta_{h_{mk} \rightarrow p(z_{mt}|\cdot)} (i+1, h_{mk}) \notag \\
& \approx p(h_{mk}) \mcl{CN}\big(h_{mk}; \hat{q}^h_{t, mk}(i), v_{t, mk}^{qh}(i)\big),
\end{align}
where
\begin{align}
\hat{q}^h_{t, mk}(i) & = \hat{q}^h_{mk}(i) - v^{qh}_{mk}(i) \hat{x}_{kt}(i) \hat{u}_{mt}(i), \\
 v_{t, mk}^{qh}(i) & \approx v_{mk}^{qh}(i) \triangleq \left( \sum_{t=1}^{T}|\hat{x}_{kt}(i)|^2 v_{mt}^u (i) \right)^{-1}, \label{qq-h-v} \\
\hat{q}^h_{mk}(i) & \triangleq \hat{h}_{kt}(i) \left( 1- v_{mk}^{qh}(i) \sum_{t=1}^{T}v_{kt}^{x}(i) v_{mt}^u(i) \right) \notag \\
& ~~~~ + v_{mk}^{qh}(i)\sum_{t=1}^{T} \hat{x}^\ast_{kt}(i) \hat{u}_{mt}(i). \label{qq-h-m}
\end{align}

\subsubsection{AMP between the outer bilinear structure and the inner bilinear structure}
The messages exchanged between the two bilinear structures involve the variable-to-factor messages $\Delta_{p(z_{mt}|\cdot) \rightarrow c_{nt}} (i, c_{nt})$, $\Delta_{p(z_{mt}|\cdot) \rightarrow x_{kt}} (i, x_{kt})$, and the factor-to-variable messages $\Delta_{x_{kt} \rightarrow p(z_{mt}|\cdot)} (i, x_{kt})$, $\Delta_{c_{nt} \rightarrow p(z_{mt}|\cdot)} (i, c_{nt})$. Analogous to \eqref{delta-pzmt-gmn3}, we can approximate $\Delta_{p(z_{mt}|\cdot) \rightarrow c_{nt}} (i, c_{nt})$ and $\Delta_{p(z_{mt}|\cdot) \rightarrow x_{kt}} (i, x_{kt})$ as
\begin{align}
& \Delta_{p(z_{mt}|\cdot) \rightarrow c_{nt}} (i, c_{nt}) \notag \\
& \approx \exp \Big\{ \big( \hat{u}_{mt} {(i)} \hat{g}_{t, mn}(i)  + v_{mt}^u(i) \left|\hat{g}_{mn}(i) \right|^2 \notag \\
&~~~~ \times \hat{c}_{nt}(i) \big) c_{nt} - \Big( v_{mt}^u(i) |\hat{g}_{mn}(i)|^2 - v^g_{mn}(i) \notag \\
&~~~~ \times \left(|\hat{u}_{mt} {(i)}|^2 - v_{mt}^u(i) \right) \Big) |c_{nt}|^2 \Big\},  \label{delta-pzmt-cmk3} \\
& \Delta_{p(z_{mt}|\cdot) \rightarrow x_{kt}} (i, x_{kt}) \notag \\
& \approx \exp \Big\{ \big( \hat{u}_{mt} {(i)} \hat{h}_{t, mk}(i)  + v_{mt}^u(i) \left|\hat{h}_{mk}(i) \right|^2 \notag \\
&~~~~ \times \hat{x}_{kt}(i) \big) x_{kt} - \Big( v_{mt}^u(i) |\hat{h}_{mk}(i)|^2 - v^h_{mk}(i) \notag \\
&~~~~ \times \left(|\hat{u}_{mt} {(i)}|^2 - v_{mt}^u(i) \right) \Big) |x_{kt}|^2 \Big\},  \label{delta-pzmt-xmk3}
\end{align}
where $\{\hat{g}_{mn}(i), v_{mn}^g (i)\}$ and $\{\hat{h}_{mk}(i),v_{mk}^h (i)\}$ represent the marginal posterior means and variances of $g_{mn}$ and $h_{mk}$ at iteration $i$, respectively; and their updates will be discussed in detail in the subsequent Subsection \ref{marginal-all}.

Using the Gaussian message \eqref{delta-pzmt-cmk3} and following  \cite[Sec. II E and F]{parker2014bilinear}, the factor-to-variable message $\Delta_{c_{nt} \rightarrow p(z_{mt}|\cdot)} (i, c_{nt})$ is approximated as
\begin{align}
& \Delta_{c_{nt} \rightarrow p(z_{mt}|\cdot)} (i, c_{nt}) \notag \\
& \approx \Delta_{ p(c_{nt}|\cdot) \rightarrow c_{nt}} (i, c_{nt})\, \prod_{t' \neq t} \Delta_{p(z_{mt'}|\cdot) \rightarrow c_{nt'}} (i, c_{nt'}), \notag \\
& \approx \Delta_{ p(c_{nt}|\cdot) \rightarrow c_{nt}} (i, c_{nt}) \,  \mcl{CN} \Big( c_{nt}; \hat{r}^c_{m, nt}(i), v^{rc}_{m, nt} (i) \Big),
\end{align}
where $\Delta_{ p(c_{nt}|\cdot) \rightarrow c_{nt}} (i, c_{nt})$ is given in \eqref{delta-pc-c} and
\begin{align}
\hat{r}^c_{m, nt}(i) & = \hat{r}^c_{nt}(i) - v^{rc}_{nt}(i) \hat{g}_{mn}(i) \hat{u}_{mt}(i),  \\
v^{rc}_{m, nt} (i) & \approx v^{rc}_{nt}(i) \triangleq s_{nt} \left( \sum_{t=1}^{T}|\hat{c}_{nt}(i)|^2 v_{mt}^u (i) \right)^{-1}, \label{rmnt-mc}\\
\hat{r}^c_{nt}(i) & \triangleq s_{nt} \left( \hat{c}_{nt}(i) \left(1- v^{rc}_{nt}(i) \sum_{m=1}^{M} v_{mn}^g(i) v_{mt}^u (i) \right) \right. \notag \\
&~~~~ \left. +v_{nt}^{rc}(i) \sum_{m=1}^{M}\hat{g}^\ast_{mn}(i)  \hat{u}_{mt}(i) \right). \label{rmnt-vc}
\end{align}
Likewise, the variable-to-factor message $\Delta_{x_{kt} \rightarrow p(z_{mt}|\cdot)} (i, x_{kt})$ is approximated as
\begin{align}
& \Delta_{x_{kt} \rightarrow p(z_{mt}|\cdot)} (i, x_{kt}) \notag \\
& \propto  p(x_{kt}) \prod_{n} \Delta_{ p(c_{nt}|\cdot) \rightarrow x_{kt}} (i, x_{kt})  \notag \\
& ~~~~\times \prod_{t'\neq t} \Delta_{ p(z_{mt'}|\cdot) \rightarrow x_{kt'}} (i, x_{kt'}), \\
& \approx  p(x_{kt}) \prod_{n} \Delta_{ p(c_{nt}|\cdot) \rightarrow x_{kt}} (i, x_{kt})  \notag \\
& ~~~~\times \mcl{CN} \Big( c_{nt}; \hat{r}^x_{m, kt}(i), v^{rx}_{m, kt} (i) \Big),
\end{align}
where $\Delta_{ p(c_{nt}|\cdot) \rightarrow x_{kt}} (i, x_{kt})$ is given in \eqref{delta-pcnt-xkt3} and
\begin{align}
\hat{r}^x_{m, kt}(i) & = \hat{r}^x_{kt}(i) - v^{rx}_{kt}(i) \hat{h}_{mk}(i) \hat{u}_{mt}(i),  \\
v^{rx}_{m, kt} (i) & \approx v^{rx}_{kt}(i) \triangleq \left( \sum_{m=1}^{M}|\hat{h}_{mk}(i)|^2 v_{mt}^u (i) \right)^{-1}, \label{rktxi-v} \\
\hat{r}^x_{kt}(i) & \triangleq  \hat{x}_{kt}(i) \left(1- v^{rx}_{kt}(i) \sum_{m=1}^{M} v_{mk}^h(i) v_{mt}^u (i) \right) \notag \\
&~~~~  +v_{kt}^{rx}(i) \sum_{m=1}^{M}\hat{h}^\ast_{mk}(i)  \hat{u}_{mt}(i). \label{rktxi-m}
\end{align}

\subsubsection{AMP within the inner bilinear structure}
Implementing AMP within the inner bilinear structure is analogous to that of the outer bilinear structure. The difference is that the inner bilinear structure involves the exploitation of the RIS on-off   information matrix $\bsm{S}$. By using the CLT argument to $c_{nt} = s_{nt} \sum_{k} f_{nk} x_{kt}$, we can approximate  $\Delta_{p(c_{nt}|\cdot) \rightarrow c_{nt}} (i, c_{nt})$ as
\begin{align}
\Delta_{p(c_{nt}|\cdot) \rightarrow c_{nt}} (i, c_{nt})  \approx \begin{cases} \mcl{CN} \left(c_{nt}; \hat{\xi}_{nt}(i), v_{nt}^{\xi} (i) \right), \!\!\!\!&s_{nt} = 1 \\
 \delta(c_{nt}), \!\!\!\!&s_{nt} = 0
  \end{cases},  \label{delta-pc-c}
\end{align}
where
\begin{align}
v_{nt}^{\xi}(i) & =\tilde{v}_{nt}^p(i)+\sum_{k=1}^{K}v_{nk}^f(i)v_{kt}^x(i) \label{xi-m}\\
\hat{\xi}_{nt}(i) & = \tilde{p}_{nt}(i)-\hat{\eta}_{nt}(i-1) \tilde{v}_{nt}^p(i) \\
\tilde{p}_{nt}(i) & \triangleq \sum_{k=1}^{K}\hat{f}_{nk}(i)\hat{x}_{kt}(i), \\
\tilde{v}_{nt}^p(i) & \triangleq \sum_{k=1}^{K}|\hat{f}_{nk}(i)|^2 v^x_{kt}(i)
+ v^f_{nk}(i) |\hat{x}_{kt}(i)|^2. \label{xi-v}
\end{align}

Analogous to the approximation \eqref{delta-pzmt-gmn3}, the factor-to-variable messages $\Delta_{p(c_{nt}|\cdot) \rightarrow f_{nk}} (i, f_{nk})$ and $\Delta_{p(c_{nt}|\cdot) \rightarrow x_{kt}} (i, x_{kt})$ can be approximated as
\begin{align}
& \Delta_{p(c_{nt}|\cdot) \rightarrow f_{nk}} (i, f_{nk}) \notag \\
& \approx \exp \Big\{ \big( \hat{\eta}_{nt}(i) \hat{x}_{n, kt}(i)  + v_{nt}^\eta (i) \left| \hat{x}_{kt}(i) \right|^2 \notag \\
&~~~~ \times \hat{f}_{nk}(i) \big) f_{nk} - \Big( v_{nt}^\eta (i)  |\hat{x}_{kt}(i)|^2 - v^x_{kt}(i) \notag \\
&~~~~ \times \left(|\hat{\eta}_{nt}(i)|^2 - v_{nt}^\eta (i)  \right) \Big) |f_{nk}|^2 \Big\},  \label{delta-pcnt-fnk3} \\
& \Delta_{p(c_{nt}|\cdot) \rightarrow x_{kt}} (i, x_{kt}) \notag \\
& \approx \exp \Big\{ \big( \hat{\eta}_{nt}(i) \hat{f}_{t, nk}(i)  + v_{nt}^\eta (i) \left| \hat{f}_{nk}(i) \right|^2 \notag \\
&~~~~ \times \hat{x}_{kt}(i) \big) x_{kt} - \Big( v_{nt}^\eta (i)  |\hat{f}_{nk}(i)|^2 - v^f_{nk}(i) \notag \\
&~~~~ \times \left(|\hat{\eta}_{nt}(i)|^2 - v_{nt}^\eta (i)  \right) \Big) |x_{kt}|^2 \Big\},   \label{delta-pcnt-xkt3}
\end{align}
respectively, where $\hat{f}_{t, nk}(i)$ and $ \hat{x}_{n, kt}(i)$ denotes the means of the messages $\Delta_{p(c_{nt}|\cdot) \rightarrow f_{nk}} (i, f_{nk})$ and $\Delta_{p(c_{nt}|\cdot) \rightarrow x_{kt}} (i, x_{kt})$, and
\begin{align}
& v_{nt}^\eta (i) = \frac{s_{nt}}{v_{nt}^{\xi}(i)} \left( 1- \frac{v_{nt}^c (i)}{v_{nt}^\xi (i)} \right), \\
& \hat{\eta}_{nt}(i)=\frac{s_{nt}}{v_{nt}^{\xi}(i)} \big( \hat{c}_{nt}(i)-\tilde{p}_{nt}(i) \big),
\end{align}
where $\hat{c}_{nt}(i)$ and $v_{nt}^c (i)$ are the posterior means and variances of $c_{nt}$, and their computations will be discussed in the subsequent Section \ref{marginal-all}.

Using the Gaussian message \eqref{delta-pcnt-fnk3} and  following  \cite[Sec. II E and F]{parker2014bilinear}, the variable-to-factor message $\Delta_{f_{nk} \rightarrow p(c_{nt}|\cdot)} (i, f_{nk})$ is approximated as
\begin{align}
& \Delta_{f_{nk} \rightarrow p(c_{nt}|\cdot)} (i, f_{nk}) \notag \\
& \propto p(f_{nk}) \prod_{t' \neq t} \Delta_{ p(c_{nt}|\cdot) \rightarrow f_{nk}} (i, f_{nk}), \notag \\
& \approx p(f_{nk})\,\mcl{CN} \left(f_{nk}; \hat{q}^f_{t, nk}(i), v_{t, nk}^{qf}(i) \right)
\end{align}
where
\begin{align}
\hat{q}^f_{t, nk}(i) & = \hat{q}^f_{nk}(i) - v^{qf}_{nk}(i) \hat{x}_{kt}(i) \hat{\eta}_{nt}(i),  \\
v_{t, nk}^{qf} (i) & \approx v_{nk}^{qf}(i)  \triangleq \left( \sum_{t=1}^{T}|\hat{x}_{kt}(i)|^2 v_{nt}^{\eta}(i) \right)^{-1}, \label{q-f-variance} \\
\hat{q}^f_{nk}(i) & \triangleq \hat{f}_{nk}(i) \left(1- v^{qf}_{nk}(i) \sum_{t=1}^{T} v_{kt}^x(i) v_{nt}^\eta (i) \right) \notag \\
&~~~~ +v_{nk}^{qf}(i) \sum_{t=1}^{T}\hat{x}^\ast_{kt}(i)  \hat{\eta}_{nt}(i). \label{q-f-mean}
\end{align}
Likewise, with the message \eqref{delta-pcnt-xkt3}, the variable-to-factor message $\Delta_{x_{kt} \rightarrow p(c_{nt}|\cdot)} (i, x_{kt})$ can be approximated as
\begin{align}
& \Delta_{x_{kt} \rightarrow p(c_{nt}|\cdot)} (i, x_{kt}) \notag \\
& \propto  p(x_{kt}) \prod_{n} \Delta_{ p(c_{nt}|\cdot) \rightarrow x_{kt}} (i, x_{kt})  \notag \\
& \approx p(x_{kt})\,\mcl{CN} \Big( x_{kt}; \hat{\gamma}^x_{n, kt}(i), v^{\gamma x}_{n, kt} (i) \Big),
\end{align}
where
\begin{align}
\hat{\gamma}^x_{n, kt}(i) & = \hat{\gamma}^x_{kt}(i) - v^{\gamma x}_{kt}(i) \hat{f}_{nk}(i) \hat{\eta}_{nt}(i),  \\
v^{\gamma x}_{n, kt} (i) & \approx v^{\gamma x}_{kt}(i) \triangleq \left( \sum_{n=1}^{N}|\hat{f}_{nk}(i)|^2 v_{nt}^\eta (i) \right)^{-1}, \label{rx-v}\\
\hat{\gamma}^x_{kt}(i) & \triangleq \hat{x}_{kt}(i) \left(1- v^{\gamma x}_{kt}(i) \sum_{n=1}^{N} v_{nk}^f(i) v_{nt}^\eta (i) \right) \notag \\
&~~~~ + v^{\gamma x}_{kt}(i) \sum_{n=1}^{N}\hat{f}^\ast_{nk}(i)  \hat{\eta}_{nt}(i). \label{rx-m}
\end{align}

\vspace{-0.08cm}
\subsubsection{Approximated posterior means and variances} \label{marginal-all}
To obtain close-loop updates of the posterior means and variances of the associated variable nodes, we follow the steps in \cite[Sec. II F]{parker2014bilinear} to simplify $\hat{p}_{mt}(i)$ in \eqref{pmthat} and $v_{mt}^p(i)$ \eqref{vpmthat} as follows
\begin{align}
\hat{p}_{mt}(i) &= \bar{p}_{mt}(i)-\hat{u}_{mt}(i-1) \bar{v}_{mt}^p(i), \label{hat-p} \\
v_{mt}^p(i) &= \bar{v}_{mt}^p(i)+\sum_{n=1}^{N} v^g_{mn}(i) v^c_{nt}(i) + \sum_{k=1}^K v^h_{mk}(i) v^x_{kt}(i), \label{v-p}
\end{align}
where
\begin{align}
\bar{p}_{mt}(i)& =\sum_{n=1}^{N}\hat{g}_{mn}(i)\hat{c}_{nt}(i) + \sum_{k=1}^K \hat{h}_{mk}(i) \hat{x}_{kt}(i), \\
\bar{v}_{mt}^p(i) &= \sum_{n=1}^{N} \big( |\hat{g}_{mn}(i)|^2 v^c_{nt}(i) + v^g_{mn}(i) |\hat{c}_{nt}(i)|^2 \big) \notag \\
&~~~~\, + \sum_{k=1}^{K} \big( |\hat{h}_{mk}(i)|^2 v^x_{kt}(i) + v^h_{mk}(i) |\hat{x}_{kt}(i)|^2 \big). \label{bar-p}
\end{align}
By taking the product of both incoming messages at the variable node $z_{mt}$, i.e., those outgoing from the factor nodes $p(z_{mt}|\cdot)$ and $p(y_{mt} | z_{mt})$, the marginal posterior of $z_{mt}$ at the $(i+1)$-th iteration, denoted as $\Delta_{mt} (i+1, z_{mt})$, can  be approximated by
\begin{align}\label{pos-z}
\Delta^z_{mt} (i, z_{mt})
& \propto p(y_{mt}|z_{mt}) \, \Delta_{p(z_{mt}|\cdot) \rightarrow z_{mt}} (i, z_{mt}) \notag \\
& \approx \mcl{CN} (y_{mt}; z_{mt}, \sigma^2)\,\mcl{CN} \left(z_{mt};\hat{p}_{mt}(i), v^p_{mt}(i)\right) \notag \\
& = \mcl{CN} \big(z_{mt};\hat{z}_{mt}(i), v^z_{mt}(i) \big)
\end{align}
where $\hat{z}_{mt}(i)$ and $v^z_{mt}(i)$ denote the posterior mean and variance of $z_{mt}$ and they are given by
\begin{align}
\hat{z}_{mt}(i) & = \frac{\sigma^2 \hat{p}_{mt}(i)+v^p_{mt} (i)  y_{mt}}{\sigma^2+v^p_{mt} (i)}. \label{z-hat} \\
v^z_{mt}(i) & = \frac{\sigma^2 v^p_{mt}(i)}{\sigma^2+v^p_{mt}(i)}, \label{v-z}
\end{align}
Similarly, by taking the product of all incoming messages at the variable nodes $\{g_{mn}\}$, $\{h_{mk}\}$, and $\{f_{nk}\}$, we obtain
\begin{align}
\Delta^g_{mn}(i+1, g_{mn}) & \propto p(g_{mn}) \prod_{t} \Delta_{p(z_{mt}) \rightarrow g_{mn}} (i, g_{mn})\notag \\
& \approx p(g_{mn}) \mcl{CN} \big(g_{mn};\hat{q}^g_{mn}(i), v_{mn}^{qg}(i) \big), \label{po-g}\\
\Delta^h_{mk}(i+1, h_{mk}) & \propto p(h_{mk}) \prod_{t} \Delta_{p(z_{mt}) \rightarrow h_{mk}} (i, h_{mk}) \notag \\
& \approx p(h_{mk}) \mcl{CN}\left(h_{mk};\hat{q}^h_{mk}(i),v_{mk}^{qh}(i)\right), \label{po-h}
\end{align}

\begin{align}
 \Delta^f_{nk}(i+1, f_{nk}) & \propto p(f_{nk}) \prod_{t} \Delta_{p(c_{nt}) \rightarrow f_{nk}} (i, f_{nk}) \notag \\
 & \approx p(f_{nk}) \mcl{CN} \left(f_{nk};\hat{q}^f_{nk}(i),v_{nk}^{qf}(i)\right). \label{po-f}
\end{align}
The approximated posterior means and variances of $g_{mn}$, $h_{mk}$, and $f_{nk}$ are given by
\begin{align}
\hat{g}_{mn}(i+1) & = \mbs{E} \big\{ g_{mn}|\hat{q}^g_{mn}(i), v_{mn}^{qg}(i) \big\}, \label{post-m-g}\\
v_{mn}^g (i+1) & = \mbs{var}\big\{ g_{mn}|\hat{q}^g_{mn}(i), v_{mn}^{qg}(i) \big\}, \label{post-v-g}\\
\hat{h}_{mk}(i+1) & =\mbs{E} \big\{ h_{mk}|\hat{q}^h_{mk}(i), v_{mk}^{qh}(i)\big\}, \label{post-m-h}\\
v_{mk}^h(i+1) & = \mbs{var} \big\{ h_{mk}| \hat{q}^h_{mk}(i), v_{mk}^{qh}(i)\big\}, \label{post-v-h}\\
\hat{f}_{nk}(i+1) & =\mbs{E} \big\{ f_{nk}|\hat{q}^f_{nk}(i), v_{nk}^{qf}(i)\big\}, \label{post-m-f}\\
v_{nk}^f(i+1) & = \mbs{var} \big\{ f_{n,k}|\hat{q}^f_{nk}(i), v_{nk}^{qf}(i) \big\}, \label{post-v-f}
\end{align}
where the expectation and variance operations are taken over the posterior distributions in \eqref{po-g}--\eqref{po-f}, respectively.

By using the messages \eqref{delta-pzmt-cmk3} and \eqref{delta-pc-c}, we obtain the approximated marginal posterior of $c_{nt}$ (when $s_{nt} = 1$):
\begin{align}
& \Delta^c_{nt} (i, c_{nt}) \notag \\
& \propto \Delta_{p(c_{nt} |\cdot) \rightarrow c_{nt}} (i, c_{nt}) \prod_{m} \Delta_{p(z_{mt}) \rightarrow c_{nt}} (i, c_{nt}) \notag \\
 & \approx   \mcl{CN} \left(c_{nt}; \hat{\xi}_{nt}(i), v_{nt}^{\xi}(i) \right) \mcl{CN} \left(c_{nt}; \hat{r}_{nt}^c (i) ,v_{nt}^{rc} (i)\right), \label{pos-final-c}
\end{align}
from which the mean and variance of $c_{nt}$ are given by
\begin{align}
\hat{c}_{nt}(i) & = s_{nt} \mbs{E} \big\{ c_{nt}|\hat{r}_{nt}^c(i),v_{nt}^{rc}(i), \hat{\xi}_{nt}(i), v_{nt}^{\xi}(i) \big\}, \label{m-c-nt} \\
v_{nt}^c(i) & = s_{nt} \mbs{var} \big\{ c_{nt}|\hat{r}_{nt}^c(i),v_{nt}^{rc}(i), \hat{\xi}_{nt}(i), v_{nt}^{\xi}(i) \big\}. \label{m-v-nt}
\end{align}

Likewise, with the messages \eqref{delta-pzmt-xmk3} and \eqref{delta-pcnt-xkt3}, the approximated marginal posterior  of $x_{kt}$ is given by
\begin{align} \label{EE-X}
& \Delta^x_{kt}(i+1, x_{kt}) \notag \\
& \propto p(x_{kt}) \prod_{m} \Delta_{p(z_{mt}) \rightarrow x_{kt}} (i, x_{kt}) \prod_{n} \Delta_{p(c_{nt}) \rightarrow x_{kt}} (i, x_{kt}) \notag \\
& \approx p(x_{kt})\,\mcl{CN}(x_{kt}; \hat{r}^{r}_{kt}, v^{r x}_{kt})\,\mcl{CN}(x_{kt}; \hat{\gamma}^x_{kt}, v^{\gamma x}_{kt}),
\end{align}
which leads to the updates of the mean and variance of $x_{kt}$:
\begin{align}
\hat{x}_{kt}(i+1) & = \mbs{E} \big\{ x_{kt}|\hat{r}^x_{kt}(i), \hat{\gamma}^x_{kt}(i), v^{rx}_{kt}(i), v^{\gamma x}_{kt}(i)\big\}, \label{pos-m-x}\\
v_{kt}^x(i+1) & = \mbs{var} \big\{ x_{kt}|\hat{r}^x_{kt}(i), \hat{\gamma}^x_{kt}(i), v^{rx}_{kt}(i), v^{\gamma x}_{kt}(i) \big\}. \label{pos-v-x}
\end{align}
As the first $T_\mathsf{p}$ columns of $\bsm{X}$ are used as pilots, i.e.,  $\bsm{X}_\mathsf{p} = \overline{\bsm{X}}_\mathsf{p}$, where $\overline{\bsm{X}}_\mathsf{p}$ is a {\it known} pilot matrix, from \eqref{X-deco} and \eqref{pilots-delta} we have
\begin{align}
p(x_{kt}) = \delta(x_{kt} - \overline{x}_{\mathsf{p},kt}), t \in \{1,\ldots,T_\mathsf{p}\}. \label{pilot2}
\end{align}
Substituting \eqref{pilot2} into \eqref{pos-m-x} and \eqref{pos-v-x} yields
\begin{align}
\hat{x}_{kt}(i+1) & = \overline{x}_{\mathsf{p},kt},\,t \in \{1,\ldots,T_\mathsf{p}\}, \\
v_{kt}^x (i+1) & = 0,\,t \in \{1,\ldots,T_\mathsf{p}\}.
\end{align}

\subsubsection{The overall algorithm}
For clarification, we summarize the overall iterative procedure above in Algorithm \ref{algorithm1}, refereed to as the trilinear AMP (Tri-AMP) algorithm. Note that Tri-AMP relies on the knowledge of the prior channel distributions and the noise variance $\sigma^2$. Nevertheless, these unknown parameters can be learned by utilizing an expectation-maximization (EM) based approach  \cite{parker2014bilinear,bayes2008variational}. Specifically, the EM approach can be implemented by integrating with the Tri-AMP algorithm in the expectation step to obtain the marginal posterior distributions and leaning the unknown parameters in the maximization step.

Algorithm \ref{algorithm1} can be reduced to the case where the direct link is absent, i.e., $\bsm{H}=\bsm{0}$ and the dashed blue lines in the factor graph of Fig. \ref{factor1} do not exist. Due to limited space, we omit the algorithmic details in this case.

\begin{spacing}{2.1}
\begin{algorithm}[t]
\vspace{0.1mm}
\caption{\textbf{\!:} \vspace{0.15mm} Tri-AMP Algorithm}
\label{algorithm1}
\begin{algorithmic}[1]
\xiaowuhao
\REQUIRE $\bsm{Y}$, $\overline{\bsm{X}}_\mathsf{p}$, $\bsm{S}$, $\sigma^2$, prior distributions $p(\bsm{G})$, $p(\bsm{F})$, $p(\bsm{H})$, and \text{~~~~~~\,}$p(\bsm{X}_\mathsf{d})$ \\
\vspace{0.01cm}
\STATE Initialization: $\forall m,n,k, t \in \{T_\mathsf{p}+1,\ldots,T \}$: choose $\hat{g}_{mn}(1)$, $v^{g}_{mn}(1) $, $\hat{f}_{nk}(1) $, $v^{f}_{nk}(1) $, $\hat{h}_{mn}(1) $, $v^{h}_{mn}(1) $, $\hat{x}_{kt}(1) $, $v^{x}_{kt}(1) $; $\forall k,t \in \{1,\ldots,T_\mathsf{p}\}$: $\hat{x}_{kt}(1) = \overline{x}_{\mathsf{p},kt}$, $v^{x}_{kt}(1) =0 $; $\forall n,k,t$: $\hat{u}_{mt}(-1) = 0$, $\hat{\eta}_{nt}(-1) = 0$, $\hat{c}_{nt}(1) = s_{nt} \sum_{k=1}^K \hat{f}_{nk} \hat{x}_{kt}$, $v^c_{nt}(1) = s_{nt} \sum_{k=1}^K v^f_{nk}v^x_{kt}$ \\
\STATE \textbf{for} $i=1,\ldots,I_{\rm max}$ \\
\STATE~~\,\,$\forall m,t$: Update $\hat{z}_{mt}(i)$ and $v_{mt}^z(i)$ via \eqref{hat-p}--\eqref{bar-p}, \eqref{z-hat}, and \eqref{v-z}; \\
\STATE~~\,\,$\forall m,n$: Update $\hat{g}_{mn}(i+1)$ and $v_{mn}^g(i+1)$ via \eqref{v-u-mt}, \eqref{m-u-mt}, \eqref{tmn-gq},  \\
~~~~~~~~~~~~~~~\eqref{g-g-mn}, \eqref{po-g}, \eqref{post-m-g}, and \eqref{post-v-g}; \\
\STATE~~\,\,$\forall m,k$: Update $\hat{h}_{mk}(i+1)$ and $v_{mk}^h(i+1)$ via \eqref{qq-h-v}--\eqref{qq-h-m}, \eqref{po-h},  \\
~~~~~~~~~~~~~~~\eqref{post-m-h}, and \eqref{post-v-h}; \\
\STATE~~\,\,$\forall n,t$: Update $\hat{c}_{nt}(i)$ and  $v_{nt}^c(i)$ via \eqref{rmnt-mc}, \eqref{rmnt-vc}, \eqref{xi-m}--\eqref{xi-v}, \eqref{pos-final-c}, \\
~~~~~~~~~~~~~~~\eqref{m-c-nt}, and \eqref{m-v-nt}; \\
\STATE~~\,\,$\forall n,k$: Update $\hat{f}_{nk}(i+1)$ and $v_{nk}^f(i+1)$ via \eqref{q-f-variance}, \eqref{q-f-mean}, \eqref{po-f}, \\
~~~~~~~~~~~~~~~\eqref{post-m-f}, and \eqref{post-v-f}; \\
\STATE~~\,\,$\forall k,t \in \{T_\mathsf{p}+1,\ldots,T \} $: Update $v_{kt}^x(i+1) $, $\hat{x}_{kt}(i+1) $ via \\
~~~~~~~~~~~~~~~\eqref{rktxi-v}, \eqref{rktxi-m}, \eqref{rx-v}, \eqref{rx-m}, and  \eqref{EE-X}--\eqref{pos-v-x}; \\ \vspace{0.02cm}
\STATE~~\,\,$\forall k,t \in \{1,\ldots,T_\mathsf{p} \}$: $v_{kt}^x(i+1) = 0$, $\hat{x}_{kt}(i+1) = \overline{x}_{\mathsf{p},kt}$;
\STATE~~\textbf{if} {\it a certain stopping criterion is met},
\textbf{stop}
\STATE\textbf{end for}
\end{algorithmic}
\end{algorithm}
\end{spacing}

\vspace{-0.5cm}
\subsection{Damping}

The approximations used in Tri-AMP are justified only in the large system limit. In other words, the approximated messages might not come close to Gaussian distributions in practice, particularly with the finite dimensions of $M,N,K,T_{\mathsf{p}},T_{\mathsf{d}}$. In this case, the  Tri-AMP algorithm appears to have some unexpected numerical issues and diverges. This is also observed in other AMP-based approaches \cite{parker2014bilinear, liu2019matrix}. In this work, we employ the damping method to improve the numerical robustness of Tri-AMP. Specifically, in each iteration we smoothen the updates of the variances and means of ${\bsm{G}}$, ${\bsm{F}}$, ${\bsm{H}}$, $\bsm{C}$, and ${\bsm{X}}_{\mathsf{d}}$ by using a convex combination of the current and previous updates. For example, the updates of the posterior variance and mean of $g_{mn}$ in Line 4 of Algorithm \ref{algorithm1} are replaced by
\begin{align}
v_{mn}^g (i+1) & \triangleq (1-\beta) v_{mn}^g (i) + \beta v_{mn}^g (i+1), \\
\hat{g}_{mn} (i+1) &\triangleq (1-\beta) \hat{g}_{mn} (i) + \beta \hat{g}_{mn} (i+1),
\end{align}
where $\beta \in [0,1]$ is the damping factor. Our experiments under Rayleigh fading channels and QPSK/Gaussian signals suggest that choosing $\beta$ within $[0.1, 0.2]$ leads to a good performance.

\subsection{Computational Complexity}

We now briefly discuss the computational complexity of the proposed Tri-AMP algorithm. Note that the total computational complexity of Tri-AMP is due to the computation in both the outer bilinear inference and the inner bilinear inference. We therefore  sketch the respective complexity as follows. First, the complexity of the outer bilinear inference (Lines $3$-$5$ of Algorithm \ref{algorithm1}) requires $\mcl{O}(M(N+K)T)$ flops per iteration. Second, the complexity of the inner bilinear inference (Lines $6$-$9$ of Algorithm \ref{algorithm1}) requires $\mcl{O}(NKT)$ flops per iteration. As a consequence, the total computational complexity of Tri-AMP is $I_{\rm max} \left( \mcl{O}(M(N+K)T) + \mcl{O}(NKT) \right)$ flops, where $I_{\rm max}$ is the maximum number of iterations required in Tri-AMP.



\section{Asymptotic Performance Analysis}
\label{Section-V}

The previous section presents an AMP-based iterative algorithm to approximately calculate the posterior mean estimators in \eqref{postetior-all}. However, it is unclear whether the proposed algorithm can approach the theoretical MSEs in \eqref{MSE-all}, which are generally difficult to evaluate. To this end, we use the \emph{replica method} \cite{SpinGlass} to evaluate the asymptotic performance of the MSEs \eqref{MSE-all}  in the large-system limit. Note that despite not being mathematically rigorous, the replica method has proven successful in analysing the asymptotic performance of the bilinear inference problems \cite{MP_BIGAMP3,wen2015bayes} and the matrix calibration inference problems \cite{liu2019matrix}. In this section, we generalize the analytical results in \cite{wen2015bayes} to the trilinear inference problem \eqref{received-matrix}, which includes the bilinear inference problem in \cite{wen2015bayes} as a special case when the user-RIS-BS link is absent. Specifically, we shall illustrate that the MSEs \eqref{MSE-all} asymptotically converge to the MSEs of scalar additive white Gaussian noise (AWGN) channels with tractable expressions.

Our asymptotic analysis is carried out under the large-system limit, i.e., $M,N,K,T_\mathsf{p},T_\mathsf{d}$ with the ratios $M/K$, $N/K$, $T_\mathsf{p}/K$, $T_\mathsf{d}/K$ being fixed and finite. For simplicity, we utilize $K\to \infty$ to denote this large-system limit. In addition, we assume that the elements of $\bsm{H}$, $\bsm{G}$, $\bsm{F}$, $\bsm{X}$, and $\bsm{W}$ are all i.i.d. variables, and scale in the same orders as those in Section IV B. The analytical result is elaborated in Proposition \ref{pro1} of Section \ref{4-B}. Before proceeding to the main result, we first introduce preliminaries involved in the main result.


\subsection{Preliminaries}\label{4-A}
 Similarly to \eqref{X-deco}, we partition $\bsm{C}$ and $\bsm{Z}$ respectively as
\begin{align}
&\bsm{C} \triangleq [\bsm{C}_\mathsf{p},\bsm{C}_\mathsf{d}]~\text{ with }~\bsm{C}_\mathsf{p} \in \mbs{C}^{N\times T_\mathsf{p}} ~ \text{and}~\bsm{C}_\mathsf{d} \in \mbs{C}^{N\times T_\mathsf{d}}, \\
& \bsm{Z} \triangleq [\bsm{Z}_\mathsf{p},\bsm{Z}_\mathsf{d}]~\text{ with }~\bsm{Z}_\mathsf{p} \in \mbs{C}^{M\times T_\mathsf{p}} ~ \text{and}~\bsm{Z}_\mathsf{d} \in \mbs{C}^{M\times T_\mathsf{d}}.
\end{align}
To facilitate the analysis, we define the following second-order moments:
\begin{subequations} \label{temp120}
\begin{align}
q_g & \triangleq \mbs{E}\{|{g_{mn}}|^2\},\,q_h \triangleq  \mbs{E}\{|{h_{mk}}|^2\},\,q_f\triangleq\mbs{E}\{|{f_{nk}}|^2\}, \\
q_{x_\mathsf{p}}& \triangleq \mbs{E}\{|{x_{\mathsf{p},kt}}|^2\},\,q_{c_\mathsf{p}} \triangleq \mbs{E}\{|{c_{\mathsf{p},nt}}|^2\}=\rho K q_f q_{x_\mathsf{p}}, \\
q_{x_\mathsf{d}}& \triangleq \mbs{E}\{|{x_{\mathsf{d},kt}}|^2\},\,q_{c_\mathsf{d}} \triangleq \mbs{E}\{|{c_{\mathsf{d},nt}}|^2\}=\rho K q_f q_{x_\mathsf{d}},
\end{align}
\end{subequations}
where the expectations are taken over the prior distributions in \eqref{g-f-prior} and \eqref{h-x-prior}. Moreover, we define the following scalar AWGN channels:
\begin{subequations}\label{Scalar_Channel}
	\begin{align}
y_g & \triangleq \sqrt{\widetilde m_g} g+w_g,\,y_f \triangleq \sqrt{\widetilde m_f}  f+w_f, \\
y_h&\triangleq \sqrt{\widetilde m_h} h+w_h,\, y_{x_\mathsf{d}} \triangleq \sqrt{\widetilde m_{x_\mathsf{d}}} {x_\mathsf{d}}+w_{x_\mathsf{d}}, \label{A12}
\end{align}
\end{subequations}
where $\{w_h,w_g,w_f,w_{x_d}\} \sim \CN(\cdot\,;0,1)$, $g \sim p(g_{mn})$, $f \sim $ $p(f_{nk})$, $h \sim p(h_{mk})$, $x_\mathsf{d} \sim p({x_{\mathsf{d},kt}})$. The parameters $\{\widetilde m_g, $ $\widetilde m_f,\widetilde m_h,\widetilde m_{x_d}\}$ will be specified later in \eqref{temp04}. By Bayes' theorem, for given $\{\widetilde m_g, \widetilde m_f,\widetilde m_h,\widetilde m_{x_\mathsf{d}}\}$, we attain the following posterior distributions:
\begin{subequations}
\begin{align}
&p(g|y_g)\propto p(y_g|g)\,p(g_{mn}),\,p(f|y_f)\propto p(y_f|f)\,p(f_{nk}) , \notag \\	
&p(h|y_h)\propto p(y_h|h)\,p(h_{mk}),\,p(x_\mathsf{d}|y_{x_d})\propto p(y_{x_\mathsf{d}}|x_\mathsf{d})\,p(x_{\mathsf{d},kt}), \notag
\end{align}
\end{subequations}
where $p(y_g|g) =\CN \big(y_g;\sqrt{\widetilde m_g} g,1\big)$, $p(y_f|f)=\CN\big(y_f; $ $\sqrt{\widetilde m_f} f,1\big)$, $p(y_h|h)=\CN \big(y_h; \sqrt{\widetilde m_h} h,1\big)$, and $p(y_{x_\mathsf{d}}|{x_\mathsf{d}})=\CN \big(y_{x_\mathsf{d}};\sqrt{\widetilde m_{x_\mathsf{d}}} {x_\mathsf{d}},1 \big)$.

The posterior mean estimators of $g,f,h$, and $x_\mathsf{d}$ with respect to the scalar AWGN channels in \eqref{Scalar_Channel} are given by
\begin{subequations}\label{A02}
	\begin{align}
		 \hat g & =\int  p(g|y_g)g{\rm{d}} g,\,\hat f  =\int  p(f|y_f)f {\rm{d}} f\\
		 \hat h & =\int  p(h|y_h)h {\rm{d}} h,\,\hat x_\mathsf{d} =\int  p(x_\mathsf{d}|y_{x_\mathsf{d}}) x_\mathsf{d} {\rm{d}} x_\mathsf{d}.
	\end{align}
\end{subequations}
Consequently, the MSEs of the estimators in \eqref{A02} are given by
\begin{align}
\text{MSE}_{g}&=\mbs{E}_{g,y_g} \left\{|g-{\hat g}|^2\right\}, \label{B04} \\
\text{MSE}_{f}&=\mbs{E}_{f,y_f}  \big\{|f-{\hat f}|^2\big\}, \label{B03} \\
\text{MSE}_{h}&=\mbs{E}_{h,y_h}  \big\{|h-{\hat h}|^2\big\}, \label{BB02} \\
\text{MSE}_{x_\mathsf{d}}&=\mbs{E}_{x_\mathsf{d},y_{x_\mathsf{d}}} \big\{|x_\mathsf{d}-{\hat x_\mathsf{d}}|^2\big]. \label{B01}
\end{align}

\subsection{Main Result} \label{4-B}

Under some commonly used assumptions in the replica analysis, we have the following large-system limit performance on the MSEs in \ref{MSE-all} of the considered semi-blind cascaded channel estimation problem.

\begin{pro} \label{pro1}  When $K \to \infty$, $\{\text{MSE}_{\bsm{G}}, $ $ \text{MSE}_{\bsm{F}},  \text{MSE}_{\bsm{H}}, $ $ \text{MSE}_{\bsm{X}_{\mathsf{d}}} \}$ converges to $\{\text{MSE}_{g},$ $ \text{MSE}_{f}, $ $\text{MSE}_{h},$ $\text{MSE}_{x_\mathsf{d}}\}$ that  corresponds to a fixed-point solution of \eqref{B04}--\eqref{B01} and the following equations:
\begin{subequations}\label{temp04}
\begin{align}
m_g&=q_g-\text{MSE}_{g},\widetilde m_g= T_\mathsf{p}m_{c_\mathsf{p}} a_\mathsf{p}+  T_\mathsf{d}m_{c_\mathsf{d}} a_\mathsf{d},\\
m_f&=q_f-\text{MSE}_{f},\widetilde m_{f}=T_\mathsf{p} q_{x_\mathsf{p}} b_\mathsf{p}+T_\mathsf{d} m_{x_\mathsf{d}} b_\mathsf{d}, \\
m_h&=q_h-\text{MSE}_{h},\widetilde m_h=T_\mathsf{p} q_{x_\mathsf{p}} a_\mathsf{p}+T_\mathsf{d} m_{x_\mathsf{d}} a_\mathsf{d}, \\
m_{x_\mathsf{d}} &=q_{x_\mathsf{d}}-\text{MSE}_{x_\mathsf{d}},\widetilde m_{x_\mathsf{d}}=Nm_f b_\mathsf{d} + Mm_h a_\mathsf{d},
\end{align}
\end{subequations}
where
\begin{subequations}\label{temp044}
\begin{align}
a_\mathsf{p}&=\frac{\rho}{\sigma^2+\rho N(q_g q_{c_\mathsf{p}}-m_gm_{c_\mathsf{p}})+Kq_{x_\mathsf{p}}(q_h-m_h)}, \\
a_\mathsf{d} &=\frac{\rho}{ \sigma^2+\rho N(q_gq_{c_\mathsf{d}}-m_gm_{c_\mathsf{d}})+K(q_hq_{x_\mathsf{d}}-m_hm_{x_\mathsf{d}})}, \\
b_\mathsf{p}&=\frac{\rho}{ 1/(M m_g a_{\mathsf{p} })+Kq_{x_\mathsf{p}}(q_f-m_f)},  \\
b_\mathsf{d} &=\frac{\rho}{ 1/ (M m_g a_{\mathsf{d} })+K(q_f q_{x_\mathsf{d}}-m_fm_{x_\mathsf{d}})}, \\
m_{c_\mathsf{o}}&=q_{c_\mathsf{o}}-\frac{K(q_{x_\mathsf{o}}q_f-m_{x_{\mathsf{o}}}m_f)}{1+MKm_ga_{\mathsf{o}}(q_{x_\mathsf{o}}q_f-m_{x_{\mathsf{o}}}m_f)}. \label{A03}
\end{align}
\end{subequations}
Herein, $\mathsf{o}\in\{\mathsf{p},\mathsf{d}\}$; $q_g$, $q_f$, $q_h$, $q_{x_\mathsf{o}}$, and $q_{c_\mathsf{o}}$ are defined in \eqref{temp120}.
\end{pro}

The proof of Proposition \ref{pro1} is given in Appendix \ref{appa0}. It follows from Proposition \ref{pro1} that the fixed-point of \eqref{B04}--\eqref{temp04} asymptotically describes the theoretical MSEs in \eqref{MSE-all}. The fixed-point solution can be efficiently calculated by an iterative algorithm that sequentially updates $\big\{\text{MSE}_{i},\widetilde m_i,m_i,i\in\{g,f,h,x_\mathsf{d}\} \big\}$ via \eqref{B04}--\eqref{temp04} until convergence.

\subsection{Case Study: Rayleigh Fading Channels and QPSK Signals} \label{4-C}

The final expressions of the asymptotic MSEs derived in Section \ref{4-B} depend on the exact forms of the prior distributions $p(g_{mn})$, $p(f_{nk})$, $p(h_{mk})$, and $p({x_{kt}})$. We can simplify the integrals involved in \eqref{B04}--\eqref{B01} into closed-form expressions once the associated prior distributions are specified. To illustrate this, we focus on a concrete case with Rayleigh fading channels and QPSK  signals. That is, we set $p(g_{mn})= $ $\CN(g_{mn};0,q_g)$,
$p(f_{nk})=\CN(f_{nk};0,q_f)$, and $p(h_{mk}) = $ $\CN(h_{mk};0,q_h)$. The training signal $\bsm{X}_\mathsf{p}$ and the data signal $\bsm{X}_\mathsf{d}$ are both drawn from a QPSK constellation with unit variance and equal probabilities.

By substituting the prior distributions specified above into \eqref{B04}--\eqref{B01}, we attain
\begin{align}
&\text{MSE}_{g}=\frac{q_g}{1+q_g\widetilde m_g}, \\
&\text{MSE}_{f}=\frac{q_f}{1+q_f\widetilde m_f}, \\
&\text{MSE}_{h}=\frac{q_h}{1+q_h\widetilde m_h}, \\
&\text{MSE}_{x_\mathsf{d}}= 1-\int {\rm{D}}\zeta \tanh(\widetilde m_{x_\mathsf{d}}+\sqrt{\widetilde m_{x_\mathsf{d}}}\zeta),
\end{align}
where ${\rm{D}}\zeta =\frac{1}{\sqrt{2\pi}}e^{-\zeta^2/2}{\rm{d}}\zeta $ is a Gaussian integration measure. By evaluating the second scalar AWGN channel in \eqref{A12} according to \cite[Page 269]{DCBOOK}, the asymptotic symbol error rate (SER) of $\bsm{X}_\mathsf{d}$ is given by
\begin{align}
\text{SER}=2\mathcal{Q}(\sqrt{\widetilde m_{x_\mathsf{d}}})-\left(\mathcal{Q}(\sqrt{\widetilde m_{x_\mathsf{d}}}) \right)^2
\end{align}
where $\mathcal{Q}(x)=\int_{x}^{\infty} {\rm{D}}\zeta$ is the Q-function.

\begin{figure}[t]
    \centering
     \includegraphics[scale=0.322]{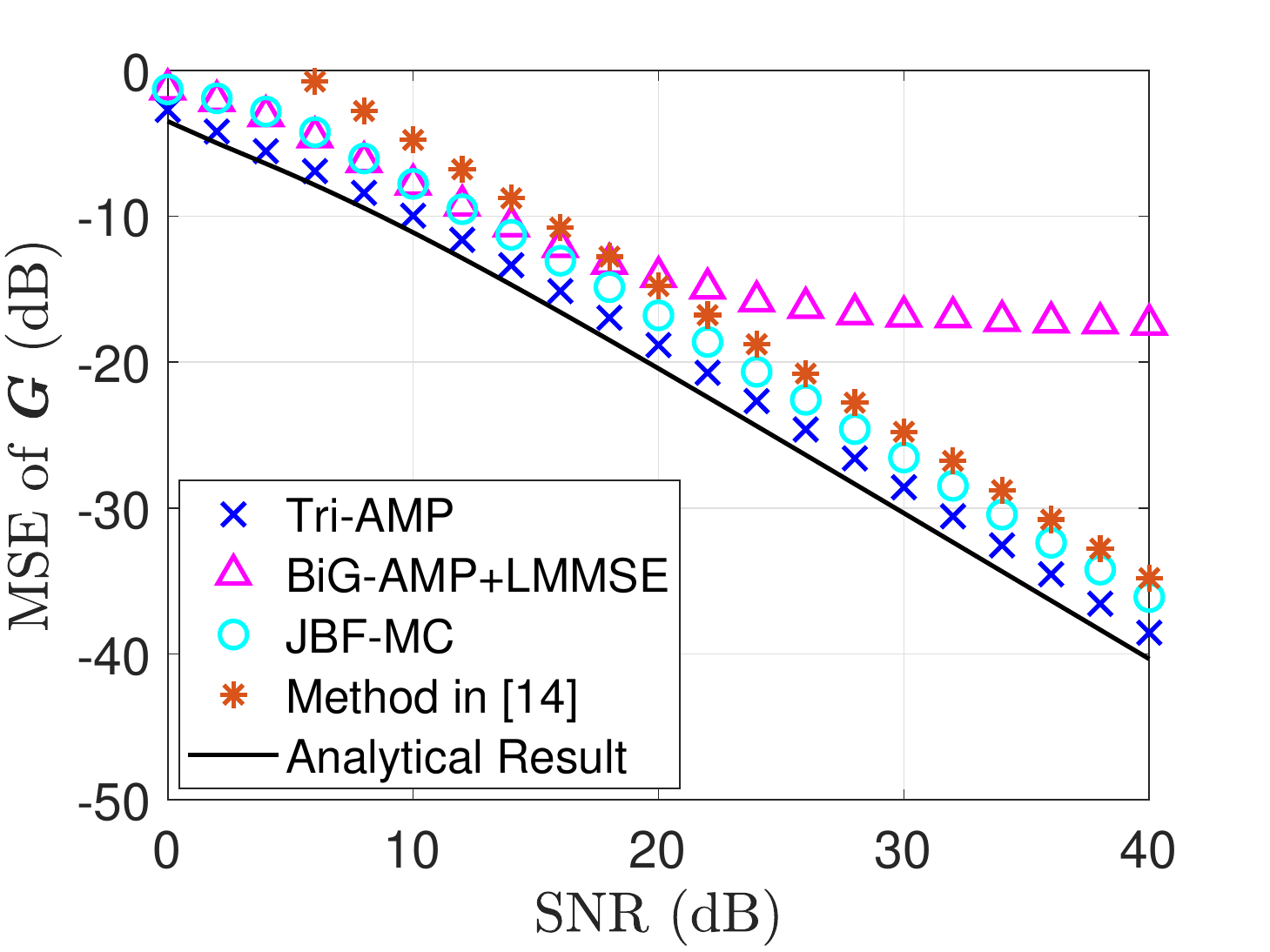}\,\,\includegraphics[scale=0.322]{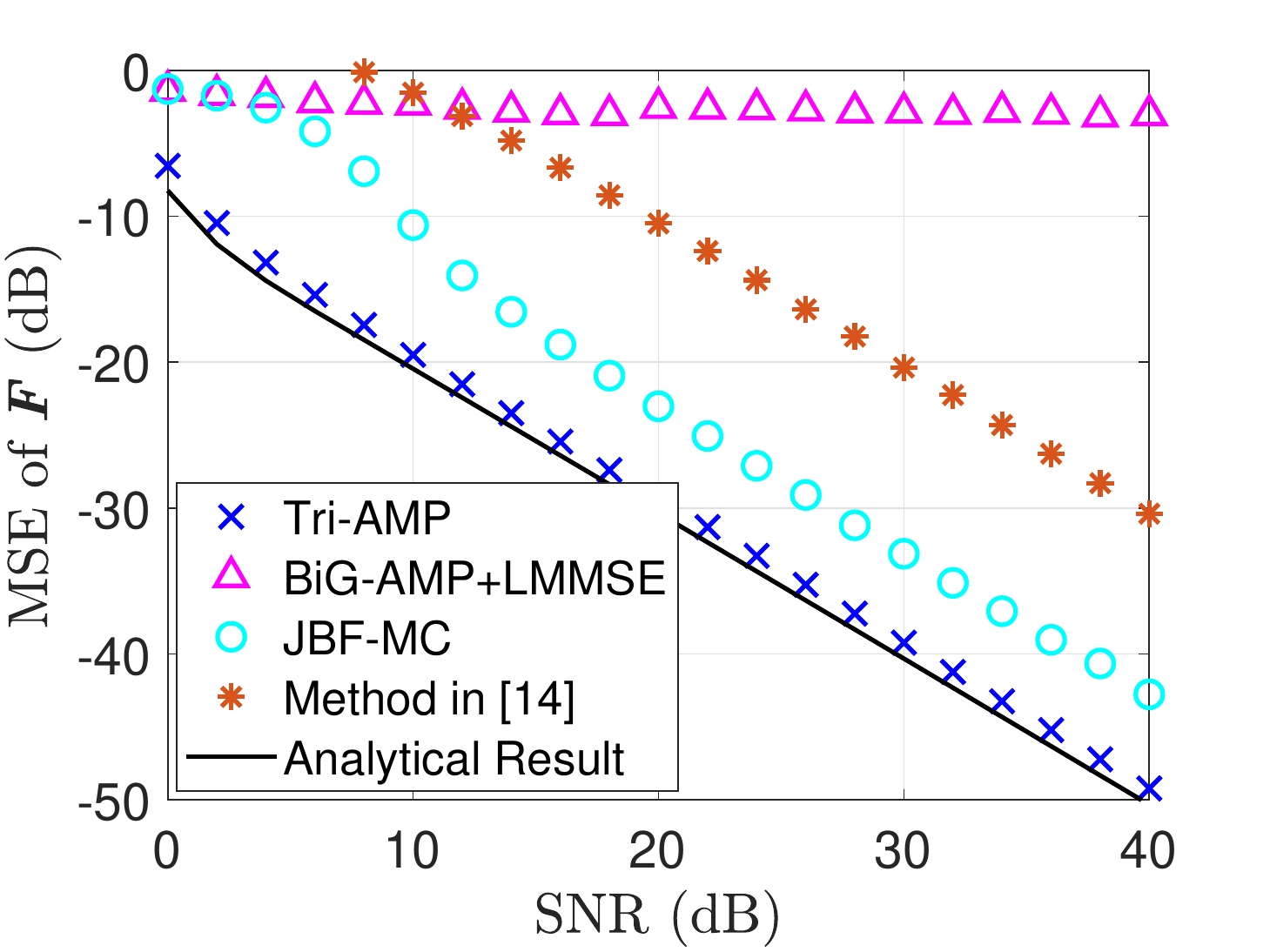} \\
    \smallskip \smallskip
    \includegraphics[scale=0.322]{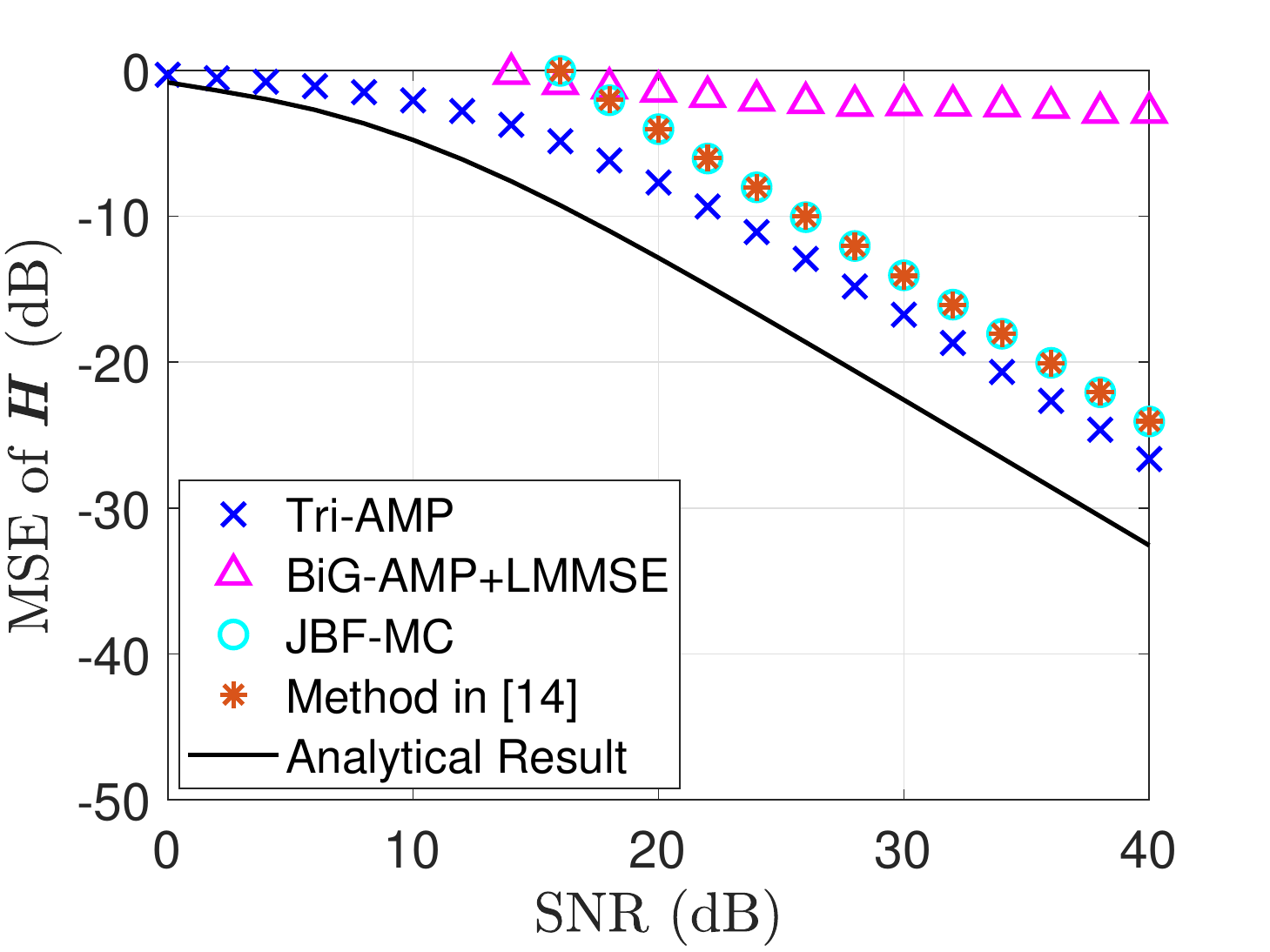}\,\,\includegraphics[scale=0.322]{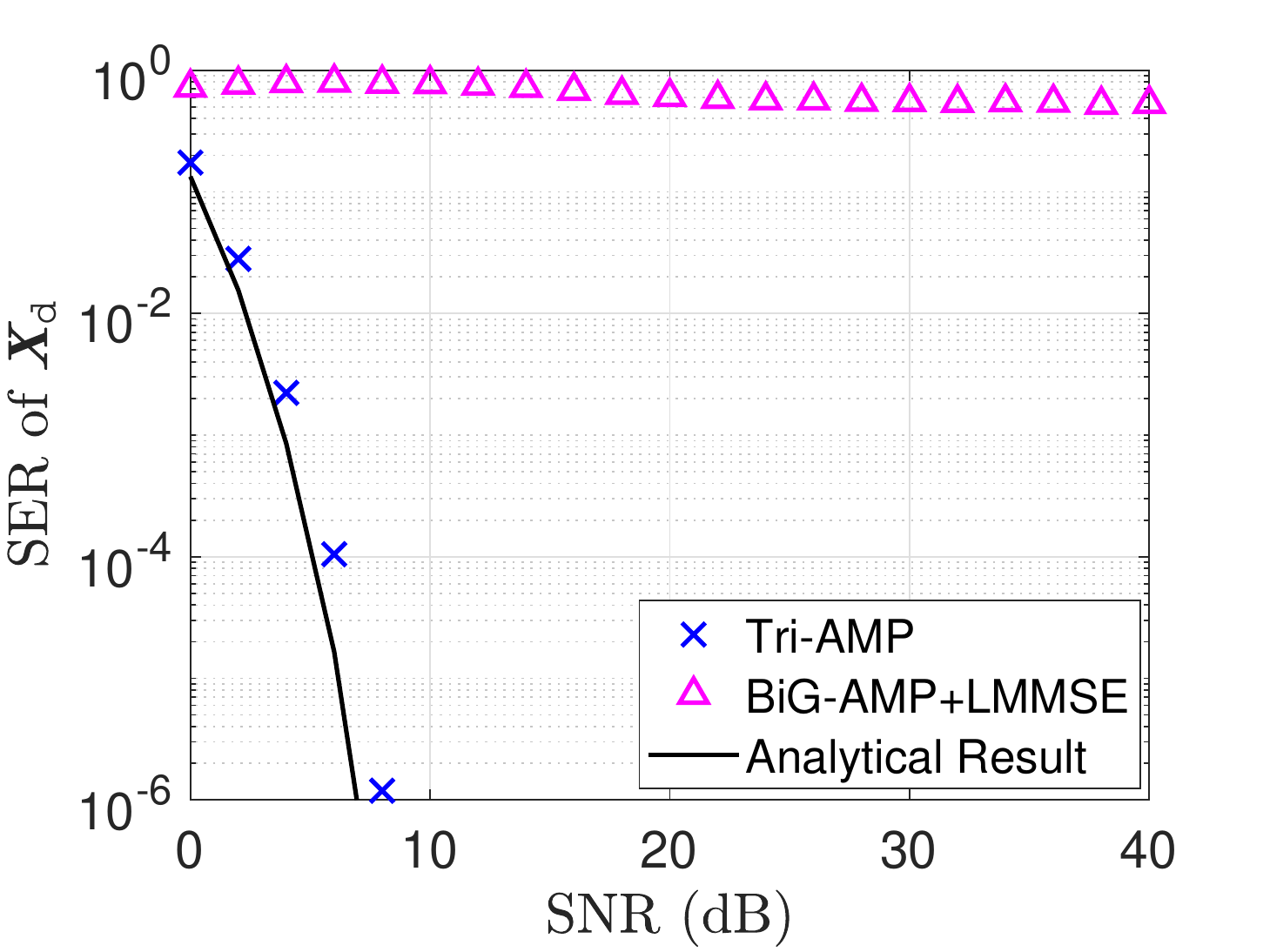}\\
    \vspace{-1mm}
    \caption{MSEs of $\bsm{G}$, $\bsm{F}$, and $\bsm{H}$, and SER of $\bsm{X}_\mathsf{d}$ versus the SNR with $M = 256$, $N=128$, $K=20$, and $T = 300$ under i.i.d. Rayleigh fading channels.}
    \label{fig3}
\end{figure}

\begin{table}[t]
\xiaowuhao
\caption{Minimum training length of respective algorithms versus the block length $T$ in the noiseless case with $M=256$, $N=128$, $K=20$ under i.i.d. Rayleigh fading channels.}
\centering
\begin{threeparttable}
\begin{tabular}{|c|c|c|c|c|c|c|c|}
\hline\hline
\multirow{2}{*}{\diagbox{\hspace{-0.25cm}\,\,Algorithm\!\!\!\!\!}{$T$\!\!}}
& \multirow{2}{*}{100} & \multirow{2}{*}{400} & \multirow{2}{*}{600} &\multirow{2}{*}{1000} & \multirow{2}{*}{2000} & \multirow{2}{*}{3000} \\
& & & & & & \\
\hline
Tri-AMP & \ding{55} & $23$ & $20$ & $20$ & $20$ & $20$  \\ \hline
\!\!BiGAMP+LMMSE\!\! & \ding{55} & \ding{55} & $160$ & $150$ & $146$ & 140 \\
\hline
JBF-MC \cite{he2019cascaded} & \ding{55} & \ding{55} & 465 & 465 & 465 & 465 \\
\hline
Method in \cite{wang2019channel} & \ding{55} & $167$ & $167$ & $167$ & $167$ & $167$ \\
\hline
Method in \cite{jensen2020optimal} & \ding{55} & \ding{55} & \ding{55} & \ding{55} & \ding{55} & $2580$ \\
\hline
PARAFAC \cite{wei2020channel} & \ding{55} & \ding{55} & \ding{55} & $800$ & $800$  & $800$ \\
\hline
Replica Result  & \ding{55} & $0$ & $0$& $0$ & $0$ & $0$ \\
\hline\hline
\end{tabular}
\vspace{-2pt}
\begin{tablenotes}
        \item[a] ``\ding{55}'' indicates that the associated scheme is infeasible even when $T_\mathsf{p} = T$.
\end{tablenotes}
\begin{tablenotes}
        \item[b] Here ``0'' in the replica result indicates that, under the specified parameter ratios of $M/K$, $N/K$, $T_\mathsf{p}/K$, $T_\mathsf{d} / K$, the asymptotic MSEs of $\bsm{G}$, $\bsm{F}$, and $\bsm{H}$ are less than $-60$ dB and the asymptotic SER of $\bsm{X}_{\mathsf{d}}$ is zero in the large system limit, even without the use of any pilot symbols. It is worth noting that the trilinear estimation problem with zero pilots suffers from the phase and permutation ambiguities when factorizing $\bsm{F}$ and $\bsm{X}$ from the product $\bsm{F} \bsm{X}$; see the detailed discussions on the phase and permutation ambiguities in \cite{zhang2018blind}. The Bayesian inference (including the replica method) cannot resolve these ambiguities. Additional reference symbols in $\bsm{X}$ are required if one needs to resolve these ambiguities \cite{zhang2018blind}.
\end{tablenotes}
\end{threeparttable}
\label{table1}
\end{table}


\section{Numerical Results}

\label{Section-VI}

This section conducts numerical experiments to corroborate the semi-blind channel estimation performance of the proposed Tri-AMP algorithm in a RIS-aided massive MIMO system. The payload data $\bsm{X}$ is generated from the i.i.d. QPSK constellation with transmit power one. All the simulation results are obtained with the setup $M=256$, $N=128$, and $K=20$ by averaging $2,000$ independent trials unless otherwise specified. The signal-to-noise ratio (SNR) is defined as
\begin{align} \label{SNR}
\text{SNR} & = \frac{\sum_{t=1}^T \mbs{E} \left\{\big\| \bsm{G} \big( \bsm{s}^{(t)} \odot (\bsm{F} \bsm{x}^{(t)} ) \big) + \bsm{H} \bsm{x}^{(t)} \big\|_2^2 \right\} }{T \sigma^2} \notag \\
& = \frac{\rho N  K \tau_g \tau_f \tau_x + K \tau_h \tau_x}{\sigma^2},
\end{align}
where $\tau_g$, $\tau_g$, $\tau_h$, and $\tau_x$ are the variances of $\bsm{G}$, $\bsm{F}$, $\bsm{H}$, and $\bsm{X}$, respectively. We use MSEs in \eqref{MSE-all} as the performance metric of the estimates of $\bsm{G}$, $\bsm{F}$, $\bsm{H}$, and evaluate them by Monte Carlo trials, whereas use the averaged SER as the performance metric for the detection of the QPSK symbol $\bsm{X}_\mathsf{d}$. A baseline method, referred to as BiG-AMP+LMMSE, is introduced for comparison. Similarly to the two-stage approach in \cite{he2019cascaded}, BiG-AMP+LMMSE is conducted in two separate stages: The first stage uses BiG-AMP \cite{parker2014bilinear} to obtain the estimates of $\bsm{G}$, $\bsm{H}$, $\bsm{C}$, and $\bsm{X}_\mathsf{d}$ from the outer matrix factorization; With the estimated $\bsm{X}_\mathsf{d}$ and $\bsm{C}$ in the first stage, the second stage utilizes the linear minimum mean square error (LMMSE) estimator to obtain the estimate of $\bsm{F}$. Beside this, JBF-MC \cite{he2019cascaded}, the methods in \cite{wang2019channel} and \cite{jensen2020optimal}, and PARAFAC \cite{wei2020channel} are also included for comparison. Note that only the estimates of $\{\bsm{G} {\rm diag}(\bsm{f}_k)\}_{k=1}^K $ are available in the methods \cite{wang2019channel} and \cite{jensen2020optimal}. For the comparison purpose, the estimates of $\bsm{G}$ and $\bsm{F}$ are uniquely obtained by assuming the perfect knowledge of $\bsm{f}_1$, i.e., the channel from the first user to the RIS. For JBF-MC \cite{he2019cascaded} and Tri-AMP, there exists diagonal ambiguities between $\bsm{G}$ and $\bsm{F}$, which are eliminated based on the true values of $\bsm{G}$ and $\bsm{F}$ in the calculation of the MSEs.

\begin{figure}[t]
    \centering
     \includegraphics[scale=0.322]{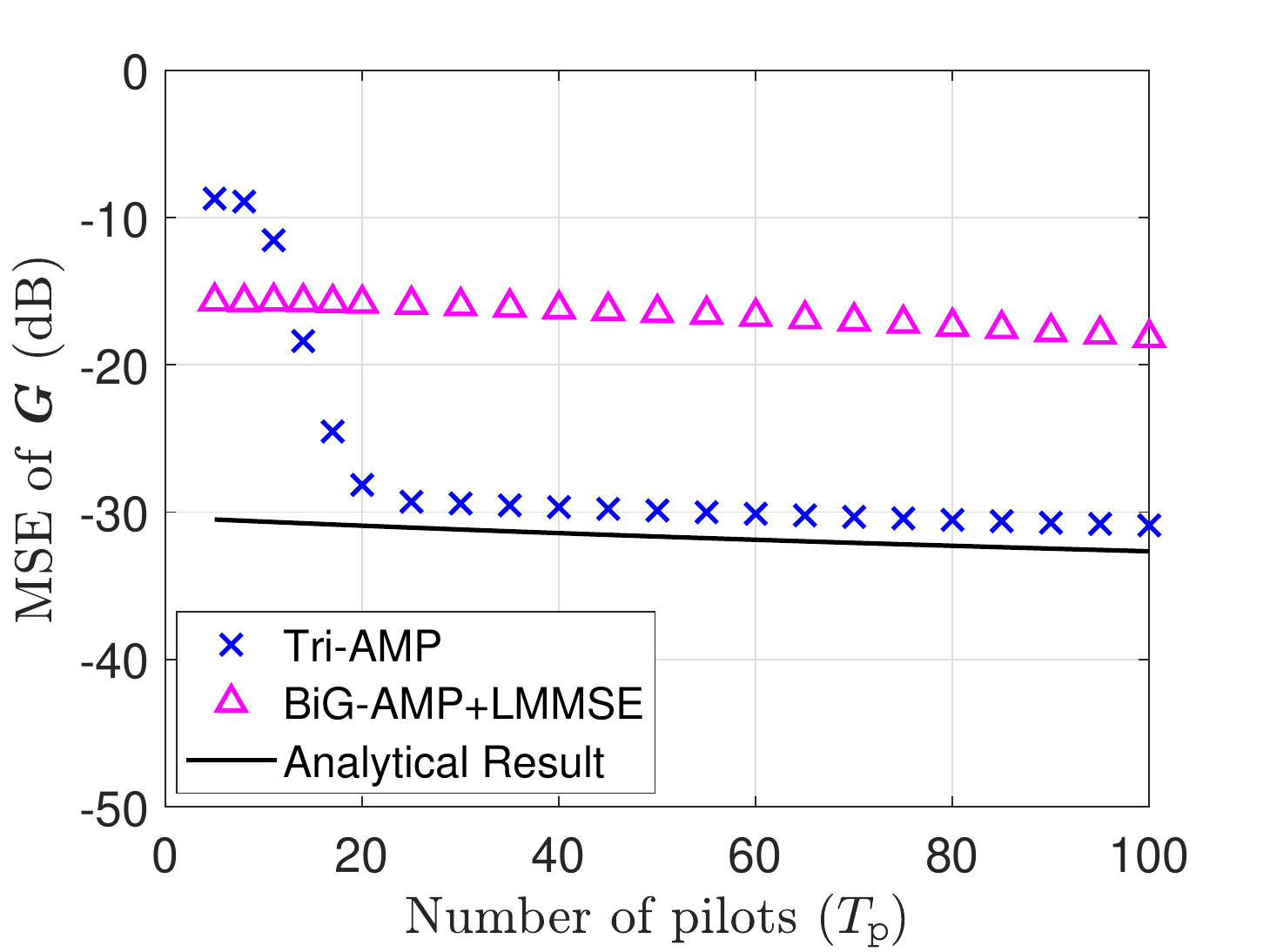}\,\,\includegraphics[scale=0.322]{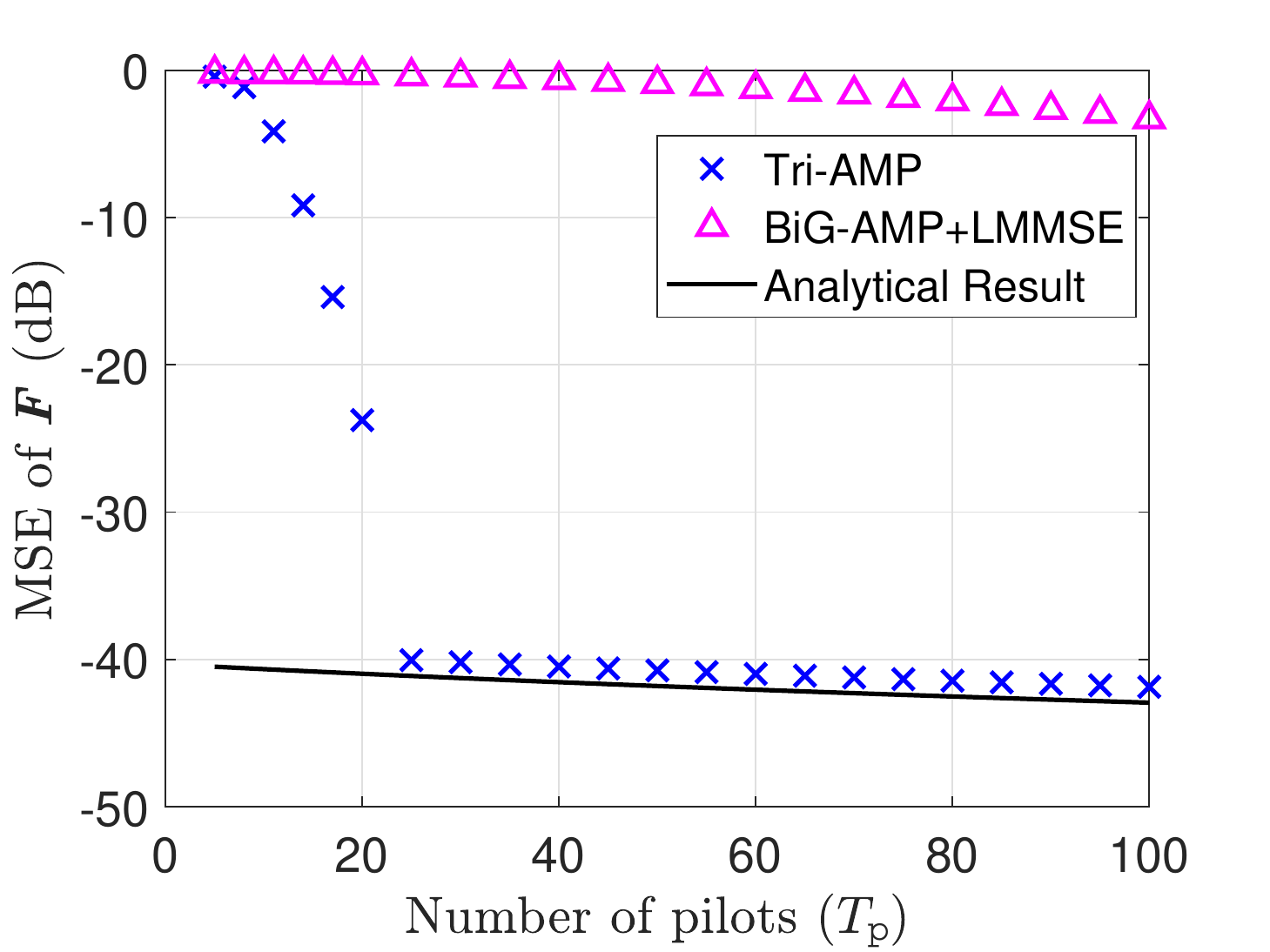} \\
    \smallskip \smallskip
    \includegraphics[scale=0.322]{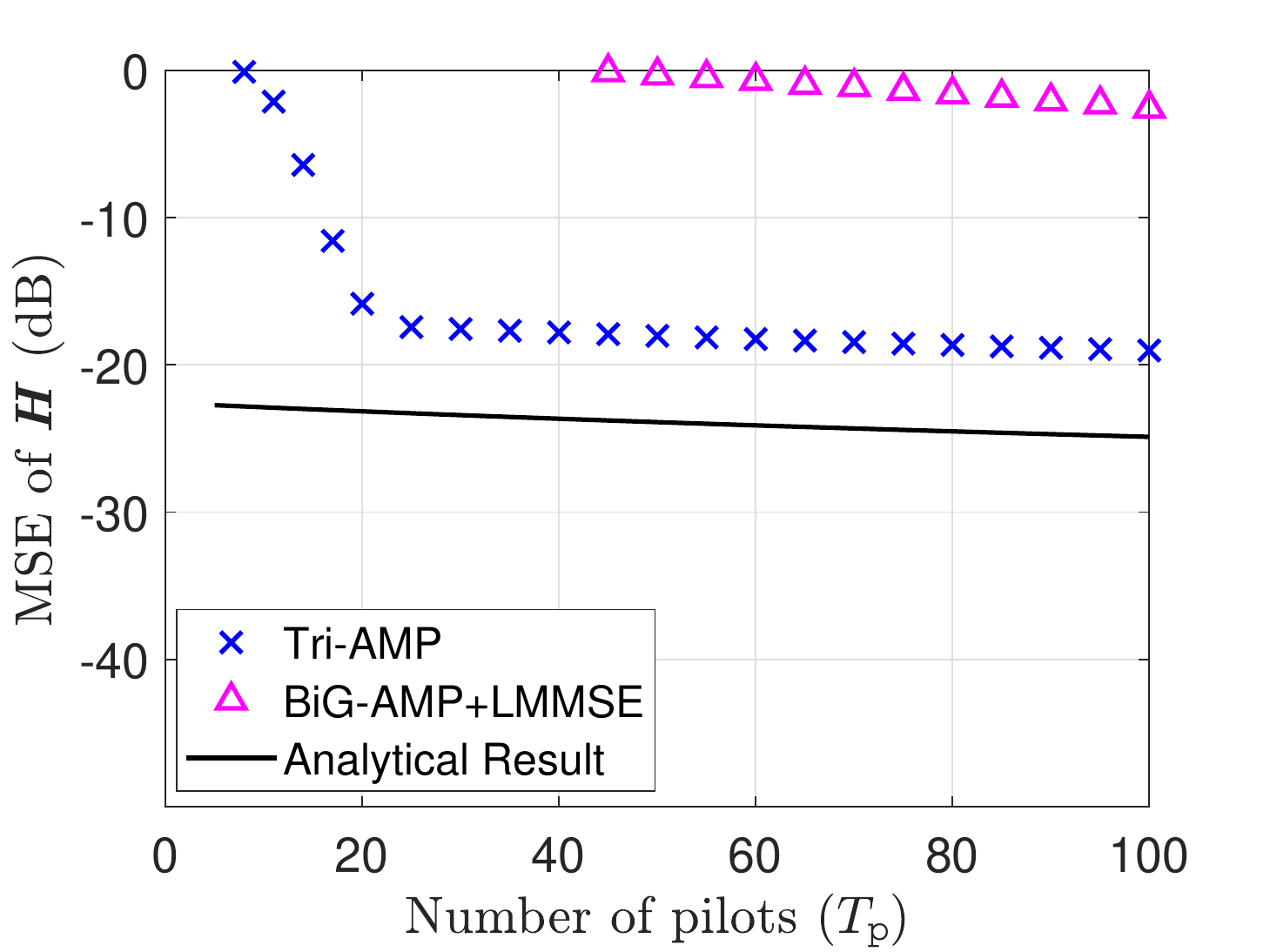}\,\,\includegraphics[scale=0.322]{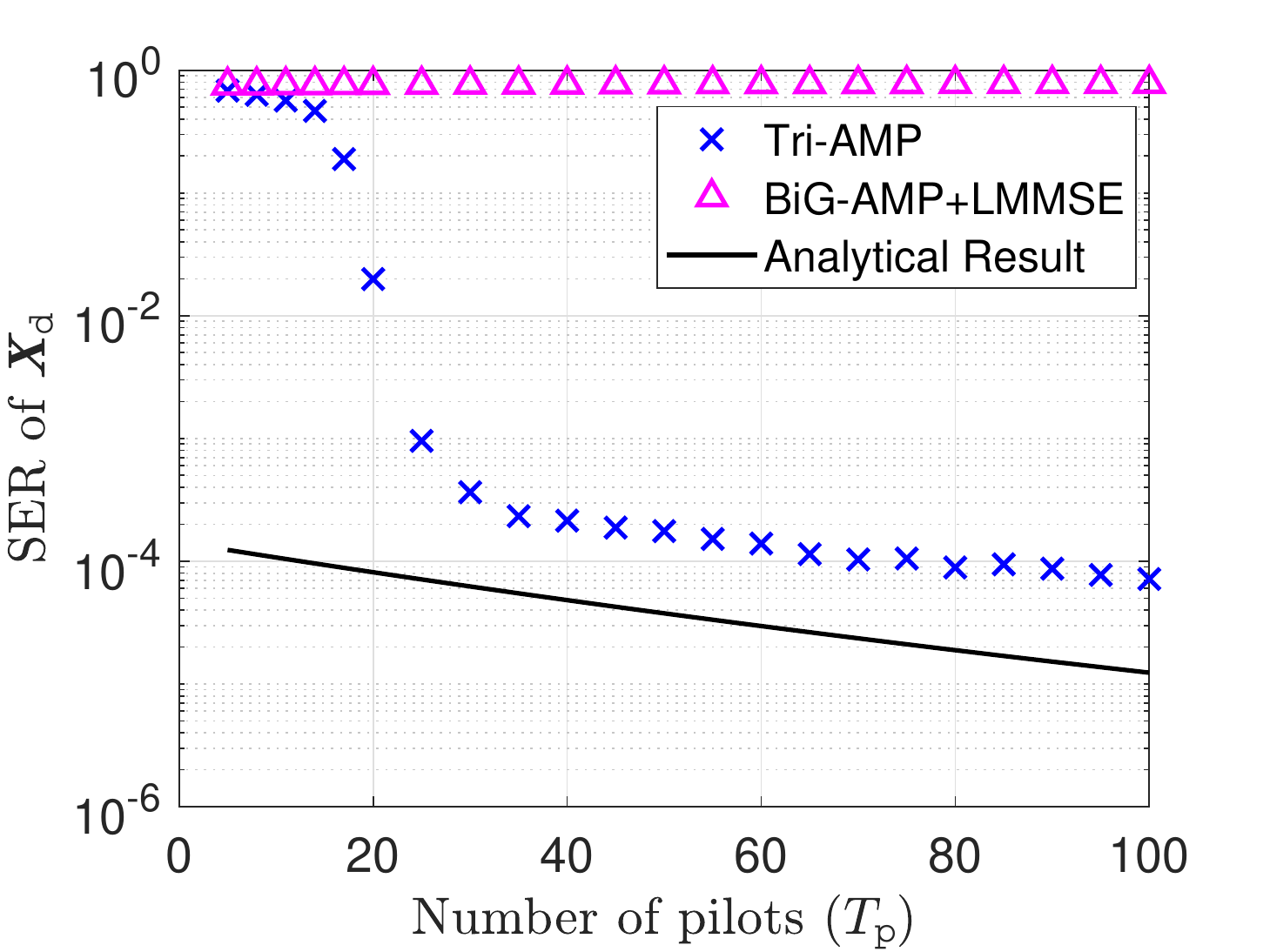}
    \vspace{-5mm}
    \caption{MSEs of $\bsm{G}$, $\bsm{F}$, and $\bsm{H}$, and SER of $\bsm{X}_\mathsf{d}$ versus the number of pilots $T_{\mathsf{p}}$ with $M = 256$, $N=128$, $K=20$, and $T_{\mathsf{d}} = 300$ under i.i.d. Rayleigh fading channels.}
    \label{fig4}
\end{figure}

\subsection{Under i.i.d. Rayleigh Fading Channels}

We first consider the case of i.i.d. Rayleigh fading channels. That is, the user-RIS channel matrix $\bsm{F}$, the RIS-BS channel matrix  $\bsm{G}$, and the user-BS channel matrix $\bsm{H}$ are generated from the i.i.d. complex circularly-symmetric normal distributions with zero mean and unit variance. We adopt the asymptotic analytical result shown in Proposition \ref{pro1} as a benchmark to validate the performance of the Tri-AMP algorithm.\footnote{Although the replica analysis in Section \ref{Section-V} is valid only in the large-system limit, we adopt the derived asymptotic MSE bound as a benchmark for finite-size systems considered in simulations. More specifically, we take the settings used in simulation to determine the values of $M/K$, $N/K$, $T_\mathsf{p}/K$, $T_\mathsf{d} / K$, and then apply Proposition \ref{pro1} with these ratios to determine the performance bound of the replica method.}

To test the minimum number of pilots required for respective algorithms, we consider the noiseless case (i.e., $\sigma^2 = 0$) with $\rho = 0.3$. We say that it is successful if the MSEs of $\bsm{G}$, $\bsm{F}$, and $\bsm{H}$ are less than $-60\,\text{dB}$ and the SER of $\bsm{X}_\mathsf{d}$ is zero. The minimum training length of the pilots of various algorithms under different block lengths $T$ is listed in Table \ref{table1} by averaging $200$ trails. The method in \cite{jensen2020optimal} considers the single user case with $T_{\mathsf{p}}\geq N+1$ and we extend it to the multiuser case with the minimum number of pilots being as $(N+1)K = 2580$. We see that  Tri-AMP requires only about $20$ pilots when $T \geq 400$, which is much smaller than the other baseline approaches. In particular, all the methods including the replica result are infeasible when $T=100$.


\begin{figure}[t]
    \centering
    \includegraphics[scale=0.319]{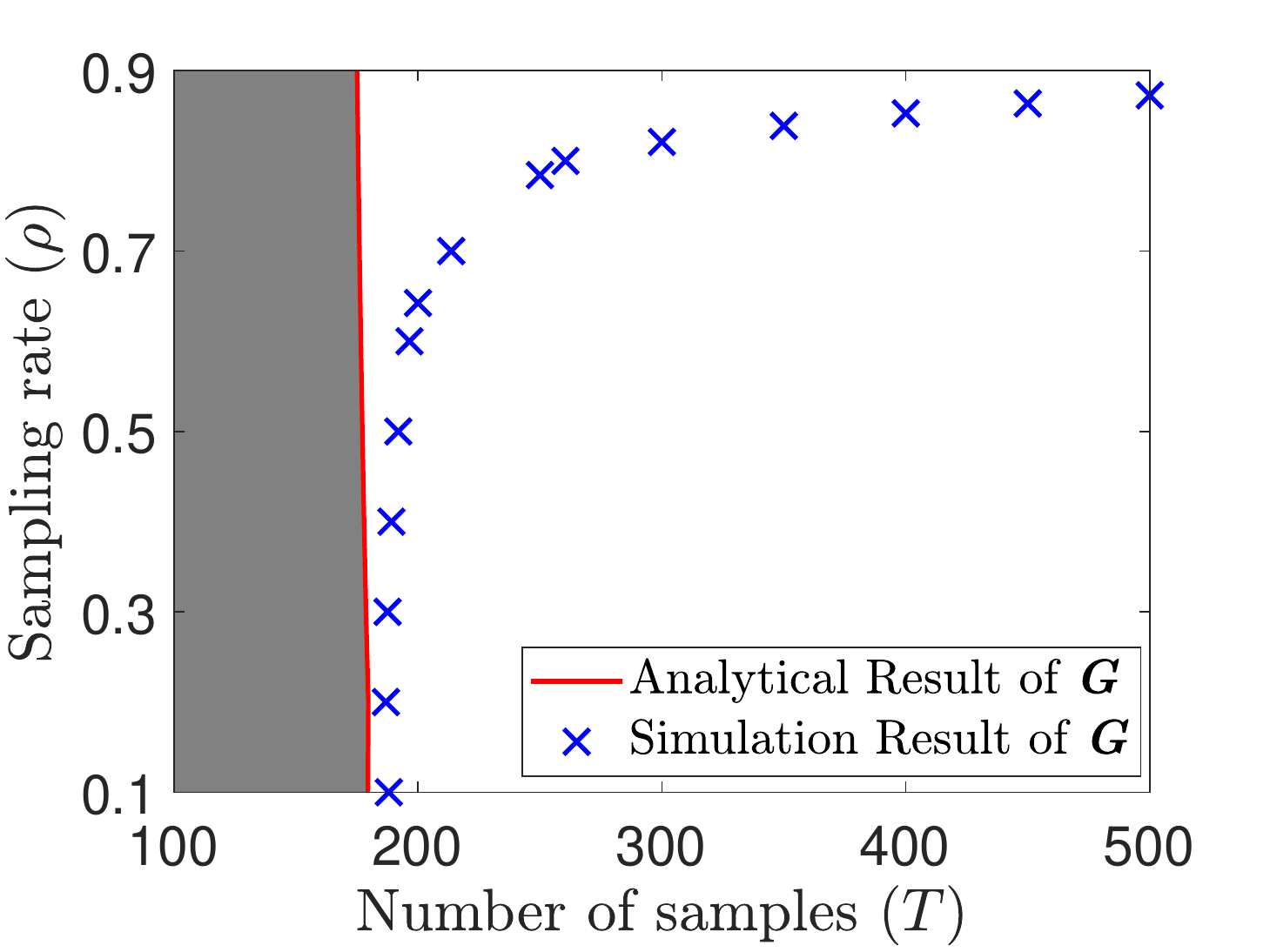}\includegraphics[scale=0.319]{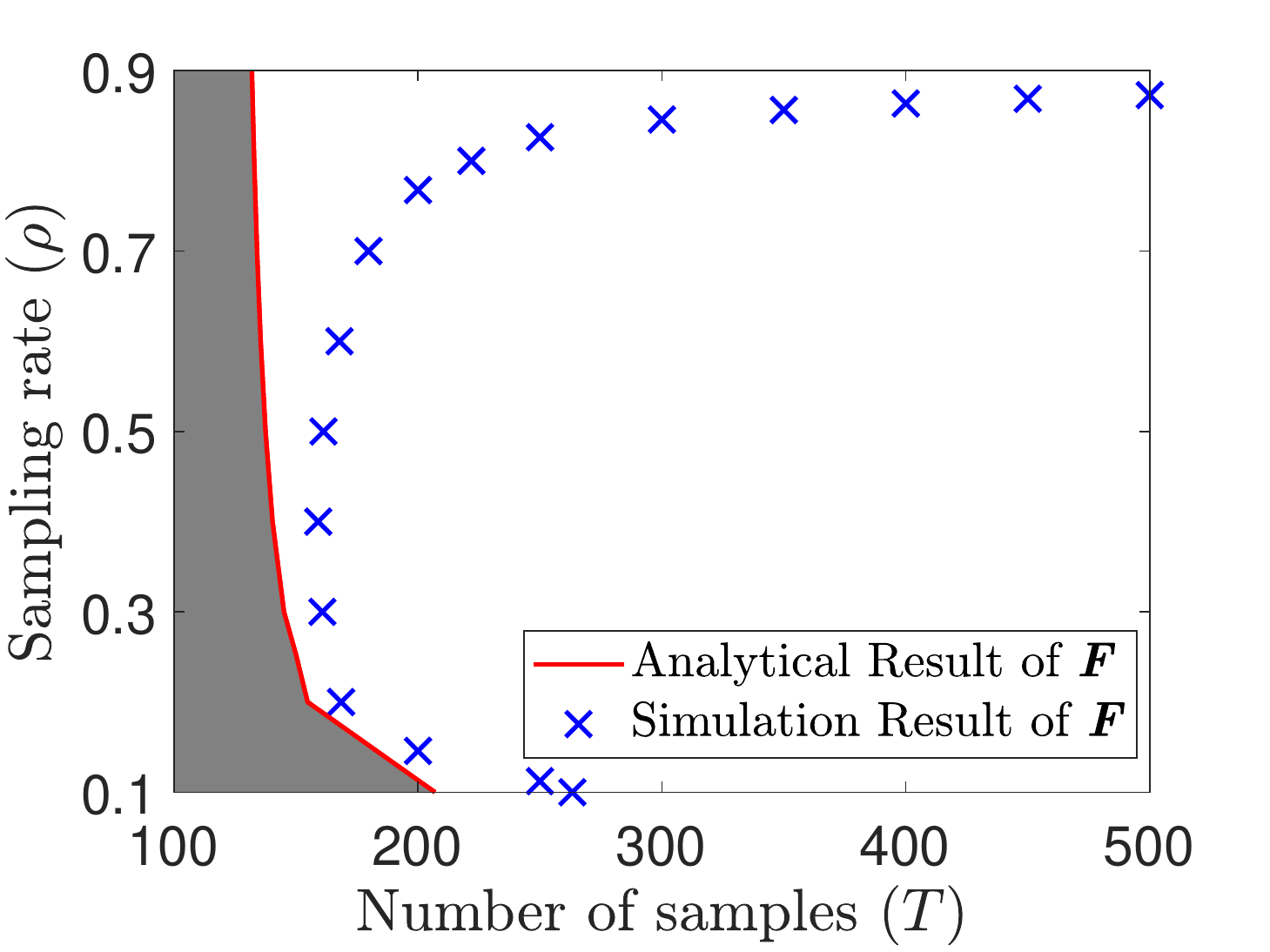} \\
    \smallskip \smallskip
    \includegraphics[scale=0.319]{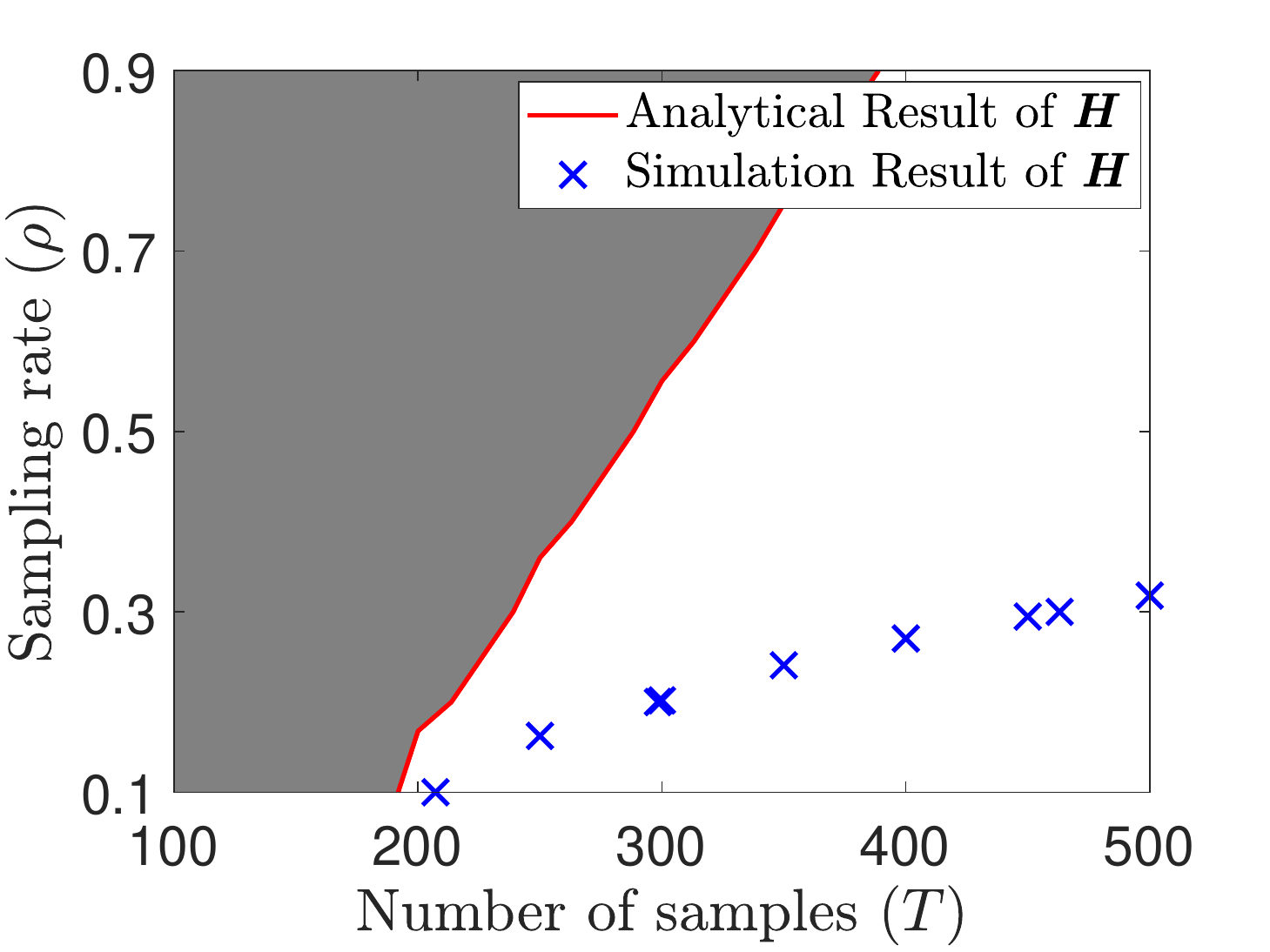}\includegraphics[scale=0.319]{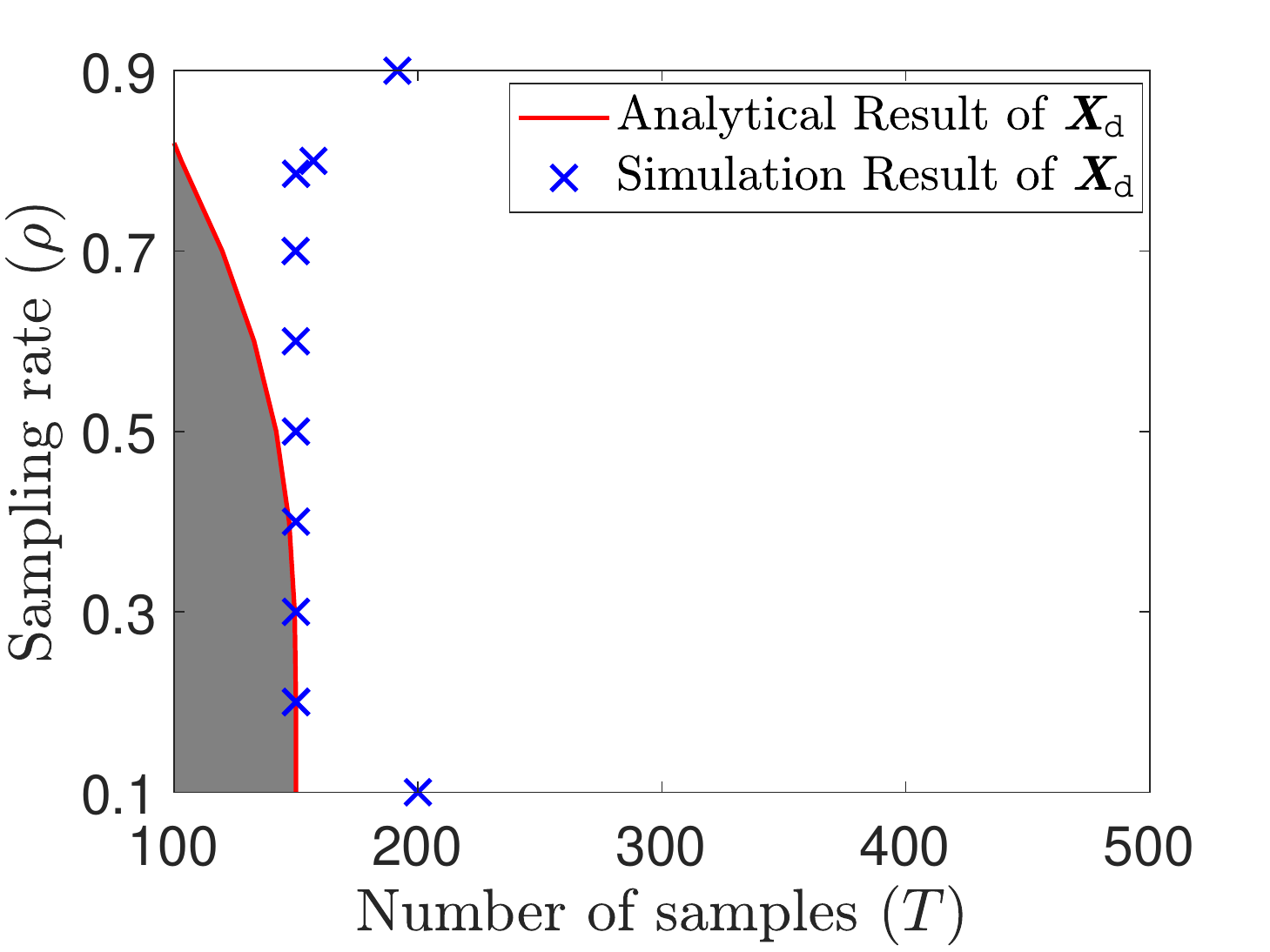}\\
    \vspace{-1mm}
    \caption{Phase diagrams of $\bsm{G}$, $\bsm{F}$, $\bsm{H}$, and $\bsm{X}_\mathsf{d}$ versus the sampling rate $\rho$ and the number of samples $T$ with $M = 256$, $N=128$, $K=20$, $\rho = 0.3$, and $T_\mathsf{p} = 90$, $\text{SNR} = 30\,\text{dB}$ under i.i.d. Rayleigh fading channels.}
    \label{fig5}
    \vspace{1mm}
\end{figure}

We now consider the noisy case with $\rho = 0.3$ and $T=300$. The methods in \cite{jensen2020optimal} and \cite{wei2020channel} are not included as it fails in the case of $T_\mathsf{p} < 300$ (see Table \ref{table1}). For Tri-AMP, the number of pilots is set to $T_{\mathsf{p}} = 90$. The method in \cite{wang2019channel} and JBF-MC \cite{he2019cascaded} use the whole transmission frame (i.e., $T = 300$ symbols) as pilots. The MSEs of $\bsm{G}$, $\bsm{F}$, and $\bsm{H}$, and the SERs of $\bsm{X}_\mathsf{d}$ versus the SNR are depicted in Fig. \ref{fig3}. The MSEs of $\bsm{G}$, $\bsm{F}$, and $\bsm{H}$ with $\text{SNR} = 30\,\text{dB}$, and the SERs of $\bsm{X}_\mathsf{d}$ with $\text{SNR} = 5\,\text{dB}$ versus the number of pilots are depicted in Fig. \ref{fig4}. As seen from Figs. \ref{fig3} and \ref{fig4}, the performance of Tri-AMP is consistently superior to the compared methods. Even in the unfair setting of Fig. \ref{fig4} ($90$ pilots for Tri-AMP and $300$ pilots for the other methods), the performance of Tri-AMP is consistently superior to the other baseline methods. The MSEs of $\bsm{G}$ and $\bsm{F}$ are very close to the analytical results. Nevertheless, there exists a gap of about $3$ dB gap between the analytical result and the simulation result with respect to the estimate of $\bsm{H}$. We conjecture that this is because the proposed message-passing algorithm perform well in sparse matrix factorization (such as the factorization of $\bsm{G} \bsm{C}$ where $\bsm{C}$ is a sparse matrix), but not so well in dense matrix factorization (such as the factorization of $\bsm{H} \bsm{X}$ where both $\bsm{H}$ and $\bsm{X}$ have non-zero elements).

Fig. \ref{fig5} depicts the phase diagrams with a varying value of the sampling rate $\rho$ and the number of samples ($T$). The boundary of the phase diagram in MSEs of $\bsm{G}$, $\bsm{F}$, and $\bsm{H}$ denotes the case with $\text{MSE} = -20\,\text{dB}$ and the gray region of the analytical result represents the region where $\text{MSE} < -20\,\text{dB}$. Likewise, the boundary of the phase diagram in evaluating the detection performance of  $\bsm{X}_\mathsf{d}$ denotes the case with $\text{SER} = 10^{-4}$. We see  that the simulation results of Tri-AMP are close to the analytical performance bound when $\rho$ is relatively low but not too low. Especially, the sparsity level $\rho$ of the on-off matrix $\bsm{S}$ is set between $0.2$ and $0.4$ to achieve a good estimation performance of the proposed Tri-AMP algorithm.

\begin{figure}[t]
    \centering
     \includegraphics[scale=0.32]{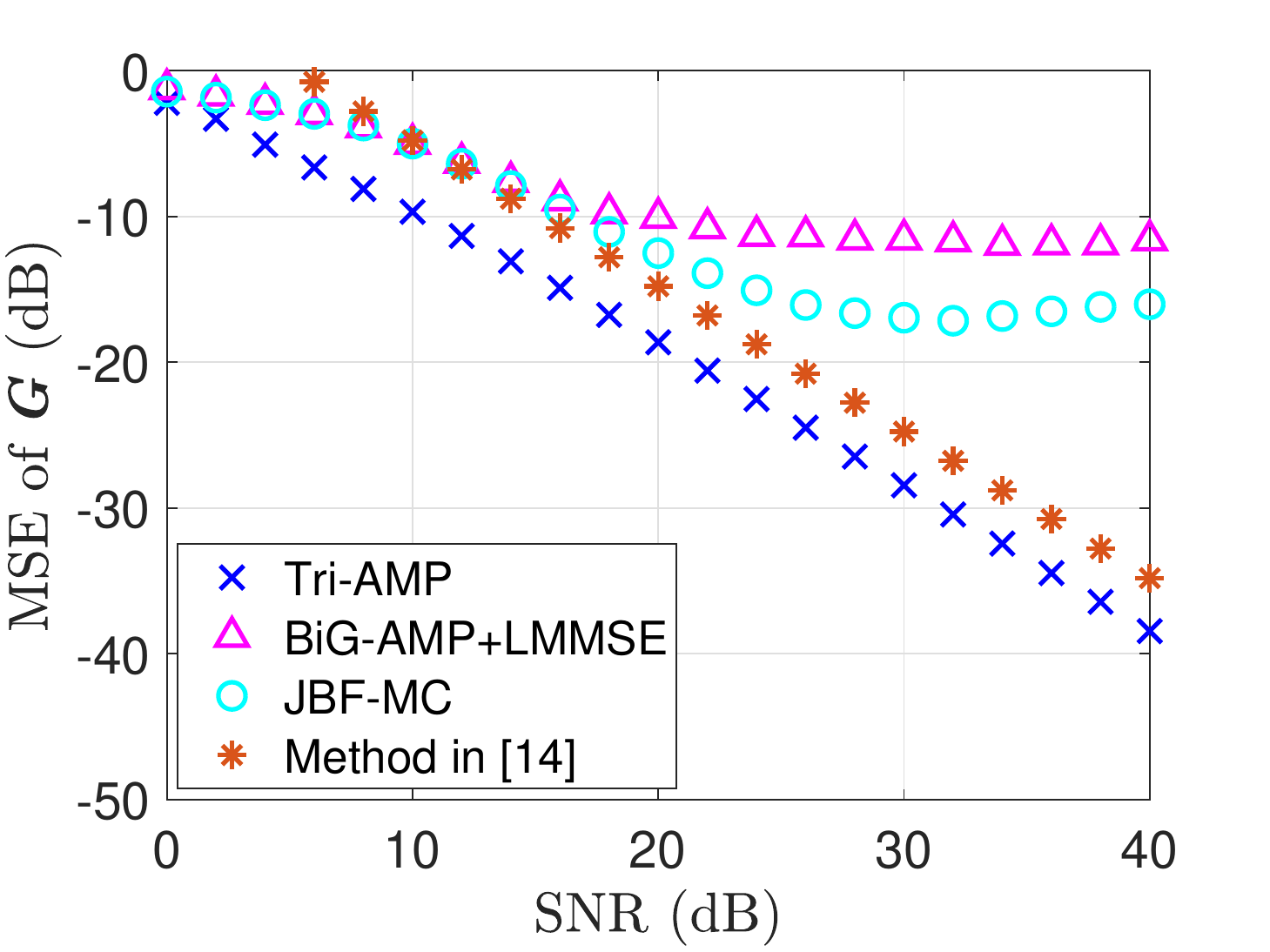}\,\,\includegraphics[scale=0.32]{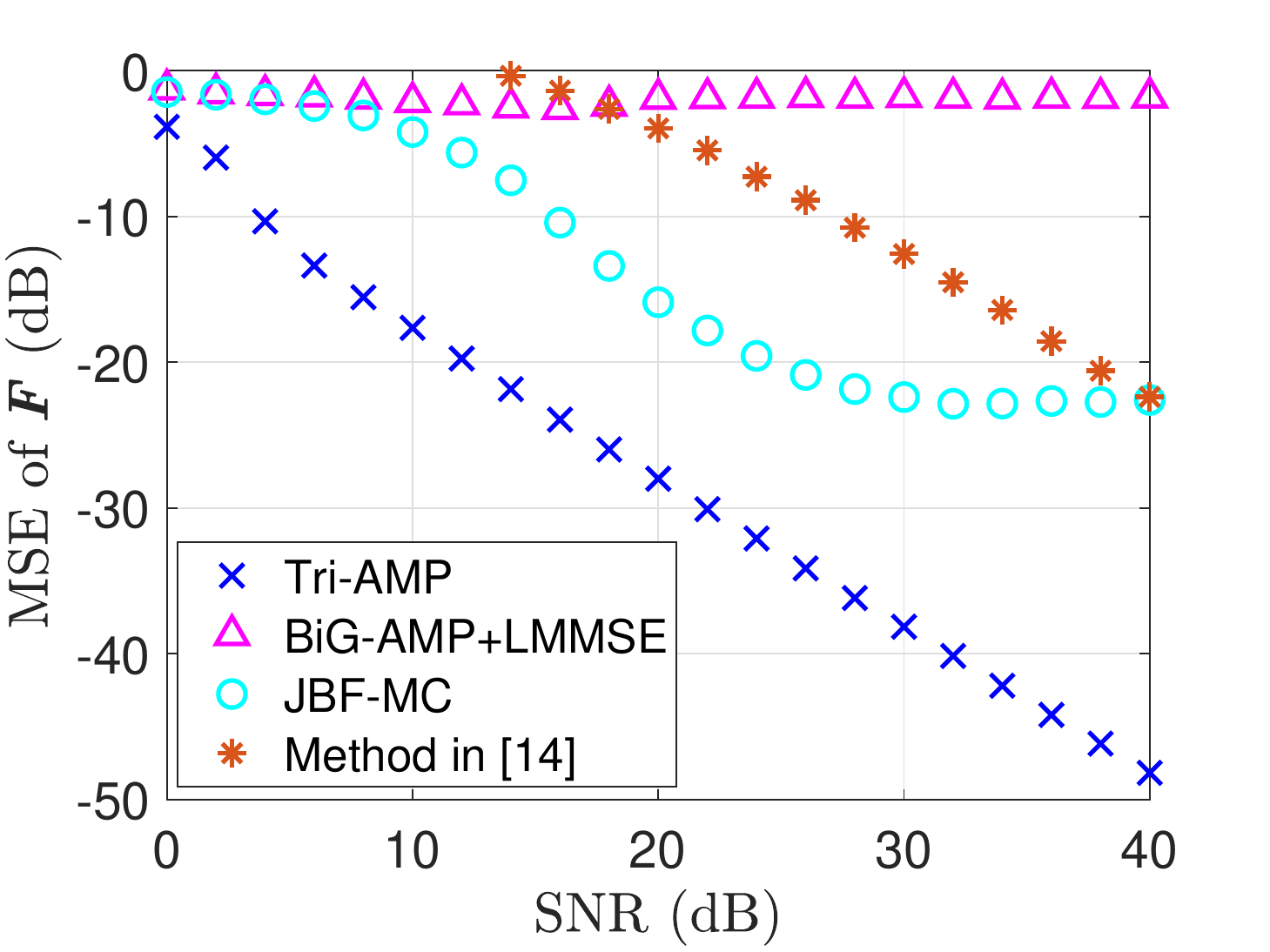} \\
    \smallskip \smallskip
    \includegraphics[scale=0.32]{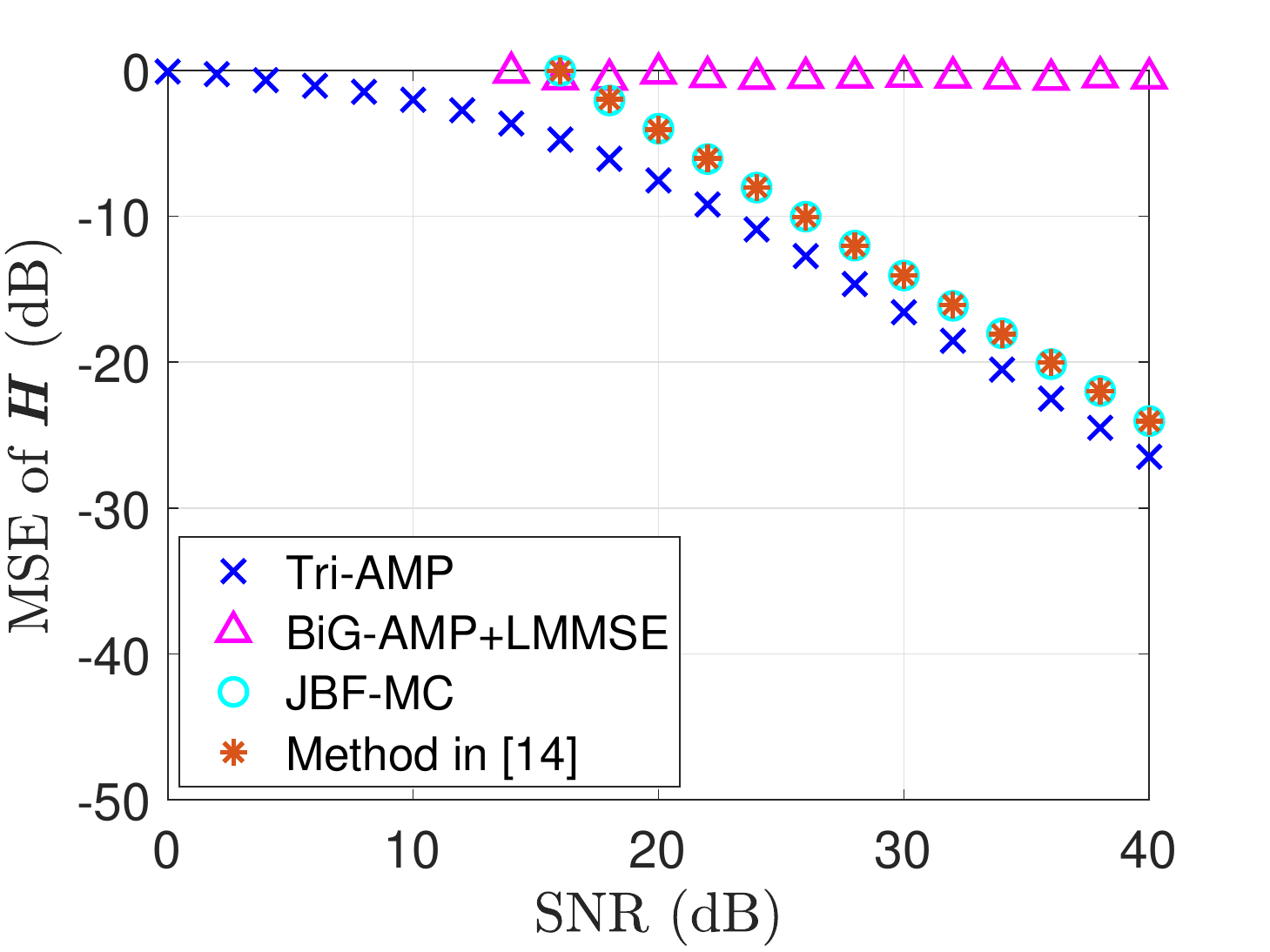}\,\,\includegraphics[scale=0.32]{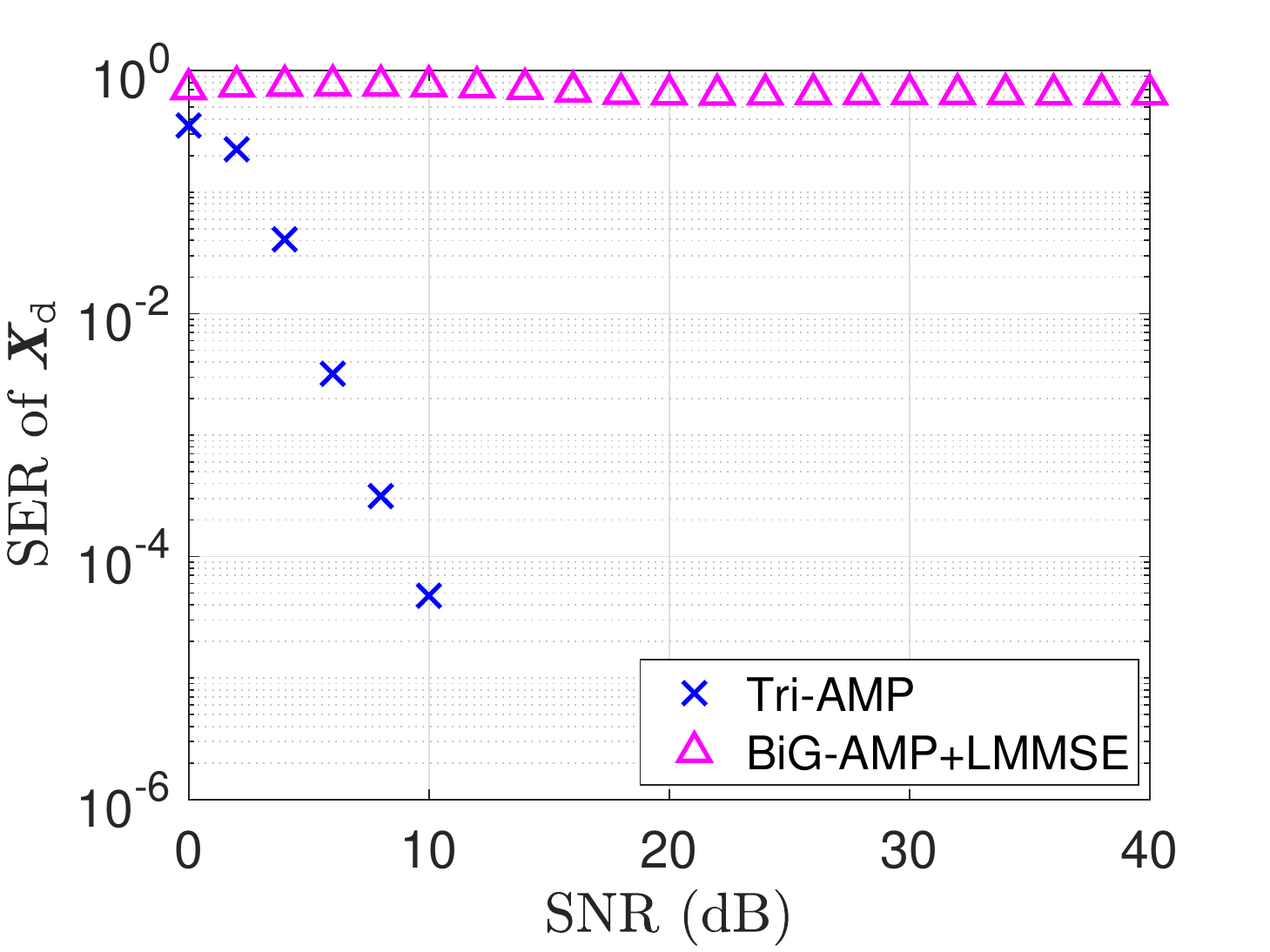}\\
    \vspace{-1mm}
    \caption{MSEs of $\bsm{G}$, $\bsm{F}$, and $\bsm{H}$, and SER of $\bsm{X}_\mathsf{d}$ versus the SNR with $M = 256$, $N=128$, $K=20$, $T_\mathsf{p} = 90$, and $T = 300$ under correlated Rayleigh fading channels.}
    \label{fig6}
\end{figure}

\begin{table}[t]
\xiaowuhao
\caption{Averaged MSE comparison versus the block length $T$ with $M=256$, $N=128$, $K=20$, and $\text{SNR} = 40\,\text{dB}$ under correlated Rayleigh fading channels.}
\vspace{-1.5mm}
\centering
\,\,\begin{tabular}{|c|c|c|c|c|} 
\hline\hline
\multirow{2}{*}{Algorithm} & \multirow{2}{*}{\diagbox{\hspace{-0.05cm} MSE \!\!\!}{$T$\!\!}} & \multirow{2}{*}{1000} & \multirow{2}{*}{2000} & \multirow{2}{*}{3000} \\
&  & & &  \\
\hline
\multirow{3}{*}{Tri-AMP}
& \!$\text{MSE}_g$ (dB)\! &$-46.28$ &$-49.71$ &$-51.55$ \\
& $\text{MSE}_f$ (dB) &$-57.46$ &$-60.87$ &$-62.98$ \\
& $\text{MSE}_h$ (dB) &$-34.29$ &$-37.72$      &$-39.48$     \\
\hline
\multirow{3}{*}{JBF-MC \cite{he2019cascaded}}
& $\text{MSE}_g$ (dB) &$-37.80$ &$-45.75$ &$-48.18$ \\
& $\text{MSE}_f$ (dB) &$-47.64$ &$-54.37$ &$-56.62$ \\
& $\text{MSE}_h$ (dB) &$-24.05$      &$-24.05$      &$-24.05$     \\
\hline
\multirow{3}{*}{\!\!Method in \cite{wang2019channel}\!\!}
& $\text{MSE}_g$ (dB) &$-39.48$       &$-41.51$     &$-42.42$ \\
& $\text{MSE}_f$ (dB) &$-22.25$       & $-22.47$    &$-22.58$ \\
& $\text{MSE}_h$ (dB) &$-24.05$      &$-24.06$      &$-24.06$     \\
\hline
\multirow{3}{*}{\!\!Method in \cite{jensen2020optimal}\!\!}
& $\text{MSE}_g$ (dB) &\ding{55} &\ding{55} &$-25.97$ \\
& $\text{MSE}_f$ (dB) &\ding{55} &\ding{55} &$-39.08$ \\
& $\text{MSE}_h$ (dB) &\ding{55} &\ding{55}      &$-34.56$     \\
\hline
\multirow{3}{*}{\!\!PARAFAC \cite{wei2020channel}\!\!}
& $\text{MSE}_g$ (dB) &$-36.33$      &$-41.08$ &$ -42.96$ \\
& $\text{MSE}_f$ (dB) &$-3.56$       &$-10.44$ &$-15.14$ \\
& $\text{MSE}_h$ (dB) &$-24.05$      &$-24.05$ &$-24.05$     \\
\hline
\hline
\end{tabular}
\begin{tablenotes}
        \item[a] ``\ding{55}'' indicates that the corresponding scheme is infeasible.
\end{tablenotes}
\label{table3}
\end{table}

\vspace{-0.3cm}
\subsection{Under Correlated Rayleigh Fading Channels}

We next provide numerical tests under the case of correlated Rayleigh fading channels and follow \cite{wang2019channel, loyka2001channel} to construct an exponential correlation matrix model. Specifically, the channel matrices $\bsm{G}$,  $\bsm{F}$, and $\bsm{H}$ are modeled as
\begin{align}
\bsm{G} \triangleq  \overline{\bsm{C}}_{gl} \overline{\bsm{G}}\,\overline{\bsm{C}}_{gr}, \bsm{F} \triangleq \overline{\bsm{C}}_f \overline{\bsm{F}}, \bsm{H} \triangleq  \overline{\bsm{C}}_h \overline{\bsm{H}},
\end{align}
respectively, where $\overline{\bsm{G}}$, $\overline{\bsm{F}}$, and $\overline{\bsm{H}}$ are i.i.d. Rayleigh fading channels and
\begin{align}
&[\overline{\bsm{C}}_{gl}]_{i,j} \triangleq  c_{gl}^{(i-j)}, [\overline{\bsm{C}}_{gr}]_{i,j} \triangleq  c_{gr}^{(i-j)}, \\
&[\overline{\bsm{C}}_f]_{i,j} \triangleq  c_f^{(i-j)},[\overline{\bsm{C}}_h]_{i,j} \triangleq  c_h^{(i-j)},
\end{align}
for $i \geq j$, where $|c_g| <1$, $|c_f| <1$, and $|c_h| <1$; for $i<j$,  $[\bsm{C}_g]_{i,j} = [\bsm{C}_g]^*_{j,i} $, $[\bsm{C}_f]_{i,j} = [\bsm{C}_f]^*_{j,i} $, $[\bsm{C}_h]_{i,j} = [\bsm{C}_h]^*_{j,i} $. The correlation coefficients $c_{gl}$, $c_{gr}$, $c_f$, and $c_h$ are set as $0.2+0.5\jmath$, $0.1+0.2\jmath$, and $0.4+0.3\jmath$, $0.3+0.1\jmath$, respectively. For Tri-AMP, we use the expectation-maximization methodology \cite{parker2014bilinear,bayes2008variational} to learn the noise variance $\sigma^2$ and the variances of the prior distribution $p(\bsm{G})$, $p(\bsm{F})$ and  $p(\bsm{H})$. Fig. \ref{fig6} depicts the MSEs of $\bsm{G}$, $\bsm{F}$, and $\bsm{H}$, and the SER of $\bsm{X}_\mathsf{d}$ versus the SNR while keeping the same setup as that of Fig. \ref{fig3} except for the channel model. We observe that Tri-AMP presents a similar performance as that of Fig. \ref{fig3} and still outperforms all the other baseline methods.

Finally, we consider relatively large transmission blocks of $T = 1000, 2000, $ and $ 3000$ with $\text{SNR} = 40\,\text{dB}$. We adopt the same parameter setup as that of Fig. \ref{fig6}, except for $T$ and $T_{\mathsf{p}}$. Particularly, Tri-AMP uses only $T_{\mathsf{p}} = 100$ pilots in the transmission frame, whereas the baseline methods use the whole transmission frame $T$ as pilots. The corresponding MSEs of $\bsm{G}$, $\bsm{F}$, and $\bsm{H}$ for respective algorithms are shown in Table \ref{table3}. We observe that even using only $100$ pilots, Tri-AMP still present a decent performance against the baseline methods which employ the whole transmission block as pilots.

\section{Conclusions}

\label{Section-VII}

We have investigated the problem of semi-blind cascaded channel estimation in RIS-aided massive MIMO systems. We formulated the semi-blind cascaded channel estimation problem as a trilinear matrix factorization task under the Bayeisan MMSE inference framework. We have developed a computationally efficient AMP based algorithm to iteratively calculate the marginal posterior distributions. We also derived the analytical MSE bounds based on the replica method in the large-system limit. Numerical examples were provided to verify that our proposed approach achieves an accurate channel estimation with a small number of pilot overhead.


{
\appendices
\section{Proof of Proposition \ref{pro1}\label{appa0}}

The key strategy for analyzing the MSEs in \eqref{MSE-all} is to evaluate the averaged free entropy \cite{tanaka2002statistical, MP_BIGAMP3}
\begin{align}\label{FreeE}
\mathcal{F}\triangleq \lim_{K\to \infty}\frac{1}{K^2}\mbs{E}_{\bsm{Y}}\left[\ln p(\bsm{Y})\right],
\end{align}
where $p(\bsm{Y}) = \int_{\bsm{G}, \bsm{F}, \bsm{H},\bsm{X}, \bsm{Z}, \bsm{C}} p(\bsm{Y}, \bsm{G}, \bsm{F}, \bsm{H},\bsm{X}, \bsm{Z}, \bsm{C})$ is the partition function. Note that directly evaluating $\mbs{E}_{\bsm{Y}}\left[\ln p(\bsm{Y})\right]$ in \eqref{FreeE} is intractable. We sidestep this issue by applying $\mbs{E} [\ln x]=\lim_{\tau\to 0}\frac{\partial}{\partial \tau} \ln\mbs{E}[x^\tau]$ in \eqref{FreeE} to transform the expectation inside the log-function, and obtain
\begin{align}\label{FreeE1}
\mathcal{F}= \lim_{K\to \infty}\frac{1}{K^2}\lim_{\tau \to 0}\frac{\partial}{\partial \tau }\ln \mbs{E}_{\bsm{Y}}\left[p^\tau(\bsm{Y})\right],
\end{align}
where $p^\tau(\bsm{Y})$ denotes $\tau$ identical replicas of $p(\bsm{Y})$ and $\tau$ is referred to as the {\it replica number}. To proceed with the calculation of $\mathcal{F}$ in \eqref{FreeE1}, the following commonly used assumptions in the replica method are made:
\begin{asum} \label{assum3}
The function
$\mbs{E}_{\bsm{Y}}\left[p^\tau(\bsm{Y})\right]$ in \eqref{FreeE1} is continuous differentiable at the vicinity of $\tau = 0$.
\end{asum}
\begin{asum} \label{assum1}
The order of the two limits $\tau \to 0$ and $K \to \infty$ in \eqref{FreeE1} can be exchanged without affecting the final result.
\end{asum}

With the above two assumptions, the calculation of $\mathcal{F}$ in  \eqref{FreeE1} becomes
\begin{align}\label{FreeE2}
\mathcal{F}= \lim_{\tau\to 0}\frac{\partial}{\partial \tau}\lim_{K\to \infty}\frac{1}{K^2} \ln \mbs{E}_{\bsm{Y}}\left[p^\tau(\bsm{Y})\right].
\end{align}
We now elaborate on how to derive the analytical expression of
\begin{align}
\lim_{K\to \infty}\frac{1}{K^2} \ln \mbs{E}_{\bsm{Y}}\left[p^\tau(\bsm{Y})\right]. \label{E-Y12}
\end{align}

We start with the calculation of $\ln \mbs{E}_{\bsm{Y}} \left[p^\tau(\bsm{Y})\right]$. For an arbitrary positive integer $\tau$, we denote the $a$-th replica of the user-BS channel matrix $\bsm{H}$ by $\bsm{H}^{(a)}$, $0\leq a\leq \tau$, which follows the same distribution as $\bsm{H}$.\footnote{For the ease of notation, we define $\bsm{H}^{(0)}\triangleq \bsm{H}$.} We further define the all $\tau+1$ replicas of $\bsm{H}$ as $\boldsymbol{\mathcal{H}}\triangleq [\vect(\bsm{H}^{(0)}),\cdots,\vect(\bsm{H}^{(\tau)})]\in \mbs{C}^{MK\times (\tau+1)}$, where $\vect(\cdot)$ is the vectorization operator. Likewise, we define the collections of all the $\tau+1$ replicas of $\bsm{G}$, $\bsm{F}$, $\bsm{X}_\mathsf{o}$, $\bsm{C}_\mathsf{o}$, and $\bsm{Z}_\mathsf{o}$ as $\boldsymbol{\mathcal{G}}$, $\boldsymbol{\mathcal{F}}$, $\boldsymbol{\mathcal{X}}_\mathsf{o}$, $\boldsymbol{\mathcal{C}}_\mathsf{o}$, and $\boldsymbol{\mathcal{Z}}_\mathsf{o}$, $\mathsf{o}\in\{\mathsf{p},\mathsf{d}\}$, respectively. Based on these definitions and the full probability formula, we express $\mbs{E}_{\bsm{Y}}\left[p^\tau(\bsm{Y})\right]$ as
\begin{align}\label{temp112}
\mbs{E}_{\bsm{Y}} \left[ p^\tau(\bsm{Y})\right]=\mbs{E}_{\bsm{\mathcal{A}}}\left[\int \mathrm{d} \bsm{Y} \prod_{a=0}^\tau p\left(\bsm{Y}|\bsm{Z}^{(a)}\right)\right],
\end{align}
where $\bsm{\mathcal{A}} \triangleq \left\{ \bsm{\mathcal{H}},\bsm{\mathcal{G}},\bsm{\mathcal{F}},
\bsm{\mathcal{X}}_\mathsf{p},\bsm{\mathcal{C}}_\mathsf{p},
\bsm{\mathcal{Z}}_\mathsf{p}, \bsm{\mathcal{X}}_\mathsf{d},\bsm{\mathcal{C}}_\mathsf{d},
\bsm{\mathcal{Z}}_\mathsf{d} \right\}$ and the expectation in the right hand side (RHS) of \eqref{temp112} is taken over $p(\bsm{\mathcal{A}})$ in \eqref{replica-E-Y}, shown at the bottom of the next page.
\begin{figure*}[b]
\vspace{-0.4cm}
\hrulefill
\centering
\begin{align}
&~~~~~ p(\bsm{\mathcal{A}})
= \prod_{a=0}^\tau p\left(\bsm{Z}^{(a)} | \bsm{H}^{(a)},\bsm{G}^{(a)},\bsm{X}^{(a)},
\bsm{C}^{(a)}\right) \,
 p\left(\bsm{C}^{(a)} | \bsm{F}^{(a)},\bsm{X}^{(a)}\right) \,
 p\left(\bsm{G}^{(a)}\right) \, p\left(\bsm{H}^{(a)}\right) p\left(\bsm{F}^{(a)}\right) \, p\left(\bsm{X}_{\mathsf{p}}^{(a)}\right)\, p\left(\bsm{X}_{\mathsf{d}}^{(a)}\right) \notag \\
&~~~~~~~~~~~~~\, = \prod_{a=0}^\tau \left( \left( \prod_{m=1}^M \prod_{t=1}^T p\left(z_{mt}^{(a)} | (\bsm{g}^{(a)}_m)^\mathsf{T},\bsm{c}^{(a)}_t ,(\bsm{h}^{(a)}_m)^\mathsf{T},\bsm{x}^{(a)}_t \right) \right) \left(\prod_{n=1}^N \prod_{t=1}^T p\left(c_{nt}^{(a)} | (\bsm{f}^{(a)}_n)^\mathsf{T},\bsm{x}^{(a)}_t\right) \right)  \left(\prod_{m=1}^M \prod_{n=1}^N p(g_{mn}^{(a)}) \right) \right. \notag \\
&~~~~~~~~~~~~~~~~~~\,\times \left. \left( \prod_{n=1}^N \prod_{k=1}^K  p(f_{nk}^{(a)}) \right) \left( \prod_{m=1}^M \prod_{k=1}^K p(h_{mk}^{(a)}) \right) \left(\prod_{k=1}^{K} \prod_{t=1}^{T_\mathsf{p}} p(x_{\mathsf{p},kt}^{(a)}) \right) \left(\prod_{k=1}^{K} \prod_{t=1}^{T_\mathsf{d}} p(x_{\mathsf{d},kt}^{(a)}) \right) \right), \label{replica-E-Y}\\
& \text{where}~p\left(z_{mt}^{(a)} | \bsm{g}^\mathsf{T}_m,\bsm{c}^{(a)}_t ,\bsm{h}^\mathsf{T}_m,\bsm{x}^{(a)}_t \right) \!=\! \delta\left(z_{mt}^{(a)}-(\bsm{g}_m^{(a)})^\mathsf{T} \bsm{c}^{(a)}_t \!-\! (\bsm{h}_m^{(a)})^\mathsf{T} \bsm{x}^{(a)}_t \right),p\left(c_{nt}^{(a)} | (\bsm{f}^{(a)}_n)^\mathsf{T},\bsm{x}^{(a)}_t\right) \!=\! \delta \left(c_{nt}^{(a)} \!-\! s_{nt} (\bsm{f}_n^{(a)})^\mathsf{T} \bsm{x}^{(a)}_t \right). \notag \\
&~~~~~~ p \left( \big\{c_{nt}^{(a)}\big\}_{0\leq a \leq \tau} |  \bsm{Q}_{x_\mathsf{p}},\bsm{Q}_{x_\mathsf{d}},\bsm{Q}_f \right)=
\left\{
\begin{aligned}
& \CN \left(\big\{ c_{nt}^{(a)} \big\}_{0\leq a \leq \tau} ; {\bf 0}, \big[ KQ_f^{ab}Q_{x_{\mathsf{p}}}^{ab} \big]_{0\leq a, b \leq \tau} \right), s_{nt} = 1~\text{and}~1 \leq t\leq T_\mathsf{p},  \\
& \CN \left(\big\{ c_{nt}^{(a)} \big\}_{0\leq a \leq \tau} ; {\bf 0}, \big[ KQ_f^{ab}Q_{x_{\mathsf{d}}}^{ab} \big]_{0\leq a, b \leq \tau} \right), s_{nt} = 1~\text{and}~T_\mathsf{p}+1\leq t\leq T, \\
& \delta(c_{nt}), s_{nt} = 0.
\end{aligned}
\right. \label{PC1}
\end{align}
\end{figure*}
To carry out the expectation over $p(\bsm{\mathcal{A}})$ in RHS of \eqref{temp112}, we introduce four $(\tau+1) \times (\tau+1)$ auxiliary matrices $\bsm{Q}_{h} \triangleq [Q^{ab}_{h}]$,  $\bsm{Q}_{g} \triangleq [Q^{ab}_{g}]$,  $\bsm{Q}_{f} \triangleq [Q^{ab}_{f}]$, and $\bsm{Q}_{x_\mathsf{o}} \triangleq [Q^{ab}_{x_\mathsf{o}}]$, whose elements are defined by
\begin{align}
Q_h^{ab} & \triangleq \frac{1}{K}\sum_{k} \big(h_{mk}^{(a)}\big)^\ast h_{mk}^{(b)}, Q_g^{ab} \triangleq \frac{1}{N}\sum_{n}\big(g_{mn}^{(a)}\big)^\ast g_{mn}^{(b)}, \notag \\
Q_f^{ab} & \triangleq \frac{1}{K} \sum_{k}\big(f_{nk}^{(a)}\big)^\ast f_{nk}^{(b)},
Q_{x_\mathsf{o}}^{ab} \triangleq \frac{1}{K}\sum_{k}\big(x_{\mathsf{o},kt}^{(a)}\big)^\ast x_{\mathsf{o},kt}^{(b)}, \notag
\end{align}
respectively, where $ 0 \leq a, b \leq \tau $ and $\mathsf{o}\in\{\mathsf{p},\mathsf{d}\}$. We first deal with the expectation with respect to $\bsm{\mathcal{C}}$. Note that $c_{nt}^{(a)} = s_{nt} \sum_{k} f_{nk}^{(a)} x_{kt}^{(a)}$. Therefore,
for $s_{nt} = 1$ and large $K$, the CLT motivates $\{c_{nt}^{(a)}\}_{a}$ as a Gaussian random vector with its PDF given in \eqref{PC1}, shown at the bottom of the next page. Accordingly, we have
\begin{align}
& \mbs{E}_{\{c_{nt}^{(a)}\}_a | \{\bsm{f}^{(a){\mathsf{T}}}_n, \bsm{x}^{(a)}_t\}_{a}} \left[ \psi(\{c_{nt}^{(a)} \}_a) \right] \notag \\
& = \mbs{E}_{\{c_{nt}^{(a)}\}_a | \bsm{Q}_{x_\mathsf{p}},\bsm{Q}_{x_\mathsf{d}},\bsm{Q}_f} \left[ \psi(\{c_{nt}^{(a)}\}_a)  \right] + \mcl{O} \left(\frac{1}{K}\right), \label{c-gaussian1}
\end{align}
where $\psi(\{c_{nt}^{(a)} \}_a) $ denotes a function of $\{c_{nt}^{(a)} \}_a$ and the expectation in \eqref{c-gaussian1} is with respect to the conditional distribution in \eqref{PC1}. Similarly, using the CLT to $z_{mt}^{(a)}$ for large $N$ and $K$, we have
\begin{align}
& \mbs{E}_{\{z_{mt}^{(a)}\}_a | \{(\bsm{g}^{(a)}_m)^\mathsf{T},\bsm{c}^{(a)}_t ,(\bsm{h}^{(a)}_m)^\mathsf{T},\bsm{x}^{(a)}_t \}_{a}} \left[ \psi(\{z_{mt}^{(a)} \}_a) \right] \notag \\
& = \mbs{E}_{\{z_{mt}^{(a)}\}_a | \bsm{Q}_{x_\mathsf{p}},\bsm{Q}_{x_\mathsf{d}},\bsm{Q}_f,\bsm{Q}_g,\bsm{Q}_h} \left[ \psi(\{z_{mt}^{(a)}\}_a) \right] + \mcl{O} \left(\frac{1}{NK}\right), \label{z-gaussian1}
\end{align}
where the expectation in \eqref{z-gaussian1} is with respect to the conditional Gaussian distribution
\begin{align}
p(\{z_{nt}^{(a)}\}_{0\leq a\leq n}|\bsm{Q}_z) \!=\! \left\{
\begin{aligned}
\CN(\{z_{nt}^{(a)}\};{\bf 0},\bsm{Q}_{z_\mathsf{d}}),& 1\leq t\leq T_\mathsf{p}, \\
\CN(\{z_{nt}^{(a)}\};{\bf 0},\bsm{Q}_{z_\mathsf{d}}),& T_\mathsf{p}+1\leq t\leq T,
\end{aligned}
\right. \notag
\end{align}
where $\bsm{Q}_{c_\mathsf{o}}=[\rho KQ_f^{ab}Q_{x_\mathsf{o}}^{ab}]_{0\leq a,b\leq n}$ and $\bsm{Q}_{z_\mathsf{o}}=[KQ_h^{ab}Q_{x_\mathsf{o}}^{ab}+NQ_g^{ab}Q_{c_\mathsf{o}}^{ab}]_{0\leq a,b\leq n}$ for  $\mathsf{o}\in\{\mathsf{p},\mathsf{d}\}$.

\begin{figure*}[!hb]
\vspace{-0.6cm}
\begin{align}
& \ln \mbs{E}_{\bsm{Y}} \left[p^\tau(\bsm{Y})\right]= \ln \left\{ \int \mathrm{d}(MK\bsm{Q}_h)\mathrm{d}(MN\bsm{Q}_g)\mathrm{d}(NK\bsm{Q}_f)
\mathrm{d}(KT_\mathsf{p}\bsm{Q}_{x_\mathsf{p}})\mathrm{d}(KT_\mathsf{d}
\bsm{Q}_{x_\mathsf{d}})\mathrm{d}(NT_\mathsf{p}\bsm{Q}_{c_\mathsf{p}})
\mathrm{d}(NT_\mathsf{d}\bsm{Q}_{c_\mathsf{d}}) \right. \nonumber\\
& \quad\qquad\qquad~~~~~~~~\, \times \left. p(\bsm{Q}_h)p(\bsm{Q}_g)p(\bsm{Q}_f)p(\bsm{Q}_{x_\mathsf{p}},\bsm{Q}_{x_\mathsf{d}})p(\bsm{Q}_{c_\mathsf{p}},\bsm{Q}_{c_\mathsf{d}})\,e^{\ln \mbs{E}_{\boldsymbol{\mathcal{Z}}_\mathsf{p},\boldsymbol{\mathcal{Z}}_\mathsf{d}} \left[\int \mathrm{d}\Yv\prod_{a=0}^\tau p(\bsm{Y}|\bsm{Z}^{(a)})\right]} \right) + \mcl{O}\left( \frac{1}{K} \right), \label{EY1}
\end{align}
\end{figure*}

Furthermore, we define the auxiliary matrix $\bsm{Q}_{c} \triangleq [Q^{ab}_{c}]$, whose elements are defined by
\begin{align}
Q_{c_\mathsf{o}}^{ab} \triangleq  \frac{1}{T_\mathsf{o}}\sum_{t}\big(c_{\mathsf{o},kt}^{(a)}\big)^\ast c_{\mathsf{o},kt}^{(b)}, \notag
\end{align}
 where $ 0 \leq a, b \leq \tau $ and $\mathsf{o}\in\{\mathsf{p},\mathsf{d}\}$. By inserting the equivalent Dirac's delta expressions of $Q_h^{ab}$, $Q_f^{ab}$, $Q_g^{ab}$, $Q_{x_\mathsf{o}}^{ab}$, and $Q_{c_\mathsf{o}}^{ab}$ into the RHS of \eqref{temp112}:
\begin{align}
1 & = K \int \prod_{m=1}^M \prod_{0\leq a\leq b}^\tau \delta\left(\sum_{k}\left(h_{mk}^{(a)}\right)^\ast h_{mk}^{(b)}-K Q_h^{ab}\right) {\rm d} Q_h^{ab}, \notag \\
1 &= N \int \prod_{m=1}^M \prod_{0\leq a\leq b}^\tau \delta\left(\sum_{n}\left(g_{mm}^{(a)}\right)^\ast g_{m n}^{(b)}-N  Q_g^{ab}\right) {\rm d} Q_g^{ab}, \notag \\
1 &= NK \int \prod_{0\leq a\leq b}^\tau \delta\left(\sum_{nk}\left(f_{nk}^{(a)}\right)^\ast f_{nk}^{(b)}-NK Q_f^{ab}\right) {\rm d} Q_f^{ab}, \notag\\
1 &= KT_{\mathsf{o}} \int \prod_{0\leq a\leq b}^\tau \delta\left(\sum_{kt}\left(x_{\mathsf{o},nt}^{(a)}\right)^\ast x_{\mathsf{o},nt}^{b} - KT_{\mathsf{o}} Q_{x_\mathsf{o}}^{ab}\right) {\rm d} Q_{x_{\mathsf{o}}}^{ab}, \notag \\
1 &= MT_{\mathsf{o}} \int \prod_{0\leq a\leq b}^\tau \delta\left(\sum_{nt}\left(c_{\mathsf{o}, nt}^{(a)}\right)^\ast c_{\mathsf{o}, nt}^{(b)}- N T_{\mathsf{o}} Q_{c_\mathsf{o}}^{ab} \right) {\rm d} Q_{c_\mathsf{o}}^{ab}, \notag
\end{align}
and by performing the integral with a change of the variables  $\bsm{\mathcal{A}}$ to $\{\bsm{Q}_{h},\bsm{Q}_{g},\bsm{Q}_{f},\bsm{Q}_{x_\mathsf{p}},
\bsm{Q}_{x_\mathsf{d}}, \bsm{Q}_{c_\mathsf{p}},
\bsm{Q}_{c_\mathsf{d}}\}$, we obtain $\ln \mbs{E}_{\bsm{Y}}\left[p^\tau(\bsm{Y})\right]$ in \eqref{EY1},
shown at the bottom of the next page, where we have used the fact that $\ln(x+\mcl{O}(1/K)) = \ln x + \mcl{O}(1/K) $ for any $x>0$ and large $K$, and the expectation in the exponential term on $(\bsm{\mcl{Z}}_{p},\bsm{\mcl{Z}}_{d})$ is conditioned on $\bsm{Q}_g$,  $\bsm{Q}_h$, $\bsm{Q}_f$, $\bsm{Q}_{x_{\mathsf{p}}}$, and $\bsm{Q}_{x_{\mathsf{d}}}$. In \eqref{EY1}, the probability measures $p(\bsm{Q}_h)$, $p(\bsm{Q}_g)$, $p(\bsm{Q}_f)$, $p(\bsm{Q}_{x_\mathsf{p}},\bsm{Q}_{x_\mathsf{d}})$, and $p(\bsm{Q}_{c_\mathsf{p}},\bsm{Q}_{c_\mathsf{d}})$ are given by
\begin{align}
& p(\bsm{Q}_h)=\mbs{E}_{\boldsymbol{\mathcal{H}}}\left\{\prod_{0\leq a\leq b}^\tau\delta\left(\sum_{mk}\left(h_{mk}^{(a)}\right)^\ast h_{mk}^{(b)}-MK Q_h^{ab}\right)\right\}, \notag
\end{align}

\begin{align}
& p(\bsm{Q}_g)=\mbs{E}_{\boldsymbol{\mathcal{G}}}\left\{\prod_{0\leq a\leq b}^\tau\delta\left(\sum_{mn}\left(g_{mn}^{(a)}\right)^\ast g_{mn}^{(b)}-MN Q_g^{ab}\right)\right\}, \notag \\
& p(\bsm{Q}_f)=\mbs{E}_{\boldsymbol{\mathcal{F}}}\left\{\prod_{0\leq a\leq b}^\tau\delta\left(\sum_{nk}\left(f_{nk}^{(a)}\right)^\ast f_{nk}^{(b)}-NK Q_f^{ab}\right)\right\}, \notag \\
&p(\bsm{Q}_{x_\mathsf{p}},\bsm{Q}_{x_\mathsf{d}})=\mbs{E}_{\boldsymbol{\mathcal{X}}_\mathsf{p},\boldsymbol{\mathcal{X}}_\mathsf{d}}\Bigg\{\prod_{\mathsf{o}\in\{\mathsf{p},\mathsf{d}\}}\prod_{0\leq a\leq b}^\tau \notag \\
& \quad \quad \quad \quad \quad \quad~ \times \delta\Bigg(\sum_{kt}\left(x_{\mathsf{o},kt}^{(a)}\right)^\ast x_{\mathsf{o},kt}^{(b)}- KT_\mathsf{o} Q_{x_\mathsf{o}}^{ab} \Bigg) \Bigg\},  \notag
\end{align}
\begin{align}
& p(\bsm{Q}_{c_\mathsf{p}},\bsm{Q}_{c_\mathsf{d}})=\mbs{E}_{\boldsymbol{\mathcal{C}}_\mathsf{p},\boldsymbol{\mathcal{C}}_\mathsf{d}}\Bigg\{ \prod_{\mathsf{o} \in\{\mathsf{p},\mathsf{d}\}}\prod_{0\leq a \leq b}^\tau \notag \\
& \quad \quad \quad \quad \quad \quad~ \times \delta\Bigg(\sum_{nt}\left(c_{\mathsf{o},nt}^{(a)}\right)^\ast  c_{\mathsf{o},nt}^{(b)}-NT_\mathsf{o} Q_{c_\mathsf{o}}^{ab} \Bigg) \Bigg\},  \notag
\end{align}
respectively, where the expectation in $p(\bsm{Q}_{c_\mathsf{p}},\bsm{Q}_{c_\mathsf{d}})$ is with respect to $(\bsm{\mcl{C}}_{\mathsf{p}}, \bsm{\mcl{C}}_{\mathsf{d}})$ conditioned on $\bsm{Q}_f$, $\bsm{Q}_{x_{\mathsf{p}}}$, and $\bsm{Q}_{x_{\mathsf{d}}}$.

We next show that according to the G$\ddot{\text{a}}$rtner-Ellis Theorem \cite[Chapter 2.3]{touchette2011basic}, the probability measures $p(\bsm{Q}_h)$, $p(\bsm{Q}_g)$, $p(\bsm{Q}_f)$, $p(\bsm{Q}_{x_\mathsf{p}},\bsm{Q}_{x_\mathsf{d}})$, and $p(\bsm{Q}_{c_\mathsf{p}},\bsm{Q}_{c_\mathsf{d}})$ satisfy the large derivation theory (LDT). We start with the calculation of the scaled cumulant generating function of $\bsm{Q}_h$:
\begin{align}
\lambda(\widetilde{\bsm{Q}}_h) & \triangleq \lim_{K \to \infty}\frac{1}{MK} \ln \mbs{E}_{\boldsymbol{\mathcal{H}}} \left\{e^{\tr(\widetilde{\bsm{Q}}_h\boldsymbol{\mathcal{H}}^\mathsf{H} \boldsymbol{\mathcal{H}})}\right\} \notag \\
& = \lim_{K \to \infty}\frac{1}{MK} \ln \left( \prod_{m} \prod_{k} \mbs{E}_{\bsm{h}_{mk}} e^{\bsm{h}^{\mathsf{H}}_{mk} \widetilde{\bsm{Q}}_h \bsm{h}_{mk}}  \right) \notag \\
& \overset{(a)}{=} \ln \mbs{E}_{\boldsymbol{h}} \left\{e^{\bsm{h}^{\mathsf{H}}\widetilde{\bsm{Q}}_h \bsm{h}} \right\}, \label{lambda-Q-h}
\end{align}
where $\widetilde{\bsm{Q}}_h \in \mbs{C}^{(\tau+1) \times (\tau+1)}$
is the dual variable of $\bsm{Q}_h$, $\bsm{h}_{mk} = \big[h^{(0)}_{mk},\ldots,h^{(\tau+1)}_{mk}\big]^{\mathsf{T}}$, $(a)$ comes by the i.i.d. assumption of the elements of $\bsm{H}$, and $\bsm{h} \triangleq [h^{(0)},\ldots,h^{(\tau+1)}]^{\mathsf{T}}$, $h^{(a)} \sim p(h)$. It follows that  $\lambda(\widetilde{\bsm{Q}}_h)$ given in \eqref{lambda-Q-h} is continuous and differentiable with respect to $\widetilde{\bsm{Q}}_h$ and the LDT can be applied and $p(\bsm{Q}_h)$ can be expressed in an exponential formula as follows
\begin{align}
p(\bsm{Q}_h) & = e^{-MK \mcl{R}(\bsm{Q}_h) + \mcl{O}(1)}, \label{LDPh}
\end{align}
where $\mcl{R}(\bsm{Q}_f)$ is called the rate function of the scaled cumulant generating function $\lambda(\widetilde{\bsm{Q}}_h)$ and given by the Legendre-Fenchel transform \cite[eqs. (129) and (130)]{tanaka2002statistical}:
\begin{align}
\mcl{R}(\bsm{Q}_h)=\inf_{\widetilde{\bsm{Q}}_h}\left\{ \mcl{I}_h \right\},
\end{align}
where
\begin{align}
\mathcal{I}_h = &\lambda(\widetilde{\bsm{Q}}_h)-\tr(\widetilde{\bsm{Q}}_h\bsm{Q}_h). \label{temp1150_H}
\end{align}
Similarly, we can also apply the LDT to the probability measures $p(\bsm{Q}_g)$, $p(\bsm{Q}_f)$, $p(\bsm{Q}_{x_\mathsf{p}},\bsm{Q}_{x_\mathsf{d}})$ and $p(\bsm{Q}_{c_\mathsf{p}},\bsm{Q}_{c_\mathsf{d}})$, which are expressed  in exponential formulas as
\begin{align}
p(\bsm{Q}_g) & = e^{-MN \mcl{R}(\bsm{Q}_g) + \mcl{O}(1)}, \label{LDPg}\\
p(\bsm{Q}_f) & = e^{-NK \mcl{R}(\bsm{Q}_f) + \mcl{O}(1)}, \label{LDPf} \end{align}
\begin{align}
p(\bsm{Q}_{x_\mathsf{p}},\bsm{Q}_{x_\mathsf{d}}) &= e^{-KT \mcl{R}(\bsm{Q}_{x_\mathsf{p}},\bsm{Q}_{x_\mathsf{d}}) + \mcl{O}(1)}, \label{LDPx} \\
p(\bsm{Q}_{c_\mathsf{p}},\bsm{Q}_{c_\mathsf{d}}) & = e^{-NT \mcl{R}(\bsm{Q}_{c_\mathsf{p}},\bsm{Q}_{c_\mathsf{d}}) + \mcl{O}(1)}. \label{LDPc}
\end{align}
Herein, $\mcl{R}(\bsm{Q}_g)$, $\mcl{R}(\bsm{Q}_f)$, and $\mcl{R}(\bsm{Q}_{x_\mathsf{p}},\bsm{Q}_{x_\mathsf{d}})$ are the corresponding rate functions and given by
\begin{align}
\mcl{R}(\bsm{Q}_g) & =\inf_{\widetilde{\bsm{Q}}_g}\left\{ \mcl{I}_g \right\}, \\
\mcl{R}(\bsm{Q}_f) & =\inf_{\widetilde{\bsm{Q}}_f}\left\{ \mcl{I}_f \right\}, \\
\mcl{R}(\bsm{Q}_{x_\mathsf{p}},\bsm{Q}_{x_\mathsf{d}}) & =
\inf_{\widetilde{\bsm{Q}}_{x_\mathsf{p}}, \widetilde{\bsm{Q}}_{x_\mathsf{d}} } \left\{ \mcl{I}_x \right\}, \\
\mcl{R}(\bsm{Q}_{c_\mathsf{p}},\bsm{Q}_{c_\mathsf{d}}) & =
\inf_{\widetilde{\bsm{Q}}_{c_\mathsf{p}}, \widetilde{\bsm{Q}}_{c_\mathsf{d}} } \left\{ \mcl{I}_c \right\},
\end{align}
respectively, where
\begin{align} \mathcal{I}_g=&\lambda(\widetilde{\bsm{Q}}_g)-\tr(\widetilde{\bsm{Q}}_g\bsm{Q}_g),\label{SGSF-g}\\
\mathcal{I}_f=&\lambda(\widetilde{\bsm{Q}}_f)-\tr(\widetilde{\bsm{Q}}_f\bsm{Q}_f),\label{SGSF-f}\\
\mathcal{I}_x=&\sum_{\mathsf{o}\in\{\mathsf{p},\mathsf{d}\}}  \left(\lambda(\widetilde{\bsm{Q}}_{x_\mathsf{o}})-
\tr(\widetilde{\bsm{Q}}_{x_\mathsf{o}}\bsm{Q}_{x_\mathsf{o}})\right), \label{SGSF-x} \\
\mathcal{I}_c=&\sum_{\mathsf{o} \in\{\mathsf{p},\mathsf{d}\}} \left(   \lambda(\widetilde{\bsm{Q}}_{c_{\mathsf{o}}})-\tr(\widetilde{\bsm{Q}}_{c_\mathsf{o}}\bsm{Q}_{c_\mathsf{o}})\right).
\label{temp118}
\end{align}
In \eqref{SGSF-g}--\eqref{temp118}, $\lambda(\widetilde{\bsm{Q}}_g)$,  $\lambda(\widetilde{\bsm{Q}}_f)$, $\lambda(\widetilde{\bsm{Q}}_{x_\mathsf{o}})$, and $\lambda(\widetilde{\bsm{Q}}_{c_{\mathsf{o}}})$ are the scaled cumulant generating functions of $\widetilde{\bsm{Q}}_g$, $\widetilde{\bsm{Q}}_f$, $\widetilde{\bsm{Q}}_{x_{\mathsf{o}}}$,  $\widetilde{\bsm{Q}}_{c_\mathsf{o}}$, and given by
\begin{align}
\lambda(\widetilde{\bsm{Q}}_g)
& = \ln \mbs{E}_{\boldsymbol{g}} \left\{e^{\bsm{g}^{\mathsf{H}}\widetilde{\bsm{Q}}_g \bsm{g}}\right\}, \\
\lambda(\widetilde{\bsm{Q}}_f)
& = \ln \mbs{E}_{\boldsymbol{f}} \left\{e^{\bsm{f}^{\mathsf{H}}\widetilde{\bsm{Q}}_f \bsm{f}}\right\}, \\
\lambda(\widetilde{\bsm{Q}}_{x_\mathsf{o}})
& = \ln \mbs{E}_{\boldsymbol{x}_{\mathsf{o}}} \left\{e^{\bsm{x}_{\mathsf{o}}^{\mathsf{H}}\widetilde{\bsm{Q}}_h \bsm{x}_{\mathsf{o}}}\right\}, \\
\lambda(\widetilde{\bsm{Q}}_{c_{\mathsf{o}}}) & = \rho \ln \left( \mbs{E}_{\boldsymbol{c}_{\mathsf{o}}|\bsm{Q}_{c_\mathsf{o}}} \left\{
e^{\tr(\boldsymbol{c}_{\mathsf{o}} \widetilde{\bsm{Q}}_{c_\mathsf{o}}
\boldsymbol{c}_{\mathsf{o}}) } \right\} \right), \label{SGSF-Q-C2}
\end{align}
respectively, where $\mathsf{o}\in\{\mathsf{p},\mathsf{d}\}$, $\bsm{g} \triangleq [g^{(0)},\ldots,g^{(\tau+1)}]^{\mathsf{T}}$, $g^{(a)} \sim p(g)$, $\bsm{f} \triangleq [f^{(0)},\ldots,f^{(\tau+1)}]^{\mathsf{T}}$, $f^{(a)} \sim p(f)$, $\bsm{x}_{\mathsf{o}} \triangleq [x^{(0)}_{\mathsf{o}},\ldots,x^{(\tau+1)}_{\mathsf{o}}]^{\mathsf{T}}$, $x_{\mathsf{o}}^{(a)} \sim p(x_{\mathsf{o}})$, and in \eqref{SGSF-Q-C2} we have additionally utilized the fact that $\rho N T_{\mathsf{o}}$ elements of $\bsm{C}_{\mathsf{o}}$ are non-zero.

By plugging \eqref{LDPh}, \eqref{LDPg}--\eqref{LDPx}, and \eqref{LDPc} into \eqref{EY1}, it then follows from the Varadhan's theorem \cite[Chapter 2.4]{touchette2011basic} and \cite[Appendix I]{tanaka2002statistical} that the integral in  \eqref{E-Y12} is dominated by the maximum argument of the exponential functions, i.e.,
\begin{align} \label{A006}
\lim_{K\to \infty} \frac{1}{K^2} \ln \mbs{E}_{\bsm{Y}}\left\{ p^\tau (\Yv) \right\} & = \sup_{\{\bsm{Q}_i\}} \inf_{\{\widetilde{\bsm{Q}}_i\}} \mathcal{F}(\tau) \notag \\
& = \underset{\{\bsm{Q}_i,\widetilde{\bsm{Q}}_i\}}{\extr} \mathcal{F}(\tau),
\end{align}
where $i \in \{h,g,f,x_\mathsf{p},x_\mathsf{d},c_\mathsf{p},c_\mathsf{d}\}$, `extr' denotes the operation of extremization,
\begin{align}
\mathcal{F}(\tau) \triangleq & \left( \frac{M T_{\mathsf{o}}}{K^2}\mathcal{I}_z+\frac{M}{K}\mathcal{I}_h+\frac{MN}{K^2}\mathcal{I}_g \right. \notag \\
& \, \left. +\frac{N}{M}\mathcal{I}_f+\frac{T_{\mathsf{o}}}{K}\mathcal{I}_x+\frac{N T_{\mathsf{o}}}{K^2}\mathcal{I}_c\right), \label{F-tau}
\end{align}
and
\begin{align}
\mathcal{I}_z = \sum_{\mathsf{o} \in\{\mathsf{p},\mathsf{d}\}}
\ln\mbs{E}_{\{z_\mathsf{o}^{(a)}\}|\bsm{Q}_z} \left\{ \int \mathrm{d}y\prod_{a=0}^n\CN\left(y;z_\mathsf{o}^{(a)},\sigma^2\right)  \right\}, \label{I-Z}
\end{align}

With \eqref{A006}, the average free energy $\mcl{F}$ in \eqref{FreeE2} becomes
\begin{align}\label{FreeE3}
\mathcal{F}= \lim_{\tau\to 0}\frac{\partial}{\partial \tau} \underset{\{\bsm{Q}_i,\widetilde{\bsm{Q}}_i\}}{\extr} \mathcal{F}(\tau),
\end{align}
which implies that the key is to obtain an analytic expression of \eqref{A006}. However, it is still prohibitive to get explicit expressions about the saddle points by setting zero gradient of $\mcl{F}(\tau)$ with respect to $\{\bsm{Q}_i,\widetilde{\bsm{Q}}_i\}$. For the sake of analytical tractability, we adopt the following replica symmetric ansatz:
\begin{asum} \label{assum2}
The saddle points of $\{\bsm{Q}_i,\widetilde{\bsm{Q}}_i\}$ of $\mcl{F}(\tau)$ follow the replica symmetric forms, i.e.,
\begin{align}
\bsm{Q}_i & = (q_i - m_i ) \bsm{I} + m_i \bsm{1} \bsm{1}^{\mathsf{T}} \label{A06} \\
\widetilde{\bsm{Q}}_i & = (\widetilde{q}_i - \widetilde{m}_i ) \bsm{I} + \widetilde{m}_i \bsm{1} \bsm{1}^{\mathsf{T}} \label{B02}
\end{align}
\end{asum}
where $i \in \{h,g,f,x_\mathsf{p},x_\mathsf{d},c_\mathsf{p},c_\mathsf{d}\}$, $\bsm{I}$ and $\bsm{1}$ denotes the $(\tau+1) \times (\tau+1)$ identity matrix and the $(\tau+1)$ column vector with all elements being $1$, respectively.

Substituting \eqref{A06} and \eqref{B02} into \eqref{F-tau}, we have
 \begin{align}\label{A07}
\mathcal{F}= \lim_{\tau\to 0}\frac{\partial}{\partial \tau}~
\underset{\{q_i,m_i,\widetilde q_i,\widetilde m_i\}}{\extr}
\mathcal{F}(\tau),
\end{align}
where $i \in \{h,g,f,x_\mathsf{p},x_\mathsf{d},c_\mathsf{p},c_\mathsf{d}\}$.  As such, the extermination with respect to $\{\bsm{Q}_i,\widetilde{\bsm{Q}}_i\}$ in \eqref{FreeE3} is reduced to those with respect to $\{q_i,m_i,\widetilde q_i,\widetilde m_i\}$ in \eqref{A07}.

Now, we compute the final expression of $\mcl{F}(\tau)$ in \eqref{F-tau} under Assumption \ref{assum2}. It suffices to simplify $\mcl{I}_z$, $\mcl{I}_h$, $\mcl{I}_g$, $\mcl{I}_f$, $\mcl{I}_x$, and $\mcl{I}_c$. By plugging \eqref{A06} and \eqref{B02} into \eqref{I-Z} and using the complex Hubbard-Stratonovich transform\footnote{For $\eta \in \mbs{C}$ and $x \in \mbs{C}$, $e^{|x|^2} =  \frac{1}{\pi} \int e^{-|\eta|^2 + 2 {\rm Re}(x \eta^\ast)} {\rm d} \eta$}, we simplify $\mathcal{I}_z$ in \eqref{I-Z} as
\begin{align}\label{A10}
\mathcal{I}_z=2M\!\! \sum_{\mathsf{o}\in\{\mathsf{p},\mathsf{d}\}} \!\! T_\mathsf{o}\ln\mbs{E}_{v,u^\prime,y} \big\{  \mbs{E}^\tau_u\left\{\Norm(y;c_{\mathsf{o},1} u+c_{\mathsf{o},2}v;\sigma^2)\right\} \big\},
\end{align}
where $c_{\mathsf{o},1}=\sqrt{q_{z_\mathsf{o}}-m_{z_\mathsf{o}}}$, $c_{\mathsf{o},2}=\sqrt{m_{z_\mathsf{o}}}$, $y\sim \Norm(y;c_{\mathsf{o},1} u^\prime+c_{\mathsf{o},2} v,\sigma^2)$, $v,u^\prime,u\sim\Norm(\cdot;0,1)$. Similarly, the expression of $\mathcal{I}_h$ in \eqref{temp1150_H} can be simplified as \eqref{temp12}, shown at the bottom of this page,
\begin{figure*}[b]
\hrulefill
\begin{align}
& \mathcal{I}_h =
\mbs{E}_{h^{\prime}}\left\{\int\mathrm{d}{y_h}e^{-|y_h-\sqrt{\widetilde m_h}h^{\prime}|^2}  \left( \mbs{E}_{h}\left[e^{-\widetilde m_h|{h}|^2+2\sqrt{\widetilde m_h}\Re(y_h^\ast h)}\right]\right)^\tau \right\} - (\tau+1) q_h\widetilde q_h - (\tau^2+\tau) \widetilde{m}_h m_h, \label{temp12} \\
& \mathcal{I}_g =
\mbs{E}_{g^{\prime}}\left\{\int\mathrm{d}{y_g}e^{-|y_g-\sqrt{\widetilde m_g}g^{\prime}|^2}  \left( \mbs{E}_{g}\left[e^{-\widetilde m_g|{g}|^2+2\sqrt{\widetilde m_g}\Re(y_g^\ast g)}\right]\right)^\tau \right\} - (\tau+1) q_g\widetilde q_g - (\tau^2+\tau) \widetilde{m}_g m_g, \label{temp13} \\
& \mathcal{I}_f =
\mbs{E}_{f^{\prime}}\left\{\int\mathrm{d}{y_f}e^{-|y_f-\sqrt{\widetilde m_f}f^{\prime}|^2}  \left( \mbs{E}_{f}\left[e^{-\widetilde m_f|{f}|^2+2\sqrt{\widetilde m_f}\Re(y_f^\ast f)}\right]\right)^\tau \right\} - (\tau+1) q_f\widetilde q_f - (\tau^2+\tau) \widetilde{m}_f m_f, \label{temp14} \\
& \mathcal{I}_{x_{\mathsf{o}}} = \sum_{\mathsf{o} \in \{\mathsf{p},\mathsf{d}\}} \left(
\mbs{E}_{{x_{\mathsf{o}}}^{\prime}}\left\{\int\mathrm{d}{y_{x_{\mathsf{o}}}}e^{-|y_{x_{\mathsf{o}}}-\sqrt{\widetilde m_{x_{\mathsf{o}}}}{x_{\mathsf{o}}}^{\prime}|^2}  \left( \mbs{E}_{{x_{\mathsf{o}}}}\left[e^{-\widetilde m_{x_{\mathsf{o}}}|{{x_{\mathsf{o}}}}|^2+2\sqrt{\widetilde m_{x_{\mathsf{o}}}}\Re(y_{x_{\mathsf{o}}}^\ast {x_{\mathsf{o}}})}\right]\right)^\tau \right\} - (\tau+1) q_{x_{\mathsf{o}}}\widetilde q_{x_{\mathsf{o}}} - (\tau^2+\tau) \widetilde{m}_{x_{\mathsf{o}}} m_{x_{\mathsf{o}}}  \right), \label{temp16} \\
& \mathcal{I}_{c_{\mathsf{o}}} = \sum_{\mathsf{o} \in \{\mathsf{p},\mathsf{d}\}} \left(
\mbs{E}_{{c_{\mathsf{o}}}^{\prime}}\left\{\int\mathrm{d}{y_{c_{\mathsf{o}}}}e^{-|y_{c_{\mathsf{o}}}-\sqrt{\widetilde m_{c_{\mathsf{o}}}}{c_{\mathsf{o}}}^{\prime}|^2}  \left( \mbs{E}_{{c_{\mathsf{o}}}}\left[e^{-\widetilde m_{c_{\mathsf{o}}}|{{x_{\mathsf{o}}}}|^2+2\sqrt{\widetilde m_{c_{\mathsf{o}}}}\Re(y_{c_{\mathsf{o}}}^\ast {c_{\mathsf{o}}})}\right]\right)^\tau \right\} - (\tau+1) q_{c_{\mathsf{o}}}\widetilde q_{c_{\mathsf{o}}} - (\tau^2+\tau) \widetilde{m}_{c_{\mathsf{o}}} m_{c_{\mathsf{o}}}  \right). \label{temp116}
\end{align}
\vspace{-0.4cm}
\begin{align}
&\mathcal{F}=\frac{2M}{K^2}\sum_{\mathsf{o}\in\{\mathsf{p},\mathsf{d}\}}T_\mathsf{o}\ln\left(\mbs{E}_{v,u^\prime} \left[\int \mathrm{d}y \Norm(y;\sqrt{q_{z_\mathsf{o}}-m_{z_\mathsf{o}}}u^\prime+\sqrt{m_{z_\mathsf{o}}}v;\sigma^2)\ln \mbs{E}_u\left\{\Norm(y;\sqrt{q_{z_\mathsf{o}}-m_{z_\mathsf{o}}}u+\sqrt{m_{z_\mathsf{o}}}v;\sigma^2)\right\} \right] \right)\nonumber\\
&~~~~~~~+\frac{M}{K}\left(\mbs{E}_{h^{\prime}}\left\{\int\mathrm{d}{y_h}e^{-|y_h-\sqrt{\widetilde m_h}h^{\prime}|^2}\ln \mbs{E}_{h}\left[e^{-\widetilde m_h|{h}|^2+2\sqrt{\widetilde m_h}\Re(y_h^\ast h)}\right]\right\}-\widetilde m_h m_h\right)\nonumber \\
&~~~~~~~+\frac{MN}{K^2}\left(\mbs{E}_{g^{\prime}}\left\{\int\mathrm{d}{y_g}e^{-|y_g-\sqrt{\widetilde m_g}g^{\prime}|^2}\ln \mbs{E}_{g}\left[e^{-\widetilde m_g|{g}|^2+2\sqrt{\widetilde m_g}\Re(y_g^\ast g)}\right]\right\}-\widetilde m_g m_g\right)\nonumber\\
&~~~~~~~+\frac{1}{K}\sum_{\mathsf{o}\in\{\mathsf{p},\mathsf{d}\}}T_\mathsf{o}\left(\mbs{E}_{x_\mathsf{o}^\prime}\left\{\int\mathrm{d}{y_{x_\mathsf{o}}}e^{-|y_{x_\mathsf{o}}-\sqrt{\widetilde m_{x_\mathsf{o}}}x_\mathsf{o}^\prime|^2}\ln \mbs{E}_{x_\mathsf{o}}\left[e^{-\widetilde m_{x_\mathsf{o}}|{x_\mathsf{o}}|^2+2\sqrt{\widetilde m_{x_\mathsf{o}}}\Re(y_{x_\mathsf{o}}^\ast x_\mathsf{o})}\right]\right\}-\widetilde m_{x_\mathsf{o}} m_{x_\mathsf{o}}\right)\nonumber\\
&~~~~~~~+\frac{N}{K^2}\sum_{\mathsf{o}\in\{\mathsf{p},\mathsf{d}\}}T_\mathsf{o}\left(\mbs{E}_{c_o^\prime}\left[\int\mathrm{d}{y_{c_o}}e^{-|y_{c_o}-\sqrt{\widetilde m_{c_\mathsf{o}}}c_\mathsf{o}^\prime|^2}\ln \mbs{E}_{c_\mathsf{o}}\left\{e^{-\widetilde m_{c_\mathsf{o}}|{c_\mathsf{o}}|^2+2\sqrt{\widetilde m_{c_\mathsf{o}}}\Re(y_{c_\mathsf{o}}^\ast c_o)}\right\}\right]-\widetilde m_{c_\mathsf{o}} m_{c_\mathsf{o}}\right). \label{A009}
\end{align}
\end{figure*}
where $h,h^\prime \sim p(h_{mk})$. The simplified expressions of $\mathcal{I}_g$, $\mathcal{I}_f$, $\mathcal{I}_x$, and $\mathcal{I}_c$ under Assumption \ref{assum2} are similar to the derivation of \eqref{temp12} and are given in \eqref{temp13}, \eqref{temp14},  \eqref{temp16}, and \eqref{temp116}, respectively.

With \eqref{A07} and \eqref{A10}, we compute the extremum of $\mathcal{F}(\tau)$ with respect to $\{q_i,m_i,\widetilde q_i,\widetilde m_i\}$, which can be obtained by setting the partial derivatives of $\mathcal{F}(\tau)$ to zero. By doing so, we find that the extreme point $\{m_i,\widetilde m_i\}$ satisfies \eqref{B04}--\eqref{B01}. Furthermore, noting that $\lim_{\tau\to 0}\mbs{E}_{\bsm{Y}}\{p^\tau(\bsm{Y})\}=1$, we find $\widetilde q_i=0$ and $q_i$ satisfies \eqref{temp120}.

Under Assumption \ref{assum3}, by taking the derivative of $\extr \mathcal{F}(\tau)$ with respect to $\tau$ and letting $\tau \to 0$, we obtain the analytical expression of $\mathcal{F}$ in \eqref{A009},
where $h,h^\prime, v,u,u^\prime,$ are defined in and \eqref{A10} and  \eqref{temp12}; $g,g^\prime \sim p(g_{mn})$; $f,f^\prime \sim p(f_{nk})$; $x_\mathsf{o},x_\mathsf{o}^\prime \sim p(x_{\mathsf{o},kt})$. This concludes the proof.
}

\renewcommand\refname{References}
\bibliographystyle{IEEEtran}
\bibliography{bib-he,ref}
\end{document}

%% file: defns.tex
\definecolor{Red}{rgb}{1,0,0}
\definecolor{Blue}{rgb}{0,0,1}
\definecolor{Olive}{rgb}{0.41,0.55,0.13}
\definecolor{Green}{rgb}{0,1,0}
\definecolor{MGreen}{rgb}{0,0.8,0}
\definecolor{DGreen}{rgb}{0,0.55,0}
\definecolor{Yellow}{rgb}{1,1,0}
\definecolor{Cyan}{rgb}{0,1,1}
\definecolor{Magenta}{rgb}{1,0,1}
\definecolor{Orange}{rgb}{1,.5,0}
\definecolor{Violet}{rgb}{.5,0,.5}
\definecolor{Purple}{rgb}{.75,0,.25}
\definecolor{Brown}{rgb}{.75,.5,.25}
\definecolor{Grey}{rgb}{.5,.5,.5}

\def\tr{\mathop{\rm tr}\nolimits}%
\def\extr{\mathop{\rm extr}\nolimits}%
\def\vect{\mathop{\rm vec}\nolimits}%
\def\Re{\mathop{\rm Re}\nolimits}%
\newcommand{\Yv}{{\bf Y}}

\def\de \mathrm{d}

\newcommand{\Norm}{\mathcal{N}}
\newcommand{\CN}{\mathcal{CN}}

\def\textiid{i.i.d.\@\xspace}
\newcommand\iid{\ifmmode\text{ i.i.d. } \else \textiid \fi}

\newcommand{\beqs}{\begin{equation*}}
\newcommand{\eeqs}{\end{equation*}}
\newcommand{\beq}{\begin{equation}}
\newcommand{\eeq}{\end{equation}}